\documentclass[longauth]{aa}
\usepackage{subcaption}
\usepackage{graphicx}
\usepackage{xcolor}

\usepackage{txfonts}
\usepackage{hyperref}
\usepackage{xspace}
\hypersetup{
    colorlinks=true,
    linkcolor=blue,
    citecolor=blue,
    filecolor=magenta,      
    urlcolor=cyan,
    pdftitle={OASIS: Mini-quenched and lulling galaxies},
    pdfpagemode=FullScreen,
    }
\usepackage{lastpage}
\makeatletter
\renewcommand*\aa@pageof{, page \thepage{} of \pageref*{LastPage}}
\makeatother

\newcommand{\nii}{[N\,\textsc{ii}]}

\newcommand{\oii}{[O\,\textsc{ii}]}
\newcommand{\oiii}{[O\,\textsc{iii}]}

\newcommand{\mstar}{\ensuremath{\text{M}_\star}\xspace}

\newcommand{\ssim}{\ensuremath{\sim\!}\xspace}

\newcommand{\Mstar}{\ensuremath{M_\star}\xspace}

\newcommand{\MSun}{\ensuremath{{\rm M}_\odot}\xspace}

\let\oldAA\AA
\renewcommand{\AA}{\oldAA\xspace}

\newcommand{\Halpha}{H$\alpha$\xspace}
\newcommand{\Hbeta}{H$\beta$\xspace}

\newcommand{\jwst}{\textit{JWST}\xspace}

\newcommand{\ppxf}{{\sc ppxf}\xspace}

\newcommand{\prospector}{{\sc prospector}\xspace}

\begin{document}

   \title{OASIS I: Spectroscopy of Mini-Quenched and Lulling Galaxies Constrains the Quiet Side of Bursty Star Formation at z > 4}

    \titlerunning{OASIS I: Mini-quenched and lulling galaxies}

\author{
Tobias J.\ Looser\inst{1}\fnmsep\thanks{tobias.looser@cfa.harvard.edu}
\and Francesco D'Eugenio\inst{2,3}
\and William M.\ Baker\inst{4}
\and Santiago Arribas\inst{5}
\and Andrew J.\ Bunker\inst{6}
\and Stefano Carniani\inst{7}
\and St\'ephane Charlot\inst{8}
\and Jacopo Chevallard\inst{6}
\and Mirko Curti\inst{9}
\and Qiao Duan\inst{2,3}
\and Kevin Hainline\inst{10}
\and Jakob M.\ Helton\inst{11}
\and Yuki Isobe\inst{2,3,12}
\and Benjamin D.\ Johnson\inst{1}
\and Pierluigi Rinaldi\inst{13,14}
\and Brant Robertson\inst{15}
\and Fengwu Sun\inst{1}
\and Sandro Tacchella\inst{2,3}
\and James A.\ A.\ Trussler\inst{1}
\and Hannah \"Ubler\inst{16}
\and Joris Witstok\inst{17,18}
\and Zihao Wu\inst{1}
}

\institute{
Center for Astrophysics $|$ Harvard \& Smithsonian, 60 Garden St., Cambridge MA 02138, USA
\and Kavli Institute for Cosmology, University of Cambridge, Madingley Road, Cambridge CB3 0HA, UK
\and Cavendish Laboratory, University of Cambridge, 19 JJ Thomson Avenue, Cambridge CB3 0HE, UK
\and DARK, Niels Bohr Institute, University of Copenhagen, Jagtvej 155A, DK-2200 Copenhagen, Denmark
\and Centro de Astrobiolog\'{\i}a (CAB), CSIC--INTA, Cra.\ de Ajalvir Km.~4, 28850 Torrej\'on de Ardoz, Madrid, Spain
\and Department of Physics, University of Oxford, Denys Wilkinson Building, Keble Road, Oxford OX1 3RH, UK
\and Scuola Normale Superiore, Piazza dei Cavalieri 7, I-56126 Pisa, Italy
\and Sorbonne Universit\'e, CNRS, UMR 7095, Institut d'Astrophysique de Paris, 98 bis bd Arago, 75014 Paris, France
\and INAF -- Osservatorio di Astrofisica e Scienza dello Spazio, Via P.\ Gobetti 93/3, I-40129 Bologna, Italy
\and Steward Observatory, University of Arizona, 933 N.\ Cherry Avenue, Tucson, AZ 85721, USA
\and Department of Astronomy \& Astrophysics, The Pennsylvania State University, University Park, PA 16802, USA
\and Waseda Research Institute for Science and Engineering, Faculty of Science and Engineering, Waseda University, 3-4-1 Okubo, Shinjuku, Tokyo 169-8555, Japan
\and Department of Astronomy, The University of Texas at Austin, Austin, TX 78712, USA
\and Cosmic Frontier Center, The University of Texas at Austin, Austin, TX 78712, USA
\and Department of Astronomy and Astrophysics, University of California, Santa Cruz, 1156 High Street, Santa Cruz, CA 95064, USA
\and Max-Planck-Institut f\"ur extraterrestrische Physik (MPE), Gie{\ss}enbachstra{\ss}e 1, 85748 Garching, Germany
\and Cosmic Dawn Center (DAWN), Copenhagen, Denmark
\and Niels Bohr Institute, University of Copenhagen, Jagtvej 128, DK-2200 Copenhagen, Denmark
}

\authorrunning{T.\ J.\ Looser et al.}
   \date{}

\abstract
{JWST has revealed that star formation in early galaxies is bursty, including episodes of violent suppression, so-called mini-quenching. But these were largely serendipitous discoveries in samples of UV-bright, emission-line-dominated galaxies, so that the quiet side of the burst cycle has remained poorly explored. With the JWST Cycle~3 program OASIS (GO~5997), we searched for such galaxies deliberately, using deep 28-hour NIRSpec/PRISM spectroscopy across the full burst cycle at $z>4$. We identify three mini-quenched galaxies with $\log_{10}(M_\star/M_\odot)\simeq8.3$--8.9 and no detectable Balmer emission, at least 1.2--1.7\,dex below the star-forming main sequence (SFMS), five lulling galaxies with weak Balmer emission above the quenching threshold, and 17 nearly-main-sequence galaxies just below the SFMS. Finding them required spectroscopy: seven of our ten photometric candidates for suppressed (mini-quenched and lulling) galaxies turned out to be nearly-MS or above the SFMS, and photometric fits overestimate their current star-formation rates and cannot distinguish lulling from mini-quenched ones. The suppressed galaxies are common: five of the eight come from tiers not selected on star-formation state, where they make up 29 per cent of the galaxies at $z\approx4$--5.5 and $\log_{10}(M_\star/M_\odot)=8$--9.5, a number density of ${\sim}1.6\times10^{-3}$\,Mpc$^{-3}$\,dex$^{-1}$. Simulations predict them to be an order of magnitude rarer at these masses and to occur mainly at lower masses. The suppressed galaxies are also old: Balmer-break strength and stellar age both increase as galaxies fall further below the SFMS, with light-weighted ages of 5\,Myr for galaxies above it, ${\sim}30$\,Myr for the lulling galaxies and above 100\,Myr for the mini-quenched ones. They show little stellar reddening ($E(B-V)_\star\lesssim0.2$), consistent with depleted gas reservoirs. Suppressed phases at $z>4$ are therefore neither rare nor brief, but a common and lasting part of early galaxy growth.}
 

   \maketitle\nolinenumbers

\section{Introduction} \label{sec:intro}
The first years of \jwst\ operations \citep{Gardner+2023,Rigby+2023} have transformed our view of galaxy formation in the early Universe (redshift $z>4$). Before JWST, the stellar mass assembly of galaxies was largely observed as a steady, secular growth process. However, JWST has revealed that early galaxies instead evolve through short, feedback-regulated episodes of intense, bursty star formation driven by rapid, stochastic processes. Multiple observational lines of evidence now support this emerging picture:
(i) large scatter around the star-forming main sequence in the star-formation rate--stellar mass (SFR--\mstar) plane \citep[e.g.,][]{Clarke+2025,Cole+2025,Simmonds+2025,Rinaldi+2025} and in the rest-frame ultraviolet magnitude--stellar mass $M_{\mathrm{UV}}$-$\Mstar$ relation \citep{Munoz+2026};
(ii) signatures of stochasticity in reconstructed star-formation histories (SFHs) \citep{Tacchella+2023,Looser+2025,Dressler+2023b,Ciesla+2024,Carvajal-Bohorquez+2025,Navarro-Carrera+2026};
(iii) deviations from the gas-phase fundamental metallicity relation \citep[e.g.,][]{Curti+2024, Langeroodi+2023,Faisst+2025};
(iv) unexpectedly bright systems relative to pre-\jwst\ UV luminosity function predictions at the highest redshifts\footnote{We note, however, that multiple mechanisms likely contribute simultaneously, and bursty star formation alone appears insufficient to account for the excess \citep{Gelli+2023}.} \citep[e.g.,][]{Harikane+2023,Finkelstein+2023,Robertson+2024,Donnan+2024,Whitler+2025,Mitsuhashi+2026,Stiavelli+2026}; (v) galaxies observed during phases of suppressed star formation \citep[e.g.,][]{Looser+2024Nat,Strait2023,Looser+2025,Baker+2025_zap,Witten+2025b,Chen+2026_SPURS_z9p3,Harikane2026}; and (vi) evidence for recent rejuvenation, i.e., strong upturns of star formation following quenched phases, in individual galaxies \citep{Witten+2025a}.
Together, these findings suggest that \emph{burstiness} -- rather than continuous growth -- dominates stellar-mass assembly at early times. However, we have only scratched the surface of this stochastic behavior, and much remains to be understood. 

Galaxies at high redshift continue to define a star-forming main sequence \citep[SFMS; e.g.,][]{Clarke+2025,Cole+2025,Merida+2025,Simmonds+2025,Rinaldi+2025}, but the SFR--\mstar\ relation exhibits a markedly larger scatter than at low redshift, even after accounting for observational uncertainties \citep[e.g.,][but see \citealt{Clarke+2025} for a different view]{Cole+2025}. This broadening is most pronounced when the SFR is probed on short (${\sim}10$\,Myr) timescales \citep[e.g.,][see also \citealt{caplar+tacchella2019,Wan+2025}]{Cole+2025}. This reflects the limited time galaxies spend in ``normal'' SFMS phases \citep{Angles-Alcazar+2017,Faucher-Giguere2018,Tacchella+2020}, as they instead cycle rapidly between starburst states above the SFMS and suppressed phases below it. A natural consequence is that a relatively large subset of these galaxies should be caught in a post-starburst state, with suppressed star formation at the epoch of observation. 

However, early \jwst\ observations have preferentially targeted the brightest ``top end'' of the SFR--\mstar\ plane: compact, UV-bright galaxies with strong emission lines that are easier to detect and prioritize for spectroscopy \citep[e.g.,][]{Caputi+2024,DiCesare+2025,Carvajal-Bohorquez+2025,Cole+2025,Clarke+2025,Navarro-Carrera+2024,Perry+2025}. This observational focus leaves the complementary, off-burst population largely unexplored. Yet off-burst systems are critical for understanding galaxy formation: they reveal their older stellar populations, whereas in starburst galaxies the young stars outshine older populations and obscure the underlying stellar-mass assembly. Moreover, theoretical models diverge most strongly \emph{below} the SFMS \citep[e.g.,][]{Dome2023,Gelli+2023}, making galaxies in decline -- rather than those in active burst phases -- the key to constraining the combination of accretion and feedback physics governing star formation.

Two classes of such off-burst populations have now emerged in the \jwst\ era. (A) \emph{Mini-quenched} galaxies lie below the redshift-dependent quenching threshold (sSFR\,$<0.2/t_\mathrm{obs}$, i.e. ${\approx}0.13$--0.17\,Gyr$^{-1}$ at $z=4$--5; Eq.~\eqref{eq:quenching-threshold}). (B) \emph{Lulling} systems, by contrast, display substantially reduced, but non-zero star-formation rates, placing them ${\geq}0.6$\,dex below the SFMS yet above the quenching threshold. In \jwst\ spectra of typical depth (a few to ${\sim}30$\,h), mini-quenched galaxies typically show no detectable Balmer lines and lulling galaxies only weak ones, but deeper or shallower observations would change their detectability. Both classes typically exhibit Balmer breaks, revealing intermediate-age populations formed in earlier bursts. However, despite their importance, only a small number of lulling and mini-quenched galaxies have been observed spectroscopically to date. Their abundance has so far been estimated from photometry \citep{Trussler+2025,Khullar+2026}, in the latter case calibrated on a small spectroscopic sample.

Throughout, we further use \emph{nearly-MS} for galaxies that lie close to,
but still below, the star-forming main sequence: systems that are neither
clearly suppressed nor above the SFMS.

We make these classes quantitative with two boundaries. The first is the
quenching threshold of Eq.~\eqref{eq:quenching-threshold}, which depends only
on the cosmic time at the redshift of observation and separates
mini-quenched from lulling systems. The second is the distance from the
main sequence, $\Delta_{\rm MS} = \log_{10}\mathrm{SFR} -
\log_{10}\mathrm{SFR_{MS}}(M_\star, z)$ at each galaxy's own mass and redshift: lulling galaxies lie
above the quenching threshold and at $\Delta_{\rm MS} \le -0.6$,
nearly-MS galaxies at $-0.6 < \Delta_{\rm MS} \le 0$, and the remainder are classified as galaxies above the SFMS. The SFMS normalization and intrinsic scatter at $z>4$
in the low-mass regime probed here remain poorly constrained, so the $\Delta_{\rm MS}$ boundary is a convention rather than a physical threshold.

The first mini-quenched galaxy was identified at $z=7.3$ by \citet{Looser+2024Nat}, with a stellar mass of $\log(M_\star/{\rm M_\odot})\approx8.7$. Shortly after, \citet{Strait2023} discovered an even lower-mass object in the lulling regime. Since then, the sample has expanded: \citet{Looser+2025}, \citet{Takahashi+2025} and \citet{Khullar+2025} increased the known population of low-mass lulling systems; \citet{Covelo-Paz+2025} and \citet{Khullar+2025} identified additional low-SFR galaxies spanning a wide mass range, $M_\star \sim 10^{7.6}$--$10^{10.5}\,M_\odot$, thereby filling in the previously unpopulated intermediate-mass regime at $10^{9}$--$10^{10}\,M_\odot$; \citet{Baker+2025_zap} and \citet{Harikane2026} reported suppressed galaxies at $z\approx8.5$ and $z=11$, respectively. Evolved Balmer-break galaxies have also been found in overdense regions at $z>4$ \citep{Morishita+2025,Witten+2025b,Baker+2025}, suggesting an early link between environment and the regulation of star formation.

These low-mass post-starburst systems inhabit a region of parameter space distinct from the massive quiescent galaxies mostly reported at slightly later epochs \citep{Weibel+2025,carnall+2023b,carnall+2023c,Setton+2024_MQG,Baker+2025_quiescent,Baker+2025_quiescent_b,deGraaff+2025,Stevenson+2025,Merlin+2025,Zhang+2025,Turner+2025}, as well as from intermediate-mass quiescent galaxies \citep{Baker+2025, Khullar+2025,Merlin+2025}. Mini-quenched and lulling galaxies have lower stellar masses ($\log M_\star/{\rm M_\odot}\lesssim9$) and their star formation declines on much shorter timescales (tens of Myr), whereas massive systems host older stellar populations and their star formation declines more gradually. Their spectra differ accordingly: post-starburst dwarfs show weaker Balmer breaks, bluer continua, brighter UV emission, and steeper UV slopes, as illustrated for example in Fig.~5 of \citet{Baker+2025}. Whether these galaxies extend a continuous mass-scaling sequence or represent qualitatively distinct objects, perhaps with different quenching pathways (e.g., stellar-feedback-driven versus AGN-driven) remains unclear. 

Further, mini-quenched and lulling galaxies are distinct from the recently identified ``Little Red Dots'' \citep[LRDs; e.g.,][]{Matthee+2024,Killi+2024,Kokorev+2024,deGraaff2025_Rubies}: sources that are unresolved at rest-frame optical wavelengths whose red colors and Balmer breaks arise from rapid black-hole growth embedded in extremely dense gas rather than evolved stellar continua \citep{Inayoshi+2025,Ji+2025,deGraaff+2025_Cliff,Setton+2024_LRD}.

State-of-the-art cosmological simulations have long predicted episodic, bursty star formation in low-mass galaxies at high redshift. Such bursty star-formation cycles, characterized by intense episodes of star formation followed by quiescent intervals, are a generic outcome of modern hydrodynamical cosmological simulations, such as \textsc{Fire} \citep{Faucher-Giguere2018,Ma+2018,Sun+2023}, \textsc{Serra} \citep{Gelli+2023}, \textsc{Thesan-Zoom} \citep{McClymont+2025}, \textsc{Megatron} \citep{Katz+2025}, the \textsc{Eagle}-derived \textsc{Flares} simulations \citep{Lovell+2023}, \textsc{Vela} \citep{Tacchella+2016}, \textsc{FirstLight} \citep{Ceverino+2018}, or \textsc{IllustrisTNG} \citep{Dome2023,Gui+2025}.

Yet, these models make very different quantitative predictions, reflecting substantial variation in their feedback implementations. Some frameworks omit black-hole feedback entirely, attributing regulation solely to stellar processes \citep{Ceverino+2018,Ma+2018,Gelli+2023}, whereas others include AGN activity even in intermediate mass \citep[e.g.,][]{Lovell+2023} and dwarf galaxies \citep[e.g.,][]{Lovell+2023,koudmani+2022,Arjona-Galvez+2024}, consistent with emerging observational evidence \citep[e.g.,][]{Greene+2004ApJ...610..722G,Mezcua+2018MNRAS.478.2576M}. On top of these fundamental differences, distinct relative importance and sub-grid treatment of stellar winds, radiation pressure, photoionization heating, and supernova-driven outflows further contribute to divergent predictions for gas expulsion, re-accretion, the intensity of bursts, the timescales over which galaxies go through burst cycles, and especially how long they remain below the SFMS. Distinguishing among these feedback prescriptions requires unbiased observational constraints on galaxies across \emph{all} phases of the burst cycle, and especially the off-burst phases.

Because feedback regulates mass growth, chemical enrichment, and environment \citep{Somerville2015, Naab2017}, constraining burstiness at high redshift \citep{Faucher-Giguere2018, Sun+2023} has far-reaching implications, including for the growth of supermassive black holes \citep{Habouzit2017} and the production and escape of ionizing photons relevant to reionization \citep{Trebitsch2017, Simmonds2024}. 

Key open questions include:
\begin{itemize}
    \item How common are galaxies below the star-forming main sequence, and at what stellar masses do they occur?
    \item What are their stellar population ages and dust content?
    \item Can photometry alone identify them?
\end{itemize}

To answer these questions, we present results from the \jwst\ Cycle~3 program \textit{Observing All phases of StochastIc Star formation} (OASIS; GO~5997; PI: T.~J.~Looser, co-PI: F.~D'Eugenio), designed explicitly to sample galaxies across the full burst cycle, and especially those below the SFMS. OASIS was conceived to \emph{correct} the prevailing \jwst\ selection bias by targeting systems not only in luminous, emission-line-dominated phases but also in lulling and mini-quenched states. Its design enables a more representative spectroscopic census of bursting, main sequence, lulling, and mini-quenched galaxies, providing direct constraints on how long systems spend above, on, and below the SFMS and thereby testing duty-cycle predictions from theoretical models.

OASIS combines deep NIRSpec/PRISM spectroscopy with NIRCam imaging to (i) detect or constrain the weak nebular emission that traces suppressed star formation, (ii) measure Balmer breaks and stellar continua, and (iii) recover stellar populations and gas conditions through full spectral fitting. The program was designed to counteract the prevailing \jwst\ selection bias toward luminous, emission-line-dominated systems by prioritizing mini-quenched and lulling candidates, while the
lower-priority tiers are selected on photometric
redshift and stellar mass, and therefore fill the masks
independently of star-formation state. The sample is consequently far less biased against suppressed galaxies than emission-line-selected samples, and comparable to photometrically selected ones \citep[e.g.,][]{Simmonds+2025,Ciesla+2024,Dressler+2023,Dressler+2023b,Santini+2025}.

This paper is organized as follows. Section~\ref{sec:Methods} describes the OASIS observations, data reduction, \ppxf\ spectral fitting, Balmer-line SFRs and Balmer-break measurements, and defines the analysis sample. We present the mini-quenched galaxies in Sect.~\ref{sec:Mini-quenched} and the lulling population in Sect.~\ref{sec:Lulling}, and assess our photometric target selection in Sect.~\ref{sec:phot_selection}. Section~\ref{sec:pop_plane} relates Balmer breaks, stellar ages and dust attenuation to position in the SFR--$M_\star$ plane and estimates the fraction and number density of suppressed galaxies, and Sect.~\ref{sec:sims} compares these results with simulations. The direction in which galaxies move through the plane, their ionization conditions and what suppresses their star formation are the subject of a companion paper \citep[][hereafter OASIS Paper~II]{Looser+2026_OASISII}, and the link between burstiness and galaxy size that of a third \citep[][hereafter OASIS Paper~III]{Baker+2026_OASISIII}. Section~\ref{sec:Summary} summarizes our conclusions.

Throughout the paper, we adopt a \citet{chabrier_galactic_2003} initial mass function (IMF) and a flat $\Lambda$CDM cosmology with $H_0=69.6$ km\,s$^{-1}$\,Mpc$^{-1}$, $\Omega_{\rm M}=0.286$, and $\Omega_\Lambda=0.714$.

\section{Data and Methods} \label{sec:Methods}

The JWST spectra used in this work were taken with the NIRSpec instrument \citep{jakobsen+2022} on board the JWST, using the PRISM disperser in combination with the micro-shutter array \citep[MSA, ][]{Ferruit2022} as part of our GO program OASIS (GO~5997; PI: T.~J.~Looser, co-PI: F.~D'Eugenio). OASIS targeted the Great Observatories Origins Deep Survey South field \citep[GOODS-S;][]{giavalisco2004} with two pointings, observed on 15--19 October 2024 and 5--7 January 2025. The first pointing targeted the Hubble Ultra Deep Field \citep[HUDF; e.g.,][]{Beckwith+2006}, and the second the JADES Origins Field \citep[JOF;][]{Eisenstein2023}. Each object received 9.3--28\,hr of PRISM exposure, depending on the number of dither pointings in which it was observed (see below), covering the wavelength range 0.6--5.3\,$\mu$m, providing spectra with a nominal wavelength-dependent spectral resolution of $R \ssim$30--330 \citep{jakobsen+2022}. 

\subsection{Target selection}\label{sec:target_selection}
For target selection we started from the JADES GOODS-S NIRCam photometric
catalog (v0.9.5; \citealt{Eisenstein2023,Rieke+2023_JADESDR1}), comprising $404\,119$ sources, using
circular-aperture (\texttt{CIRC1}) fluxes throughout, together with EAZY
photometric redshifts \citetext{\citealp{Brammer+2008}, measured as described in \citealt{Hainline2026b}} and the \textsc{Prospector}
\citep{Johnson2021ApJS..254...22J} stellar-population properties of \citet{Simmonds+2024},
restricted to fits with reduced $\chi^{2}\leq1$.

Candidate mini-quenched and lulling galaxies were identified in two steps.
First, two deliberately permissive and complementary screens reduced the
catalog to a shortlist. The photometric screen requires detections at
$\mathrm{S/N}>5$ in both F200W and F444W, $f_{\mathrm{F444W}} >
20\,\mathrm{nJy}$, $f_{\mathrm{F410M}} > 10\,\mathrm{nJy}$,
$z_{\mathrm{phot}} > 5$, and consistency between the F410M medium band and
the F444W broad band to within 20 per cent of the medium-band flux,
$|f_{\mathrm{F444W}}-f_{\mathrm{F410M}}|/f_{\mathrm{F410M}} < 0.2$. In
addition, we require \emph{either} that the source be bright in the
short-wavelength channel, $f_{\mathrm{F200W}} > 10\,\mathrm{nJy}$,
\emph{or} that the flatness extend to a second color,
$|f_{\mathrm{F356W}}-f_{\mathrm{F410M}}|/f_{\mathrm{F410M}} < 0.3$. Either
alternative on its own is intended to ensure that the flat long-wavelength
color reflects a genuine continuum rather than a coincidence or artifact:
the first because the spectral energy distribution (SED) is then measured across the full wavelength
range, the second because two consecutive line-free colors are difficult
to produce by chance. The flatness alternative is also complementary in
redshift, since at $5.5 \lesssim z \lesssim 6.6$ the [O\,\textsc{iii}] and
\Hbeta\ lines fall in F356W, blueward of both F410M and F444W, where the
F410M/F444W consistency criterion is insensitive to them. The physical screen, independent of the colors, requires Prospector $\log(\mathrm{sSFR/yr^{-1}})<-8$, $M_\star>10^{7}\,\mathrm{M_\odot}$, $z>5$, and $f_{\mathrm{F444W}}>20\,\mathrm{nJy}$. Second, every shortlisted source was inspected visually, using its EAZY SED and image cutouts, and assigned to Pri~1, 2, or 3; the lower tiers were populated automatically.

This yields a tiered prioritization scheme with eight priority levels:
\begin{itemize}
    \item \textbf{Pri~1:} Highest-priority, hand-selected targets that determined the pointing layout.  
    This tier comprises the eight most robust
    mini-quenched candidates, as identified by visual
    inspection of NIRCam and \textit{HST} photometry. Additionally, we added to this tier a small number
    of exceptional sources from the literature (e.g.\ the $z\approx14$ galaxies
    JADES-GS-z14-0 and JADES-GS-z14-1, \citealt{Carniani2024}; the $z\approx13$
    Ly$\alpha$ emitter JADES-GS-z13-1, \citealt{Witstok+2025_z13}; the MIRI Extremely Red Objects \textit{Cerberus} and \textit{Virgil} \citep{perez-gonzales+2024,iani+2025}; and the brown
    dwarf JADES-GS-BD-5, \citealt{Hainline+2024_BD}). Their OASIS spectra have already led to dedicated studies: of the extended C\,\textsc{iii}] emission and outflow of JADES-GS-z14-0 \citep{Carniani+2026_z14}, the weak metal lines of JADES-GS-z14-1 \citep{Wu+2025_z14-1}, the Ly$\alpha$ emission of JADES-GS-z13-1 \citep{Witstok2026}, the spectrum of the obscured AGN in \textit{Virgil} \citep{rinaldi+2025b}, and the atmosphere of JADES-GS-BD-5 \citep{Hainline2026}.  

    \item \textbf{Pri~2:} 133 hand-selected targets, comprising 110 further
    robust mini-quenched candidates at $z_{\mathrm{phot}} = 5.0$--$7.0$, ten high-priority lulling candidates, and 13 sources with a possible
    Balmer break at $z_{\mathrm{phot}} = 5.0$--$9.5$.

    \item \textbf{Pri~3:} 144 hand-selected lulling-galaxy candidates at
    $z_{\mathrm{phot}} = 5.0$--$8.4$ showing excess in some
    photometric filters, indicating the presence of emission lines.

    \item \textbf{Pri~4:} 634 galaxies with $F444W > 20\,\mathrm{nJy}$ and
    Prospector-inferred redshifts of $z>7$ and stellar masses
    $\Mstar > 10^{7}\,\MSun$.

    \item \textbf{Pri~5:} 6686 galaxies with $F444W > 20\,\mathrm{nJy}$,
    Prospector-inferred redshifts of $z>5$, and stellar masses
    $\Mstar > 10^{7}\,\MSun$.

    \item \textbf{Pri~6:} 3456 galaxies with $F444W > 20\,\mathrm{nJy}$ and
    EAZY photometric redshifts of $z>5$, independent of stellar mass.

    \item \textbf{Pri~7:} All 48\,063 remaining galaxies with
    $F444W > 20\,\mathrm{nJy}$.

    \item \textbf{Pri~8:} All remaining objects.
\end{itemize}

This prioritization was designed to construct a more representative sample spanning all phases of stochastic star-formation histories, enabling
a comprehensive view of early galaxy evolution beyond the burst-dominated samples targeted in previous JWST programs.

\subsection{NIRSpec observations and data reduction}\label{sec:reduction}

Optimizing each observation with the \texttt{eMPT} software \citep{Bonaventura+2023}, the program observed a total of 332 galaxies across the two observations, from which 260 secure spectroscopic redshifts could be obtained, 65 of them at $z\gtrsim4$. We note that, although the priority classes were defined using a
$z_{\mathrm{phot}}>5$ criterion, the analysis below is extended to $z>4$,
since the majority of shutters were filled with lower-priority and filler
targets, which carry no redshift criterion, and a substantial fraction of
these turn out to lie at $4 < z < 5$.
Each observation comprised three distinct dithering pointings, each using a different MSA configuration and employing a three-shutter slitlet for every target. 
The dithering strategy was chosen to place the spectra at different detector locations, mitigate detector artifacts and gaps, maximize exposure time for high-priority targets, and increase target density. 

At every dithering pointing, four sequences of three-nod patterns along the slit were executed, collecting two integrations with 19 groups each at every nod position, using NRSIRS2 readout \citep{Rauscher2012}. 
Each three-nod pattern had an exposure time of 8403\,s, resulting in a total integration time of 33{,}612\,s (9.3\,hr) per dithering pointing. 
Depending on whether a galaxy was observed in one, two, or all three dithering pointings, the total on-source exposure time therefore ranged from 9.3\,hr to 28\,hr.

Spectral extraction and flux calibration were carried out with dedicated reduction pipelines created by the ESA NIRSpec Science Operations Team and the NIRSpec GTO Team \citep{Scholtz2025_JADES_DR4}.
Preliminary redshifts were assessed from the overall continuum shape and identifiable emission features, before being refined using \ppxf, see below.

Due to the effective 0.2 arcsec slit width, the OASIS spectra exhibit wavelength-dependent aperture losses. We corrected these losses under the assumption of a point-source profile, which results in a systematic underestimation of the fluxes for the most extended, lowest-redshift galaxies. However, the higher-redshift systems that are the focus of this work are largely unaffected by this limitation \citep{Scholtz2025_JADES_DR4}.

\subsection{NIRCam parallel imaging}\label{sec:nircam_parallel}

For both pointings, we employed NIRCam \citep{Rieke2005SPIE.5904....1R,Rieke2023} in parallel mode. 
In the first pointing, the NIRCam parallel field fell, by design, into the JOF region, for which deep NIRCam wide-band imaging was already available. 
We therefore used medium-band filters to complete the filter coverage in that area. 

Specifically, we employed the NIRCam short-wavelength filters
\textit{F150W}, \textit{F162M}, \textit{F150W2}, \textit{F182M}, and
\textit{F210M}, together with the long-wavelength filters
\textit{F250M}, \textit{F300M}, \textit{F335M}, and \textit{F480M}. The
imaging was obtained with the \texttt{DEEP8} readout pattern using
7 groups per integration, 2 integrations per exposure and 3 dithers. This yields
$2.48 \times 10^{4}$~s (6.9~hr) per filter. The medium bands sample the rest-frame Balmer break
region at $z > 4$ more finely than broad bands alone, which we expect to
improve the photometric identification of faint mini-quenched and lulling
candidates in future surveys.

In the second pointing, where the NIRCam parallel field covered a mostly blank or only shallowly imaged region, we used a standard wide-filter configuration. 
We employed the NIRCam short-wavelength filters \textit{F090W}, \textit{F115W}, \textit{F150W}, and \textit{F200W}, together with the long-wavelength filters \textit{F277W}, \textit{F356W}, \textit{F410M}, and \textit{F444W}. As
in the first pointing, the imaging was obtained with the \texttt{DEEP8}
readout pattern using 7 groups per integration, 2 integrations per
exposure and 3 dithers, giving $2.48 \times 10^{4}$~s (6.9~hr) per filter.

\subsection{Emission-line and SFH fitting with \ppxf} \label{ppxf_methodology}

To model the R100/PRISM spectra, we employ the penalized PiXel-Fitting framework\footnote{\url{https://pypi.org/project/ppxf/}} \ppxf \citep{Cappellari2017,cappellari2022}, combined with bootstrap resampling to estimate uncertainties and derive physical parameters \citep{Looser+2024_FMR}. Overall, we follow very closely the methodology of \citet{Looser+2025}. The stellar continuum is represented by a non-negative linear combination of simple stellar population (SSP) templates constructed from C3K synthetic atmospheres \citep{Conroy2019} at $R=10{,}000$, adopting MIST isochrones \citep{Choi2016}, solar-abundance ratios, and a Salpeter IMF.\footnote{Templates available upon request from C.~Conroy.}

All templates are subject to a common \citet{Calzetti2000} attenuation law with $E(B-V)$ as the sole dust parameter. Masses are converted from a Salpeter to a Chabrier IMF using $M_{*,C}=0.58\,M_{*,S}$ following \citet{speagle+2014}.

The SSP library spans ages from $10^{6.0}$ to $10^{10.3}$~yr and metallicities between [M/H] $=-2.5$ and $0.0$. For each object, we restrict the age grid to remain compatible with the cosmic age at the galaxy redshift, allowing the fit to exceed that limit by 0.1--0.2~dex (26--58\%); this choice is necessary because in our non-Bayesian framework, any noise would bias the age measurements to smaller values than the hard cutoff \citep{mcdermid+2015}. Nebular lines are included by fitting Gaussian components simultaneously with the stellar continuum. The SSP models do not contain intrinsic nebular emission, and the stellar
continuum and the emission lines are therefore fitted independently.
Because this study focuses on systems with suppressed star-formation
activity, in which the nebular contribution to the continuum is expected
to be small, we neglect the nebular continuum.

Below, we outline the \ppxf-based fitting methodology applied to each OASIS spectrum, following closely the procedure of \citet{Looser+2025}:
\begin{enumerate}
    \item We first convolve the C3K SSP templates to the wavelength-dependent resolution of the data. The effective spectral resolution also depends on slit filling---that is, on the ratio between the source size and the 0.2\arcsec\ MSA microshutter width---and can vary by up to a factor of two \citep{deGraaff+2024_msafit}. Because \ppxf\ can compensate for an overestimated resolution (through velocity broadening) but cannot compensate for an underestimated one, we adopt a conservative uniform increase of the resolution by a factor of 1.7 for all targets. This factor is determined empirically from the observed widths of the emission lines.

    \item We renormalize the observed spectrum and templates by their median per-pixel flux to stabilize the numerical optimization and to enable regularization in \ppxf\ (via the \texttt{regul} keyword), thereby enforcing smoothness in the SSP weight distribution \citep{Cappellari2017}.

    \item We perform an initial fit with \texttt{regul}=5 to obtain a first-pass model and identify spectral pixel outliers, which we remove using $4\sigma$ clipping.

    \item We then generate wild bootstrap realizations by perturbing the spectrum $S$ with the residuals $R$ from the initial fit:
    \[
        S^*(\lambda) = S(\lambda) \pm R(\lambda'),
    \]
    where for each pixel $\lambda$, the residual value $R(\lambda')$ is drawn randomly from within $\pm 50$ pixels of $\lambda$, excluding $\sigma$-clipped pixels.

    \item Each perturbed spectrum $S^*(\lambda)$ is refit with \ppxf\ using the same regularization (\texttt{regul}=5).

    \item Steps 4--5 are repeated 100 times.

    \item The resulting ensemble of 100 SSP weight grids samples the underlying distribution of solutions. We average these grids to obtain a nonparametric SFH that reflects the noise properties of the data.
\end{enumerate}

This procedure yields an inferred star-formation history for each object, obtained in an archaeological sense by interpreting the stellar populations identified at the epoch of observation as the galaxy's fossil record. To derive the stellar mass \Mstar\ for each galaxy, we directly sum the individual mass-weighted SSP contributions returned by \ppxf.

Since the fitting relies solely on the spectra, any emission falling outside the MSA shutters is ignored, preventing us from capturing color gradients or clumpy morphologies; as for the slit losses (Sect.~\ref{sec:reduction}), this matters mainly for the lowest-redshift, most extended systems (see OASIS Paper~III for detailed size measurements).

A key strength of our \ppxf-based methodology is that fitting each spectrum as a
superposition of independent SSPs, on a grid of stellar ages and
metallicities, together with gas templates, avoids both enforcing a
parametric form for the SFH and fixing the stellar metallicity to a single
value or functional form \citep[cf.][]{Robotham2020}. As a result, the inferred SFHs and
chemical-enrichment histories (CEHs) remain agnostic to assumptions about
the physical drivers of galaxy evolution. Bayesian codes have moved in
this direction, for instance by adopting nonparametric SFHs with bursty
priors \citep{Tacchella2022}, but they generally retain an assumed constant or functional form for the
chemical evolution.

The price of dropping these priors in our methodology is a highly degenerate solution space,
so the inferred histories require careful validation and cautious
interpretation, particularly for individual objects, as with any SFH-fitting code. For example, a small
amount of light attributed, possibly spuriously, to old stellar
populations that barely register in the light-weighted SFH can translate
into a substantial mass formed at early times, because of the high
mass-to-light ratio of such populations. Such features must be recognized
as spurious at the interpretation stage. The essential difference is that
our approach does not impose a physical picture on the data during the
fitting, but allows the results to be interpreted in the light of physical
understanding afterwards. With these caveats in mind, the approach has been
benchmarked on NIRSpec data \citep{Looser+2025}, on LEGA-C spectroscopy at
$z \sim 1$ \citep{barone+2021,cappellari2022}, and on spatially resolved
MaNGA observations \citep[][]{Bundy2015}, where it recovers meaningful
average star-formation and chemical-enrichment histories and the stellar
fundamental metallicity relation \citep{Looser+2024_FMR}. Importantly, any scaling relation recovered from
the fitting therefore cannot be an artifact of an imposed functional form.

Applying our methodology to the OASIS data, we find that while the stellar ages, which are driven by the UV slope, the Balmer break, and the overall continuum shape, are generally robust, the spectral resolution of the NIRSpec/PRISM data is often insufficient to constrain the metallicity of the underlying stellar populations. Breaking the well-known age--metallicity degeneracy requires spectrally resolving individual hydrogen and metal absorption lines, which the PRISM resolution does not
allow, particularly for young and intermediate-age ($<1$~Gyr)
populations. Further degeneracies, introduced by dust attenuation,
flux-calibration uncertainties, and the wavelength-dependent aperture
losses affecting extended objects, compound these uncertainties and render
individual SSP weights unreliable. The bootstrap procedure described above provides an effective way to assess the stability of these weights: it yields a scatter distribution for every SSP grid point as well as for any derived quantity, such as the stellar ages. These tests show that age estimates are trustworthy when grouped into conservative 1-dex bins in $\log_{10}(\mathrm{Age/yr})$ over the range 6.0--10.0. Ages quoted at finer resolution below, such as class medians, are meant for orientation only and should be read as back-of-the-envelope estimates. 

Accordingly, we do not use the inferred metallicities quantitatively
anywhere in this work; the age--metallicity grids shown in Sects.~\ref{sec:Mini-quenched}
and~\ref{sec:Lulling} are shown only to illustrate where the light and mass weights fall, and all reported stellar
population quantities are marginalized over metallicity. Conservatively,
we also do not interpret the detailed shape of the recovered SFHs: the
weights of adjacent age bins trade off against each other, so that their
ratio is unstable even where their sum is well constrained. From the SFHs, we therefore use only the mean stellar ages in this paper; further reliable quantities that integrate the SFH over these bins, such as the SFR averaged over the last 100\,Myr, are described and used in OASIS Paper~II.

\subsection{SFR from nebular Balmer lines}
\label{sec:mass_sfr}

We derived star-formation rates (SFRs) from Balmer emission lines using the
\citet{shapley2023} calibration, which converts dust-corrected H$\alpha$
luminosities into SFRs via
$\mathrm{SFR}\,[\mathrm{M_{\odot}\,yr^{-1}}] = 10^{-41.67}\,L_{\mathrm{H}\alpha}\,[\mathrm{erg\,s^{-1}}]$.
Throughout this work, we adopted an intrinsic Balmer decrement ratio
$\mathrm{H}\alpha/\mathrm{H}\beta$ of 2.86 \citep[Case~B recombination, under standard temperature and density conditions $T_\mathrm{e}=10\,000$\,K and $n_\mathrm{e}=100\,\mathrm{cm}^{-3}$;][]{osterbrock+ferland2006} and a
nebular SMC bar extinction curve with $R_V = 2.74$ \citep{Gordon+2003} to correct for dust. For spectra for which the Balmer decrement could not be measured -- either because one or both lines were undetected,
or because H$\alpha$ was outside the spectral range -- we used stellar continuum reddening,
fitted with \ppxf, as a dust tracer. In all such cases,
$E(B{-}V)_\star$ was converted into a nebular attenuation via
$A_V = E(B{-}V)_\star\, f_{\rm neb}\, R_V$, where $f_{\rm neb} \equiv E(B{-}V)_{\rm neb}/E(B{-}V)_\star$ allows for
nebular reddening in excess of that of the stellar continuum. The canonical
value $f_{\rm neb} = 2.27$ \citep{Calzetti2000} is calibrated on massive local
starbursts and is not appropriate for our sample: \citet{Song+2026} find
$E(B{-}V)_\star/E(B{-}V)_{\rm neb} \rightarrow 1$ below a transition mass
of $M_{\star} \approx 10^{9}\,\mathrm{M_{\odot}}$. Independently,
\citet{Karthikeyan+2026} measure
$E(B{-}V)_{\rm neb} = E(B{-}V)_\star + 0.100$ at $4 < z < 5$ and
$E(B{-}V)_{\rm neb} = E(B{-}V)_\star + 0.053$ at $5 < z < 7$ in JADES
under an SMC law; at the median $E(B{-}V)_\star = 0.134$ of the $z>4$ galaxies
to which we apply this correction, these correspond to
$f_{\rm neb} \simeq 1.75$ and $1.4$, respectively. For simplicity we adopted a
single, redshift-independent value of $f_{\rm neb} = 1.3$, intermediate between
the \citet{Song+2026} and \citet{Karthikeyan+2026} determinations. Varying $f_{\rm neb}$ over the full range 1.0--2.27 changes the inferred SFRs
by $0.15$~dex at that median reddening.

For each galaxy, we applied the following mutually exclusive hierarchy:

\begin{enumerate}
    \item \textbf{Both H$\alpha$ and H$\beta$ detected:} The Balmer decrement yields
    a direct estimate of $A_V$, enabling fully dust-corrected H$\alpha$ SFRs.
    Unphysical solutions ($A_V < 0$) are reset to $A_V = 0$.

    \item \textbf{Only H$\alpha$ detected:} We apply stellar-based dust corrections
    using $E(B{-}V)_\star$.

    \item \textbf{H$\alpha$ out of range ($z>7.1$) but H$\beta$ detected:} We infer
    an effective H$\alpha$ luminosity by scaling the H$\beta$ luminosity by 2.86,
    and apply stellar-based dust corrections.

    \item \textbf{Non-detections (upper limits):} We compute H$\alpha$-based SFR
    upper limits using H$\alpha$ flux constraints and stellar reddening. If
    H$\alpha$ is out of range, H$\beta$ upper limits are converted to equivalent
    H$\alpha$ luminosities and treated analogously.
\end{enumerate}

\subsection{Balmer break measurements}
\label{sec:balmerbreak}

Balmer-break strengths are measured directly from the spectra, independently
of the \ppxf\ fitting, as the ratio of the median $f_\nu$ in two rest-frame
windows straddling the Balmer limit at 3646\,\AA,
\begin{equation}
    \label{eq:balmerbreak}
    R \equiv \frac{\mathrm{median}\left[f_\nu\right]_{3750\textrm{--}4100\,
                   \textrm{\AA}}}
                  {\mathrm{median}\left[f_\nu\right]_{3400\textrm{--}3600\,
                   \textrm{\AA}}},
\end{equation}
so that $R=1$ corresponds to the absence of a break. Uncertainties are
obtained by Monte Carlo, re-measuring the index on 100 realizations of the
spectrum in which every pixel is perturbed independently according to its
variance, and taking the robust standard deviation of the resulting
distribution. Because the index is a flux ratio measured on the data alone, it
provides a constraint on the stellar population that is independent of both
the \ppxf-derived ages (Sect.~\ref{ppxf_methodology}) and the Balmer-line SFRs
(Sect.~\ref{sec:mass_sfr}).

\begin{figure*}[h!]
  \centering
  \includegraphics[width=\textwidth]{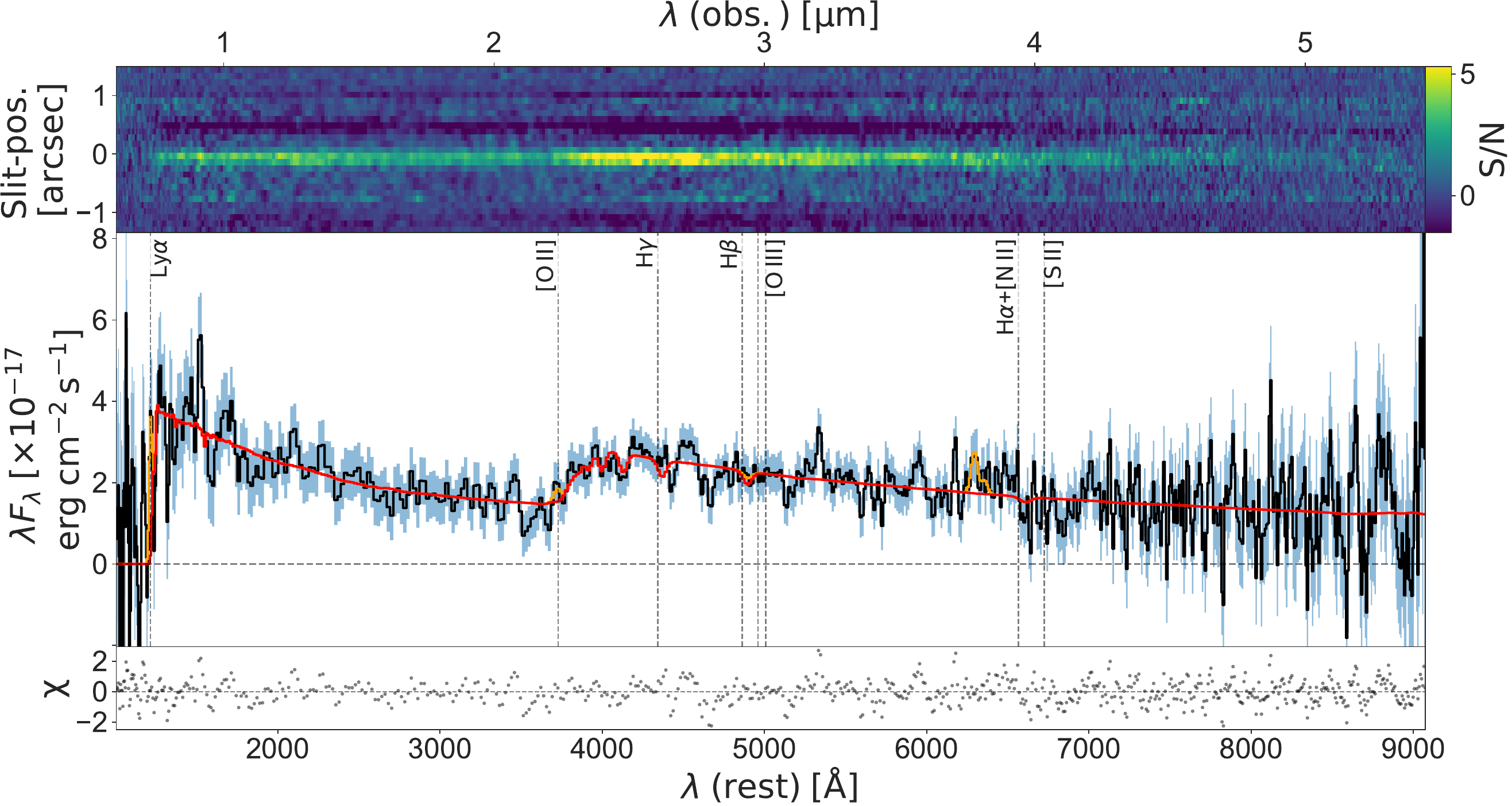}
  \caption{\textbf{NIRSpec/PRISM spectrum of the mini-quenched galaxy with ID~181834 in the OASIS sample at $z=5.0$.} The spectrum shows a clear Ly$\alpha$ drop, a pronounced Balmer break, and no emission lines. In particular, the absence of any Balmer emission lines indicates no detectable star formation activity at the epoch of observation. The best-fit \ppxf model spectrum is shown in red, and the best-fit nebular emission lines are shown in yellow. The noise is indicated by the shaded blue regions. The upper panel shows the S/N of the combined 2D spectrum. The bottom panel indicates the reduced residuals of the fits. The vertical dashed lines mark the rest-frame wavelengths of strong nebular emission lines.}
  \label{fig:mq-181834}
\end{figure*}

\begin{figure*}[h!]
  \centering
  \includegraphics[width=\textwidth]{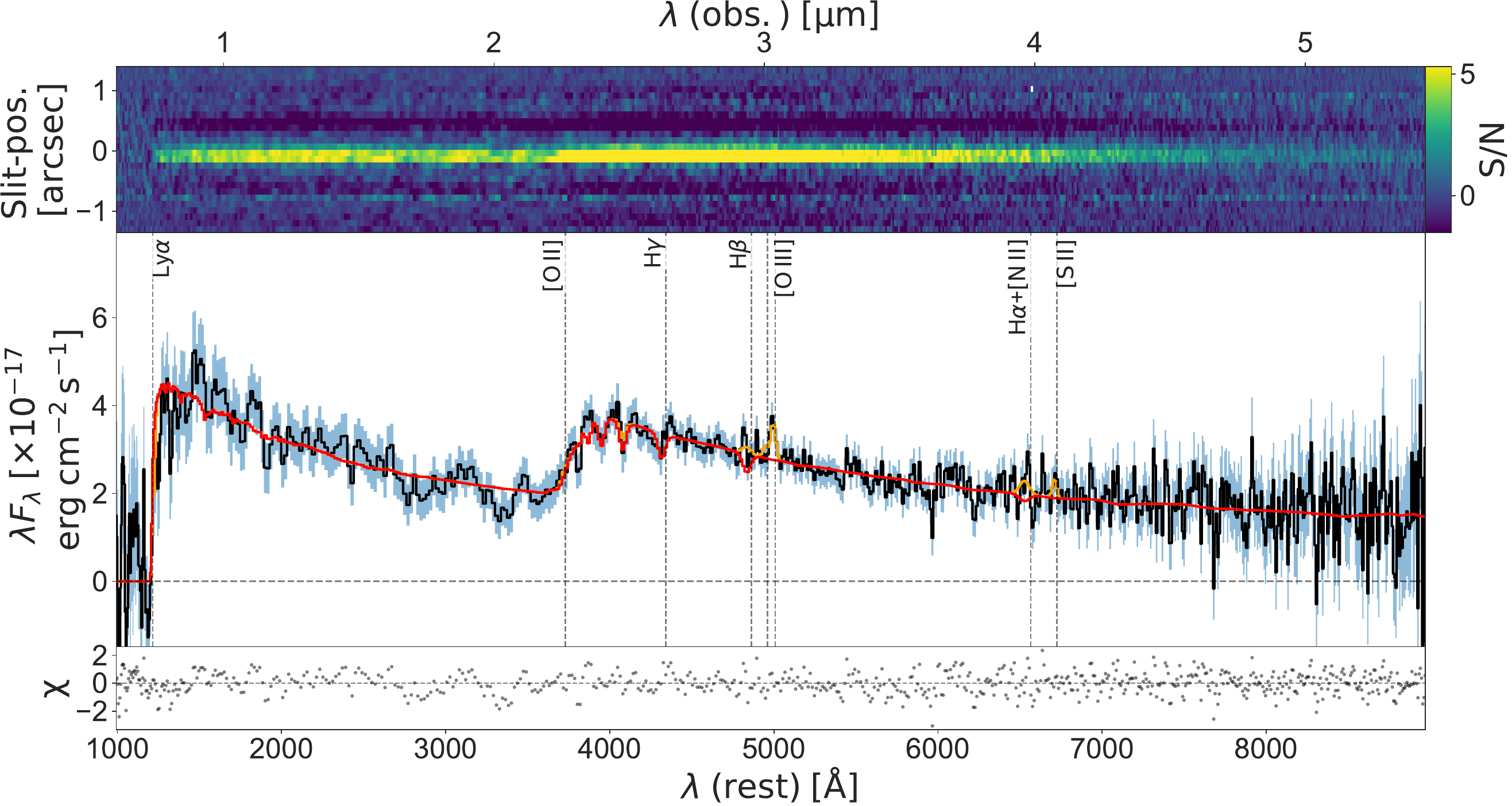}
  \caption{\textbf{NIRSpec/PRISM spectrum of the mini-quenched galaxy with ID~168352 in the OASIS sample at $z=5.1$.} The spectrum shows a clear Ly$\alpha$ drop, a pronounced Balmer break, and no Balmer emission lines, indicating no detectable star formation activity at the epoch of observation, but very weak, clearly detected \oiii\ emission. See Fig.~\ref{fig:mq-181834} for more details.}
  \label{fig:mq-168352}
\end{figure*}

\newcommand{\SFHpair}[3]{%
  \begin{figure*}[t]
    \centering
    \begin{subfigure}{0.48\textwidth}
      \centering
      \includegraphics[width=\textwidth]{SFH_#1/SFH_light_weights_boots_n_itr_100.pdf}
      \caption{Light-weighted SFH for ID~#1.}
      \label{fig:SFH-#1-light}
    \end{subfigure}
    \hfill
    \begin{subfigure}{0.48\textwidth}
      \centering
      \includegraphics[width=\textwidth]{SFH_#1/SFH_mass_weights_boots_n_itr_100.pdf}
      \caption{Mass-weighted SFH for ID~#1.}
      \label{fig:SFH-#1-mass}
    \end{subfigure}

    \caption{Light- and mass-weighted star-formation histories for the #2 galaxy ID~#1, showing 
the stellar age--metallicity grid inferred by \ppxf. The $x$-axis shows the 
age of the stellar population grid, while the $y$-axis represents the stellar 
metallicity, and the color-coding indicates the weight assigned to each stellar 
population in the \ppxf\ fit (black: zero weight). #3
}
    \label{fig:SFH-#1}
  \end{figure*}%
}

\SFHpair{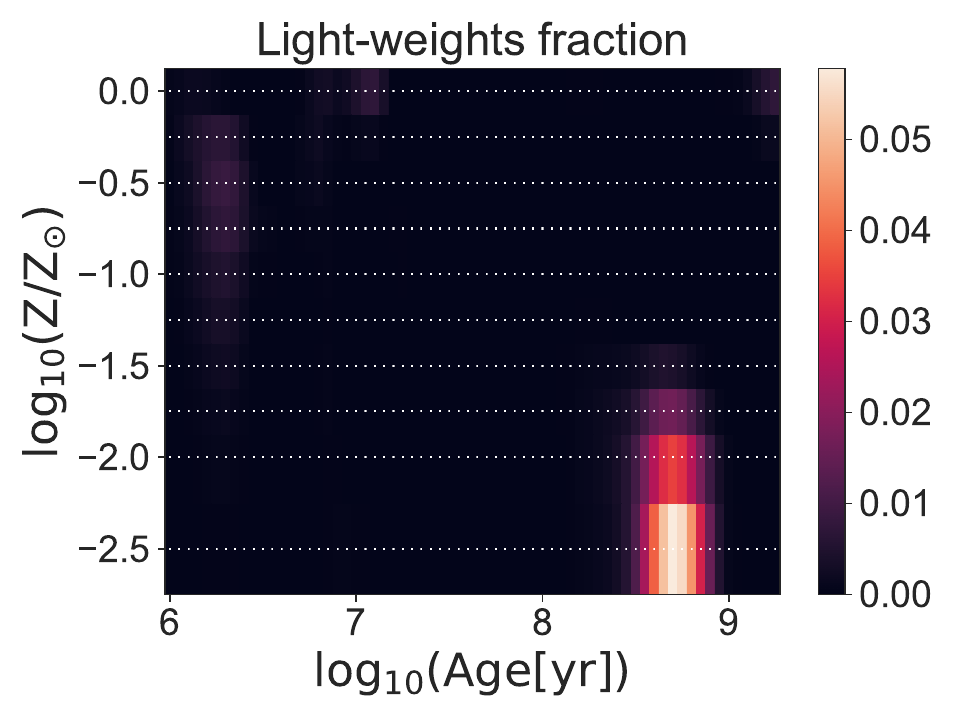}{mini-quenched}{The stellar mass of this galaxy is dominated by a
$10^{8.5\text{--}9}$~yr population (mass-weighted age $10^{8.7}$\,yr;
light-weighted $10^{8.1}$\,yr) that is nominally metal-poor. 
However, due to the low spectral resolution of the \textsc{prism}, 
metallicities are difficult to constrain reliably. The metal-rich, old stellar 
populations visible in the mass-weighted right-hand plot are likely spurious, 
as evidenced by their negligible contribution in the light-weighted SFH shown 
in the left plot. See \citet{Looser+2025} for more details.}

\SFHpair{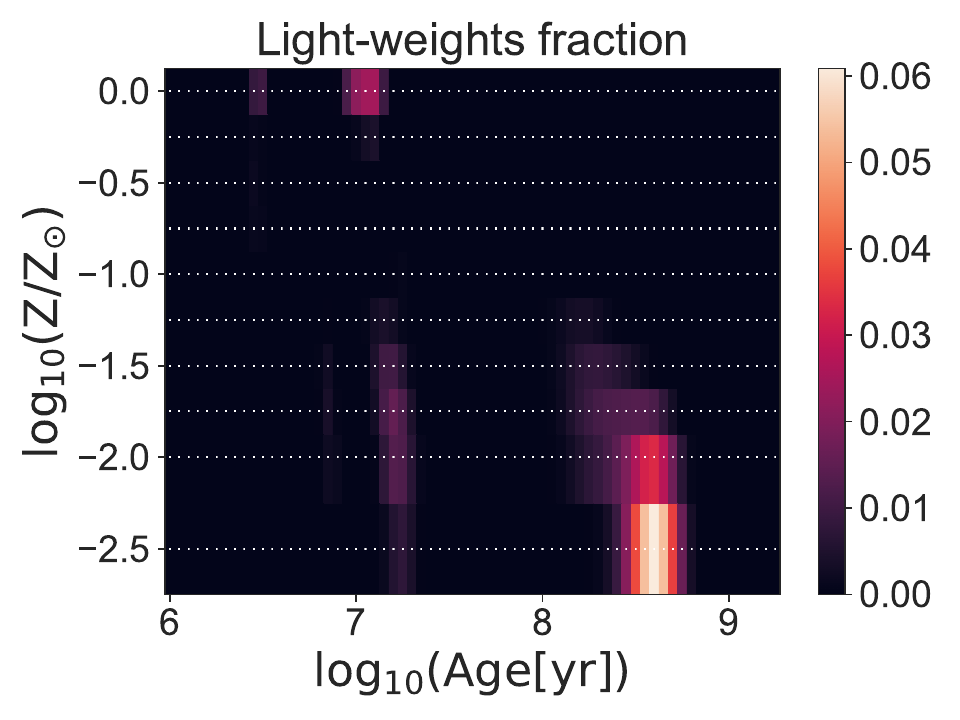}{mini-quenched}{}

\subsection{The $z>4$ analysis sample}\label{sec:analysis_sample}

Of the 65 OASIS galaxies with a secure redshift $z\gtrsim4$, 54 form the analysis sample used throughout Sects.~\ref{sec:pop_plane} and \ref{sec:sims}. The analysis requires a stellar mass and
stellar-population properties from the \ppxf\ continuum fit, a Balmer-break measurement and a
nebular SFR, and the remaining 11 galaxies lack at least one of these. Throughout, $\Delta_{\rm MS}$ is measured against the SFMS of \citet{Simmonds+2025} at each galaxy's own redshift. Three lie at $z>11$
(13731, 70836 and 329195), where the Balmer lines fall beyond the PRISM wavelength range, so
that no nebular SFR can be measured. Seven have continua too faint for a meaningful fit, or
spectra or fits that are otherwise unusable. The eleventh, 216135, has a failed \ppxf\ fit: its
mass and SFRs disagree with the photometric Prospector fit by 2.6--3.7\,dex, and its bootstrap
uncertainties are unphysically large; Appendix~\ref{appendix:faint_low_sfr} discusses it
using its Prospector mass. With this mass it is also shown in Fig.~\ref{fig:SFR-mass}, drawn as
a lulling galaxy because its SFR limit places it at least 0.7\,dex below the SFMS. Since the
limit lies above the quenching threshold, it could equally be mini-quenched, and we do not count
it among the five lulling galaxies.

\section{Results: Mini-quenched galaxies} \label{sec:Mini-quenched}

\begin{figure*}
  \begin{subfigure}[t]{0.48\textwidth}
    \centering
    \includegraphics[width=\textwidth]{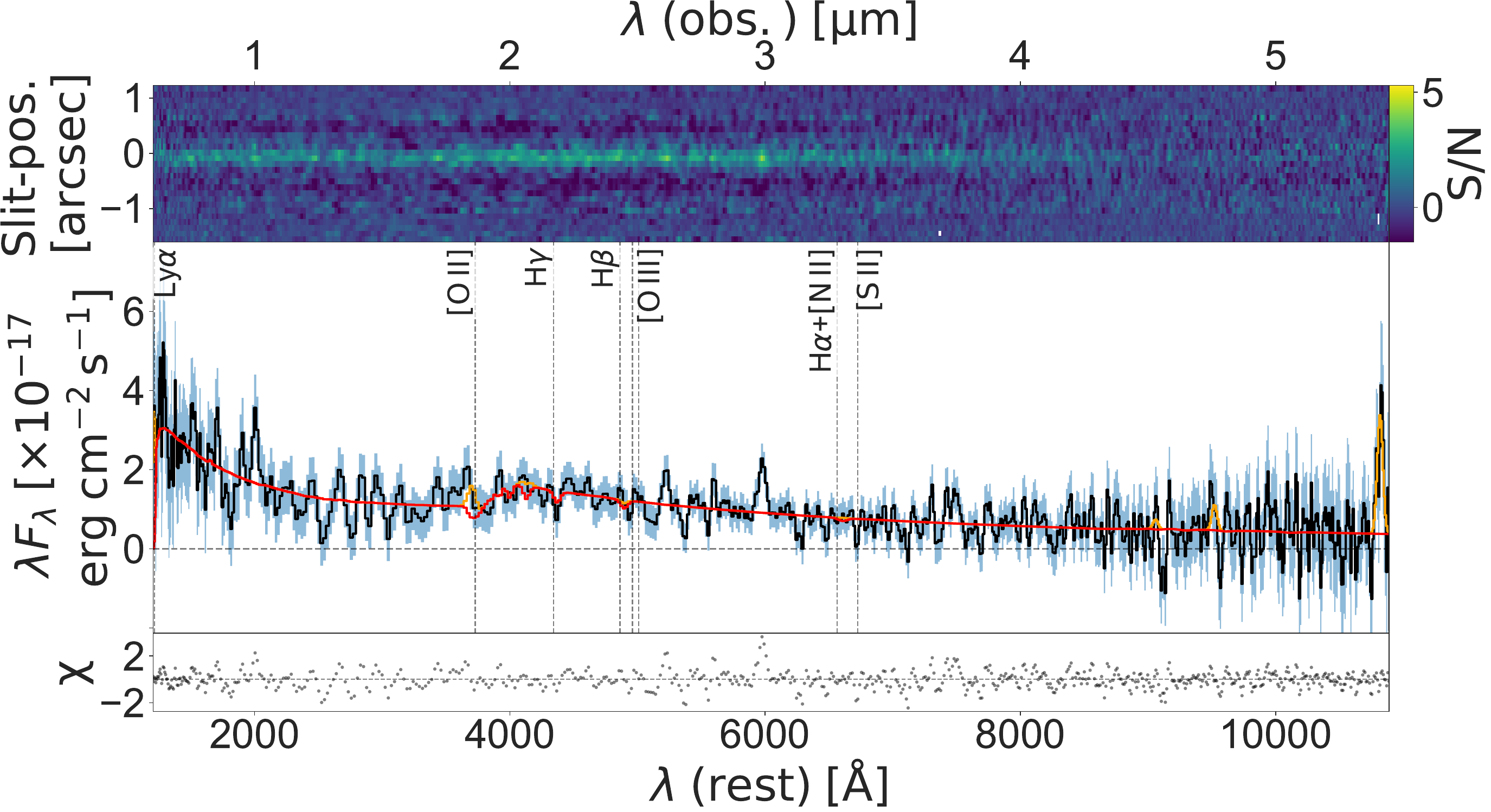}
    \caption{NIRSpec/PRISM spectrum of the mini-quenched galaxy ID~206569 at $z=4.0$. No emission lines are detected; the Balmer break is not significant ($R=1.01\pm0.24$).}
    \label{fig:mq-206569}
  \end{subfigure}
  \hfill
  \begin{subfigure}[t]{0.48\textwidth}
    \centering
    \includegraphics[width=\textwidth]{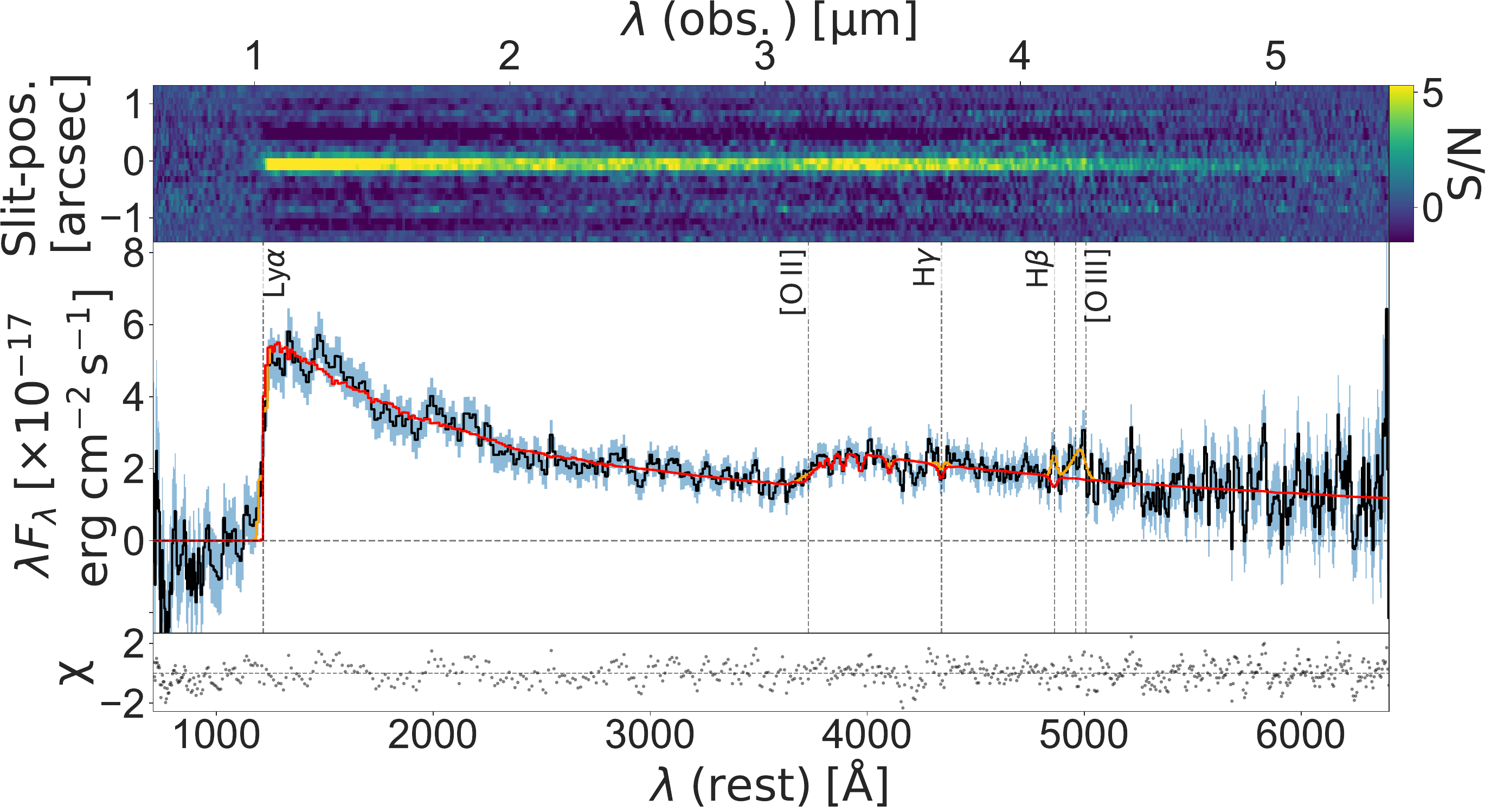}
    \caption{NIRSpec/PRISM spectrum of the lulling galaxy ID~114769 at $z=7.5$. The spectrum shows a clearly detected Ly$\alpha$ drop and a pronounced Balmer break, as well as a weak H$\beta$ emission line ($3.3\sigma$), indicating a low but non-zero star-formation rate. Weak \oiii\ emission is also detected ($3.5\sigma$).
}
    \label{fig:lulling-114769}
  \end{subfigure}

  \caption{Spectra for the mini-quenched galaxy ID~206569 and the lulling galaxy ID~114769.}
\end{figure*}

We identify three mini-quenched galaxies at $z>4$ in OASIS (Figs.~\ref{fig:mq-181834}, \ref{fig:mq-168352}, and \ref{fig:mq-206569}) with stellar masses in the range $10^{8.3}$--$10^{8.9} M_\odot$. This substantially expands the number of spectroscopically confirmed mini-quenched galaxies beyond the handful of systems reported to date (cf.\ Sect.~\ref{sec:intro}). All three OASIS objects pass the redshift-dependent quenching threshold 

\begin{equation}
    \label{eq:quenching-threshold}
    \mathrm{sSFR} < \frac{0.2}{t_\mathrm{obs}(z)}
\end{equation}

commonly adopted in the literature \citep[e.g.,][]{Gallazzi2014, pacifici+2016, carnall+2023b, Looser+2025}, where $t_\mathrm{obs}(z)$ is the cosmic time at redshift $z$. Their key properties are summarized in Table~\ref{tab:oasis_ids_coords_m444w_pri_z_mass_sfr_age_redd_balmer_burst}. All uncertainties quoted are statistical. For stellar masses and ages they are the 16th--84th percentiles of the \ppxf\ bootstrap realizations (Sect.~\ref{ppxf_methodology}); for the SFRs, the 16th--84th percentiles of a Monte Carlo over the fluxes of the lines that enter each galaxy's SFR and dust correction (Sect.~\ref{sec:mass_sfr}), with $E(B{-}V)_\star$ held fixed where it is used; and for the Balmer breaks, the robust standard deviation of the Monte Carlo of Sect.~\ref{sec:balmerbreak}.

All three systems are noticeably older than previously known mini-quenched systems \citep[e.g.,][]{Looser+2024Nat,Looser+2025}, with light-weighted stellar ages exceeding $100~\mathrm{Myr}$. Their light-weighted star-formation histories\footnote{Light-weighted quantities weight each SSP-weight by its flux,
$\langle\log t\rangle_{L}=\sum_i L_i \log t_i/\sum_i L_i$; mass-weighted
quantities weight by mass, $\langle\log t\rangle_{M}=\sum_i M_i \log t_i/\sum_i
M_i$. The two are related by the SSP mass-to-light ratio,
$L_i=M_i/(M/L)_i$.} show evidence for no or only little recent activity (Figs.~\ref{fig:SFH-181834}, \ref{fig:SFH-168352}, and \ref{fig:SFH-206569}), while their mass-weighted SFHs indicate that almost no stellar mass has formed within the past $\sim300~\mathrm{Myr}$. In these galaxies, outshining is far less severe than in starburst galaxies, so that their SFHs are better constrained (Sect.~\ref{sec:ages}). All three galaxies exhibit very low dust attenuation, and 181834 and 168352 display pronounced Balmer breaks (Table~\ref{tab:oasis_ids_coords_m444w_pri_z_mass_sfr_age_redd_balmer_burst}). ID~181834 and ID~206569 show no detectable emission lines. The corresponding
$3\sigma$ upper limits, in units of $10^{-20}\,{\rm erg\,s^{-1}\,cm^{-2}}$, are
$F({\rm H}\beta)<9.2$ and $<10.4$, $F([\ion{O}{iii}]\lambda5007)<8.2$ and
$<9.9$, and $F({\rm H}\alpha+[\ion{N}{ii}])<9.8$ and $<7.6$, respectively.

ID~168352 exhibits extremely weak but significant $[\ion{O}{iii}]\lambda5007$
emission, $F=7.4\pm1.8$ ($4.0\sigma$), while
H$\alpha+[\ion{N}{ii}]$ remain undetected at $<6.1$. A formal $3.7\sigma$ H$\beta$ feature ($F=7.5\pm2.0$) is returned for
ID~168352, but it is inconsistent with the H$\alpha$ constraint: Case~B
recombination requires $F({\rm H}\alpha)\geq2.86\,F({\rm H}\beta)=21.4\pm5.8$,
a factor ${\sim}3.5$ above the measured $3\sigma$ upper limit of
$F({\rm H}\alpha+[\ion{N}{ii}])<6.1$. Since the H$\alpha$ measurement is a
blend and dust attenuation can only increase the required H$\alpha$ flux, both
effects strengthen the discrepancy. We
therefore regard the H$\beta$ feature as spurious -- consistent with visual
inspection of the fit (Fig.~\ref{fig:mq-168352}), where it coincides with
residual continuum structure -- and treat H$\beta$ as undetected throughout.

The absence of the Balmer lines points to the lack of 
recent star formation (on timescales of $\lesssim$10~Myr). This is independently confirmed by the SFH, which is based on the stellar continuum alone. Equally important, the missing Balmer lines 
indicate the absence of strong shocks or recent (unobscured)
AGN activity, unlike what is seen in massive quiescent galaxies at
$1.5<z<5$ \citep[e.g.,][]{newman+2018a,carnall+2023b,DEugenio+2024NatAs...8.1443D,Belli+2024,deGraaff+2025,Baker+2025_quiescent}. Only the weak [O\,\textsc{iii}] emission of 168352 leaves room for a minor contribution from either.

The three mini-quenched galaxies are located at least 1.2--1.7\,dex below the star-forming main sequence, within the quenched regime (see Fig.~\ref{fig:SFR-mass}, red squares). The spectrum-based upper limits on the Balmer emission lines, which place these galaxies in the quenched region, could not have been robustly constrained using photometry alone (Sect.~\ref{sec:phot_selection} and Appendix~\ref{app:phot_spec}).

In contrast to other studies investigating suppressed star formation \citep[e.g.,][]{Helton+2024a,Morishita+2025,Witten+2025b}, we do not find evidence for an environmental connection. Our mini-quenched galaxies are isolated systems that do not reside in overdense regions, as verified by inspecting deep NIRCam catalogs of their surroundings. We find no nearby galaxies at comparable redshifts, either spectroscopic or photometric, and in particular no massive neighbors that could plausibly act as an environmental driver of their quenching. Moreover, none of their redshifts coincide with the well-known overdensities in the field \citep[e.g.,][]{Helton+2024b}. This suggests that internal feedback processes, rather than environmental effects, are responsible for their quenching.

\SFHpair{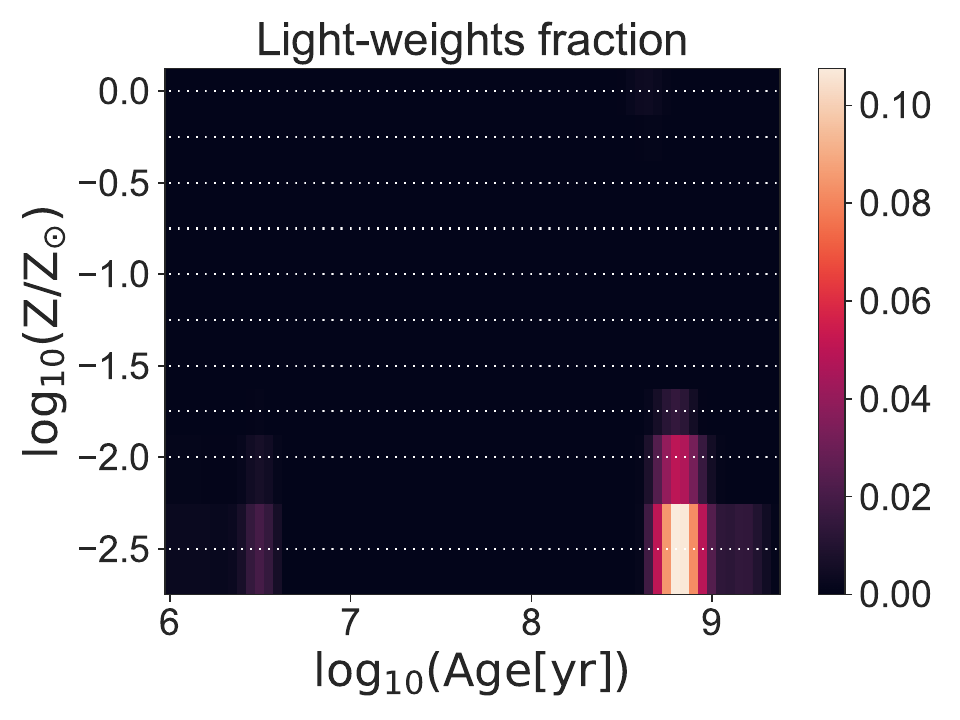}{mini-quenched}{}

\SFHpair{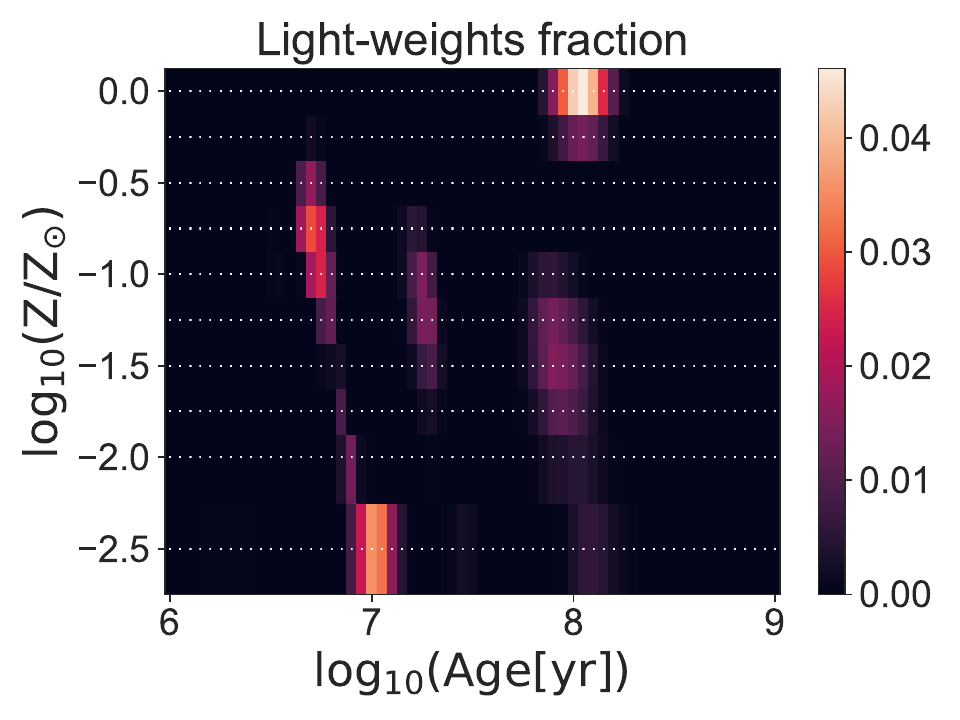}{lulling}{}

\section{Results: Lulling galaxies} \label{sec:Lulling}

\begin{figure*}[t]
  \centering
  \begin{subfigure}[t]{0.48\textwidth}
    \centering
    \includegraphics[width=\linewidth]{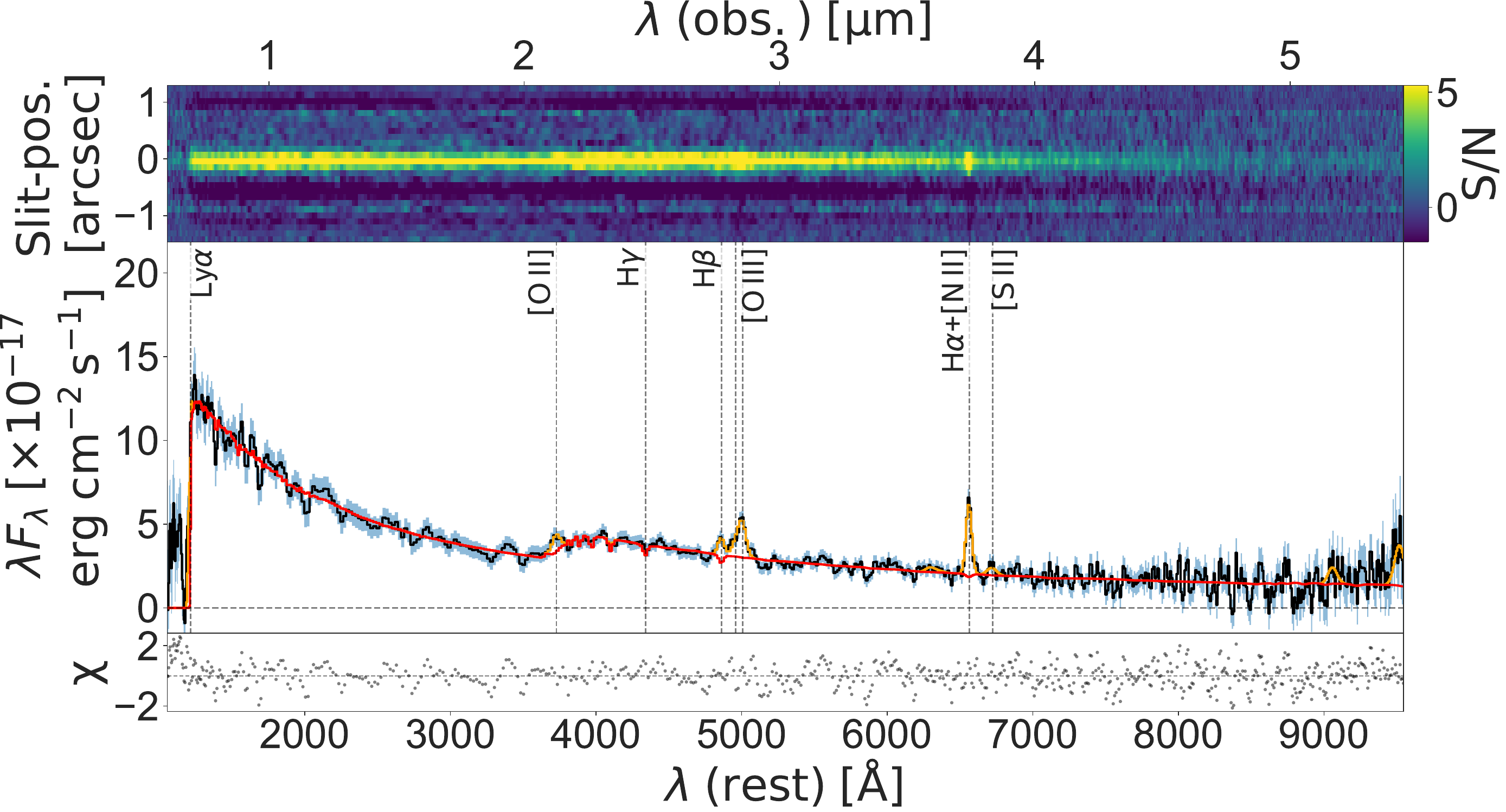}
    \caption{NIRSpec/PRISM spectrum of the lulling galaxy ID 309077, showing a clear Balmer break together with weak emission lines.}\label{fig:lulling-309077}
  \end{subfigure}\hfill
  \begin{subfigure}[t]{0.48\textwidth}
    \centering
    \includegraphics[width=\linewidth]{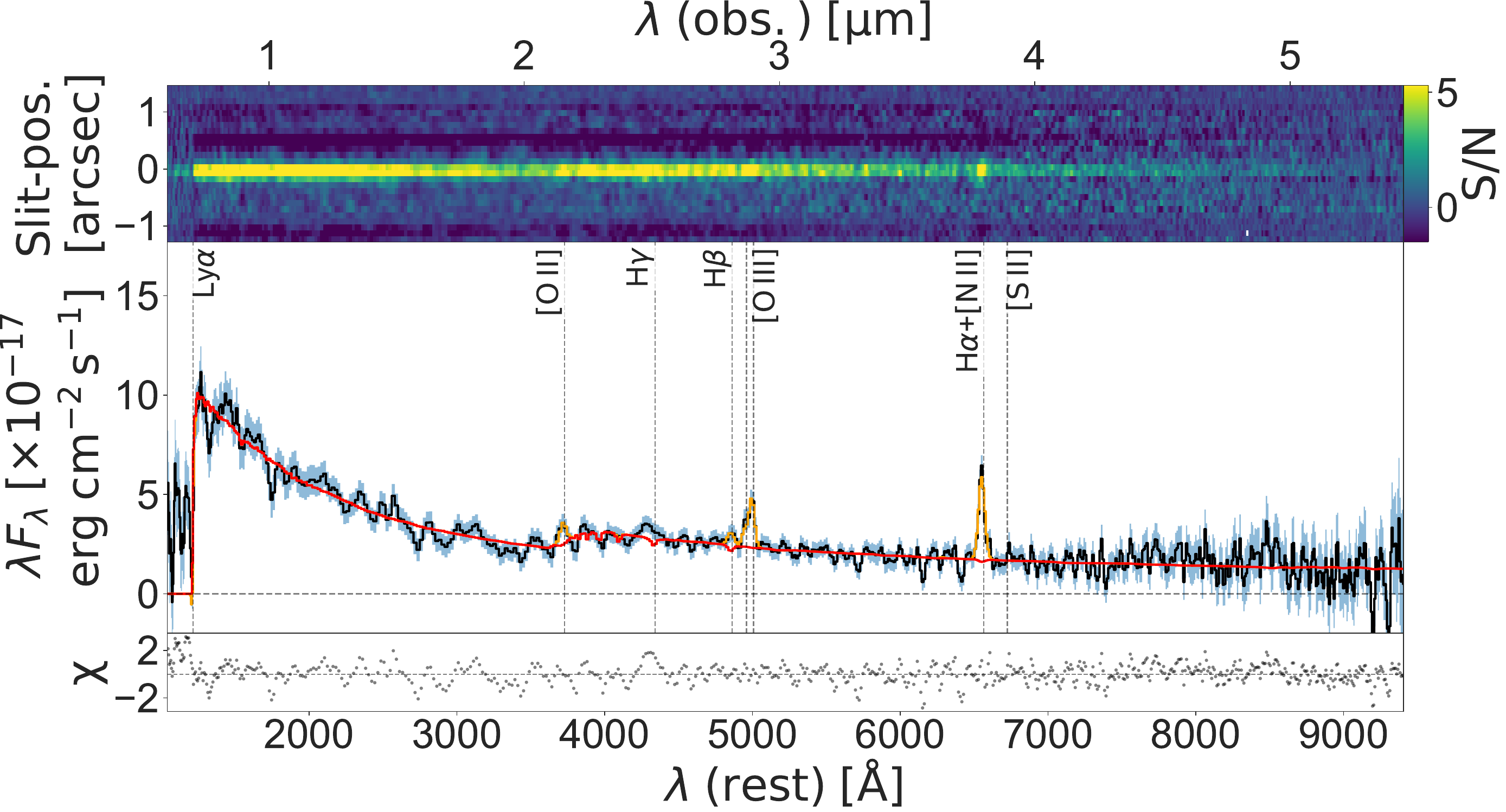}
    \caption{NIRSpec/PRISM spectrum of the lulling galaxy ID 120014, showing a clear Balmer break together with weak emission lines.}\label{fig:lulling-120014}
  \end{subfigure}

  \begin{subfigure}[t]{0.48\textwidth}
    \centering
    \includegraphics[width=\linewidth]{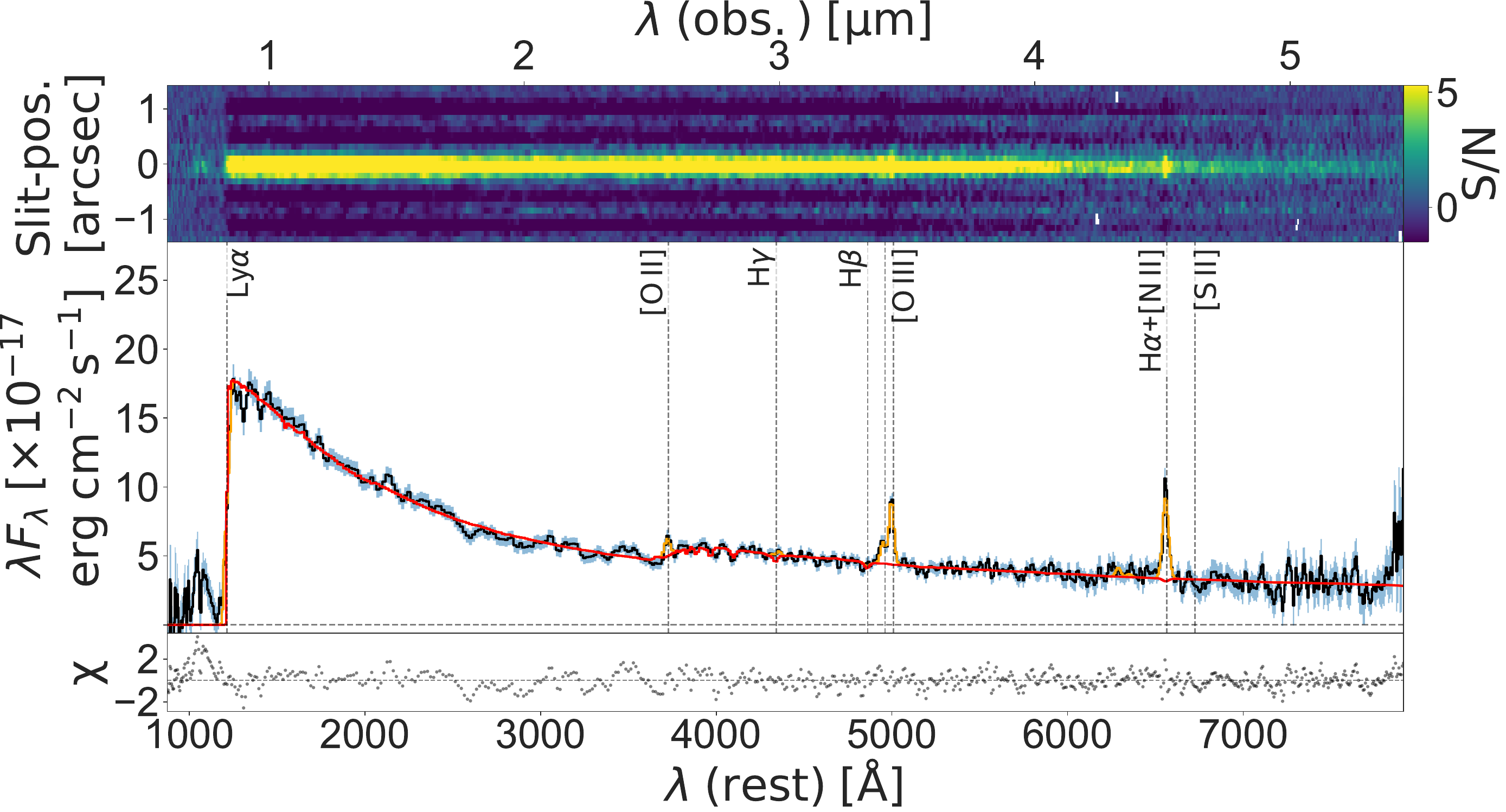}
    \caption{NIRSpec/PRISM spectrum of the lulling galaxy ID 202438, showing a weak Balmer break together with weak emission lines.}\label{fig:lulling-202438}
  \end{subfigure}\hfill
  \begin{subfigure}[t]{0.48\textwidth}
    \centering
    \includegraphics[width=\linewidth]{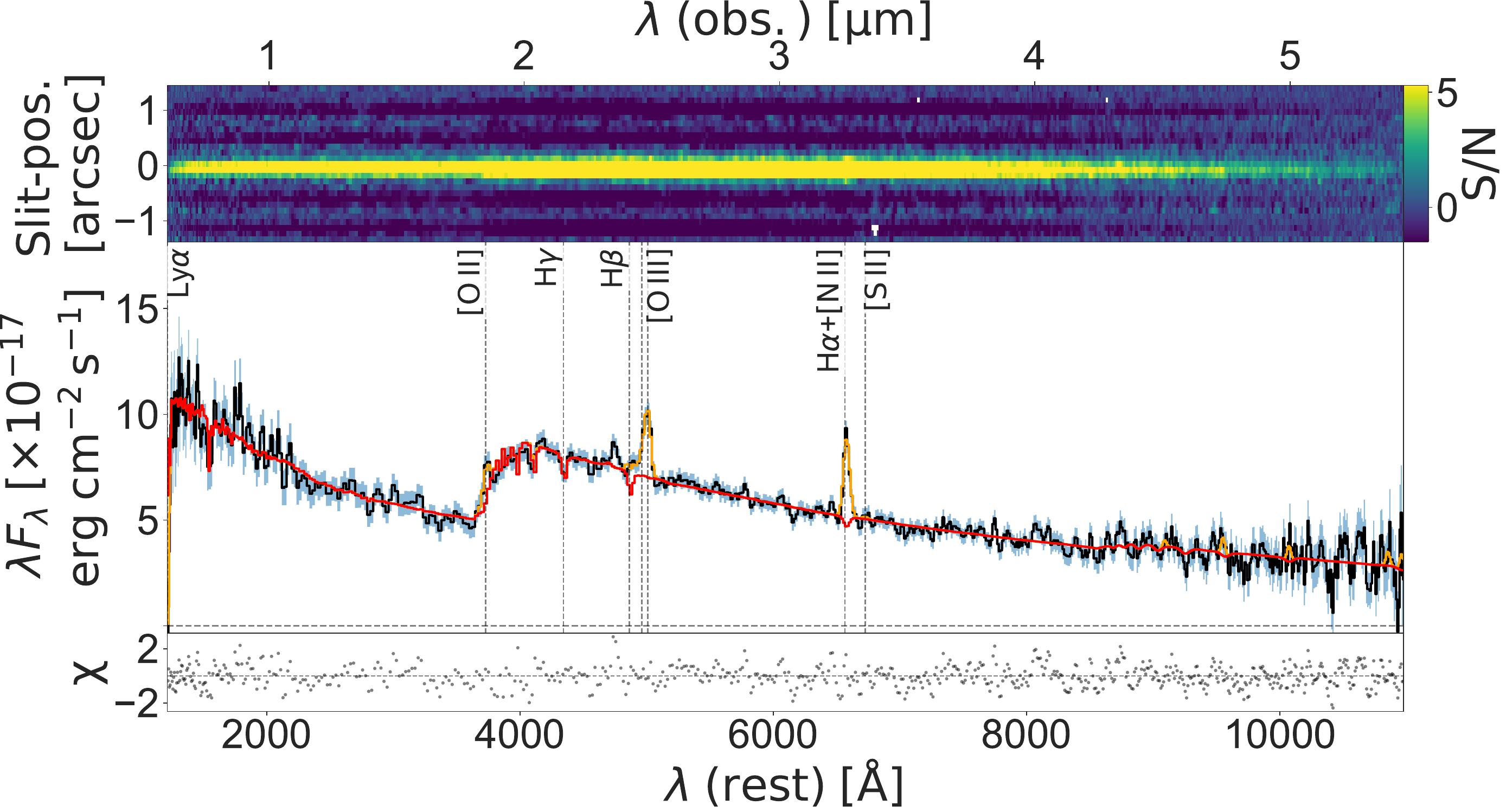}
    \caption{NIRSpec/PRISM spectrum of the lulling galaxy ID 108487. This system exhibits a deeper Balmer break and shallower UV slope than the other lulling galaxies.}\label{fig:lulling2-108487}
  \end{subfigure}

  \caption{OASIS spectra of lulling galaxies at $z>4$. Panels and colors as in Fig.~\ref{fig:mq-181834}.}
  \label{fig:lulling_all4}
\end{figure*}

\begin{figure*}[t]
  \centering

  \begin{subfigure}[t]{0.48\textwidth}
    \centering
    \includegraphics[width=\linewidth]{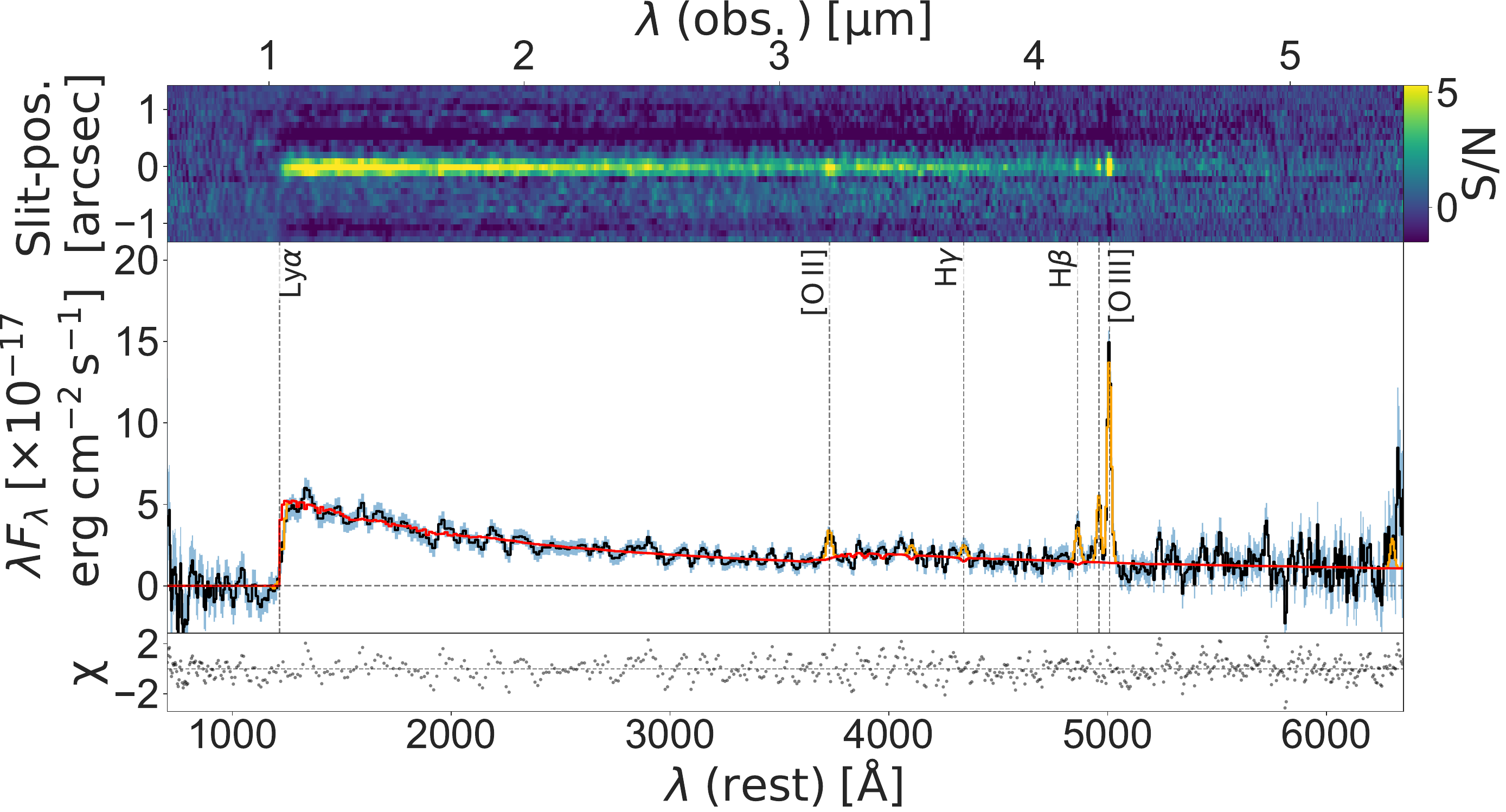}
    \caption{NIRSpec/PRISM spectrum of the nearly main-sequence galaxy ID 138085.
This source was originally selected with Pri~1 as a mini-quenched candidate. However, the spectrum reveals clear emission lines. Strong \oiii\ emission mimics a Balmer break in the photometry (Sect.~\ref{sec:phot_selection}): the galaxy is a genuine nearly-MS galaxy at $z=7.6$, misclassified as a mini-quenched candidate.}\label{fig:nearlyMS-138085}
  \end{subfigure}\hfill
  \begin{subfigure}[t]{0.48\textwidth}
    \centering
    \includegraphics[width=\linewidth]{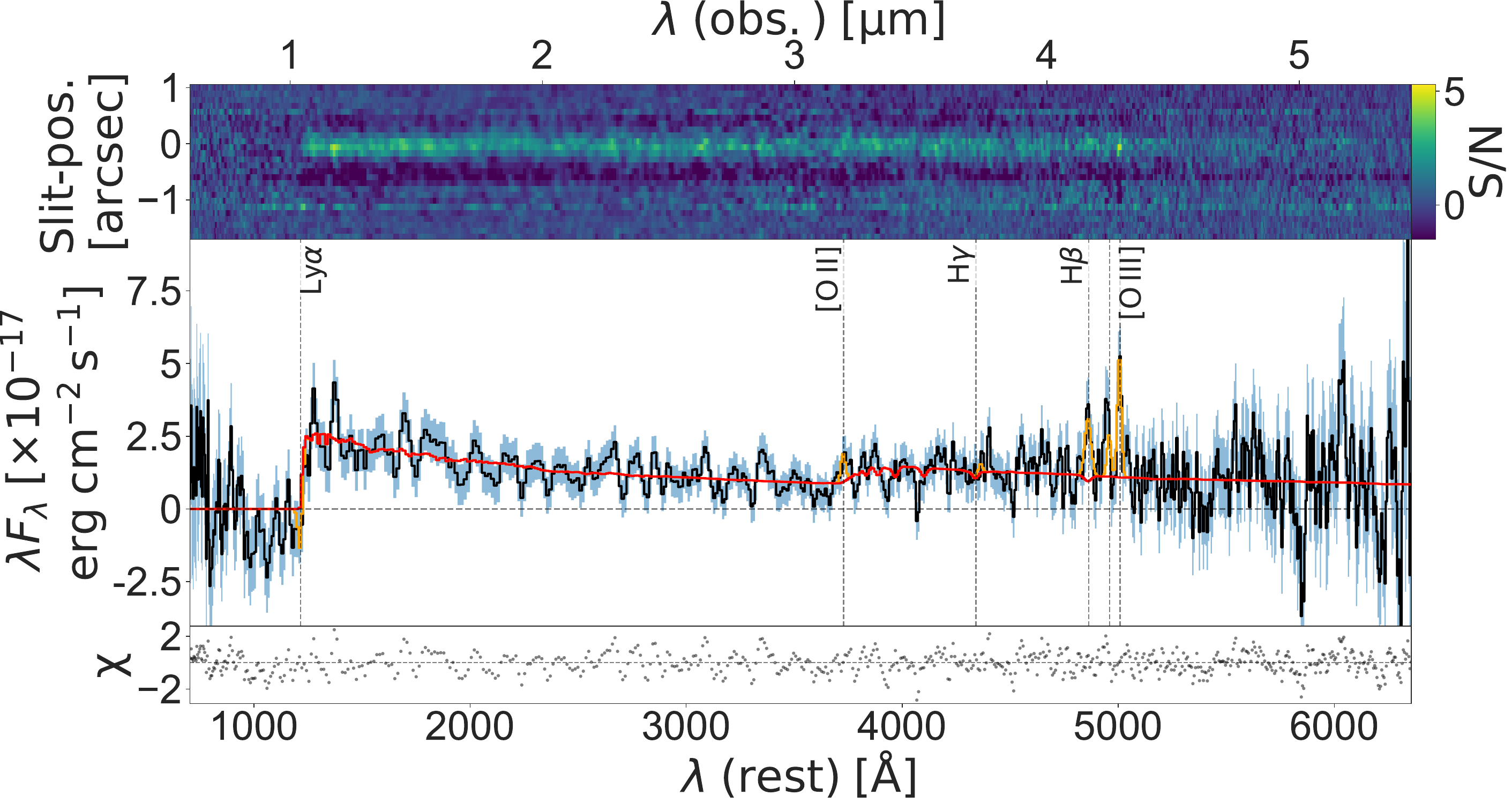}
    \caption{NIRSpec/PRISM spectrum of the nearly main-sequence galaxy ID 136522.
The spectrum shows weak emission lines, placing the galaxy close to the SFMS. Given its UV faintness and the presence of a noticeable Balmer break, this source may be a rejuvenating system, or at least one that has only recently increased its star-formation activity following a lull. However, based on visual inspection of the photometry, this object was assigned Pri~2 as a mini-quenched or more strongly lulling candidate.}\label{fig:nearlyMS-136522}
  \end{subfigure}

  \caption{Spectra of two nearly main-sequence galaxies at $z>4$.
Based on photometric inspection, both sources were expected to be less actively star-forming and to exhibit stronger Balmer breaks.}
  \label{fig:nearlyMS_spectra}
\end{figure*}

\begin{table*}
    \centering
    \caption{Properties of the $z>4$ mini-quenched, lulling and nearly-MS
    galaxies (classes defined in Sect.~\ref{sec:intro}): coordinates (RA,
    Dec; deg), F444W magnitudes ($m_{\mathrm{F444W}}$; AB), target priorities
    (Pri), redshifts ($z$), stellar masses ($\log_{10}(M_\star/M_\odot)$),
    nebular star-formation rates
    ($\log_{10}(\mathrm{SFR}/[M_\odot\,\mathrm{yr}^{-1}])$; $3\sigma$ upper
    limits marked by $<$), light- and mass-weighted stellar ages
    ($\langle\log_{10}(\mathrm{Age/yr})\rangle$), stellar reddening
    ($E(B{-}V)_\star$; mag) and Balmer-break strengths ($R$,
    Eq.~\eqref{eq:balmerbreak}). Within each class, galaxies are ordered by
    decreasing $\Delta_{\rm MS}$. Uncertainties are statistical 16th--84th
    percentiles, and robust standard deviations for $R$ (Sect.~\ref{sec:Mini-quenched}); n.m.: not measurable
    ($\sigma_R>1$). $^{\dagger}$\,Hand-selected photometric mini-quenched or
    lulling candidate (Pri~1--3; Sect.~\ref{sec:phot_selection}). The last block lists the
    hand-selected candidates that the spectroscopy places above the SFMS.}
    \label{tab:oasis_ids_coords_m444w_pri_z_mass_sfr_age_redd_balmer_burst}
    \renewcommand{\arraystretch}{1.35}
    \setlength{\tabcolsep}{4pt}
    \begin{tabular}{lllcccllllll}
        \hline\hline
        ID & RA & Dec & $m_{\mathrm{F444W}}$ & Pri & $z$ & $M_\star$ & SFR & Age$_{\rm lw}$ & Age$_{\rm mw}$ & $E(B{-}V)_\star$ & $R$ \\
        \hline
        \multicolumn{12}{c}{\textbf{Mini-quenched galaxies}}\\
        \hline
        206569 & 53.14567 & $-27.78524$ & 29.55 & 8 & 4.0 & $8.32^{+0.03}_{-0.02}$ & $<-1.58$ & $8.57^{+0.03}_{-0.03}$ & $8.86^{+0.05}_{-0.04}$ & 0.00 & $1.01\pm0.24$ \\
        181834 & 53.12796 & $-27.85753$ & 28.22 & 8 & 5.0 & $8.94^{+0.04}_{-0.05}$ & $<-1.09$ & $8.12^{+0.06}_{-0.06}$ & $8.72^{+0.05}_{-0.05}$ & 0.13 & $2.07\pm0.36$ \\
        168352$^{\dagger}$ & 53.07772 & $-27.87958$ & 27.81 & 1 & 5.1 & $8.85^{+0.04}_{-0.04}$ & $<-1.32$ & $8.09^{+0.04}_{-0.03}$ & $8.45^{+0.06}_{-0.05}$ & 0.10 & $1.96\pm0.18$ \\
        \hline
        \multicolumn{12}{c}{\textbf{Lulling galaxies}}\\
        \hline
        120014 & 53.18916 & $-27.78795$ & 28.13 & 5 & 4.8 & $8.64^{+0.08}_{-0.11}$ & $-0.62^{+0.25}_{-0.12}$ & $7.45^{+0.09}_{-0.09}$ & $8.48^{+0.18}_{-0.19}$ & 0.07 & $1.40\pm0.15$ \\
        309077 & 53.20316 & $-27.80017$ & 28.00 & 7 & 4.7 & $8.65^{+0.06}_{-0.06}$ & $-0.78^{+0.03}_{-0.03}$ & $7.58^{+0.07}_{-0.07}$ & $8.33^{+0.13}_{-0.13}$ & 0.09 & $1.39\pm0.11$ \\
        114769$^{\dagger}$ & 53.18632 & $-27.79561$ & 27.94 & 1 & 7.5 & $8.61^{+0.04}_{-0.04}$ & $-0.54^{+0.12}_{-0.16}$ & $7.53^{+0.07}_{-0.08}$ & $7.91^{+0.05}_{-0.04}$ & 0.06 & $1.56\pm0.13$ \\
        108487 & 53.14888 & $-27.80565$ & 27.20 & 7 & 4.0 & $8.95^{+0.04}_{-0.04}$ & $-0.86^{+0.02}_{-0.02}$ & $7.38^{+0.06}_{-0.07}$ & $8.63^{+0.05}_{-0.04}$ & 0.19 & $1.74\pm0.09$ \\
        202438$^{\dagger}$ & 53.15999 & $-27.79974$ & 27.39 & 3 & 5.9 & $9.36^{+0.05}_{-0.04}$ & $-0.29^{+0.03}_{-0.03}$ & $7.38^{+0.11}_{-0.15}$ & $8.85^{+0.06}_{-0.05}$ & 0.14 & $1.17\pm0.04$ \\
        \hline
        \multicolumn{12}{c}{\textbf{Nearly-MS galaxies}}\\
        \hline
        140132 & 53.15464 & $-27.76406$ & 28.14 & 7 & 4.8 & $8.43^{+0.06}_{-0.06}$ & $-0.15^{+0.10}_{-0.07}$ & $7.20^{+0.10}_{-0.10}$ & $7.40^{+0.13}_{-0.10}$ & 0.16 & $1.23\pm0.13$ \\
        189400 & 53.08544 & $-27.84246$ & 27.68 & 7 & 4.3 & $8.40^{+0.15}_{-0.13}$ & $-0.26^{+0.08}_{-0.07}$ & $7.08^{+0.06}_{-0.07}$ & $7.42^{+0.62}_{-0.30}$ & 0.21 & $1.45\pm0.10$ \\
        186744 & 53.06360 & $-27.84873$ & 25.37 & 7 & 4.2 & $9.25^{+0.04}_{-0.06}$ & $0.47^{+0.02}_{-0.02}$ & $6.95^{+0.05}_{-0.06}$ & $7.17^{+0.20}_{-0.18}$ & 0.26 & $1.44\pm0.02$ \\
        112587 & 53.19885 & $-27.79920$ & 30.58 & 8 & 5.9 & $7.39^{+0.11}_{-0.15}$ & $<-1.16$ & $7.07^{+0.14}_{-0.16}$ & $7.13^{+0.12}_{-0.12}$ & 0.16 & n.m. \\
        116340 & 53.16690 & $-27.79306$ & 31.78 & 8 & 4.4 & $7.97^{+0.08}_{-0.09}$ & $-0.84^{+0.02}_{-0.03}$ & $7.56^{+0.16}_{-0.14}$ & $8.75^{+0.12}_{-0.11}$ & 0.12 & n.m. \\
        214959 & 53.19311 & $-27.75699$ & 27.93 & 5 & 6.2 & $9.01^{+0.03}_{-0.03}$ & $0.44^{+0.04}_{-0.04}$ & $6.82^{+0.07}_{-0.06}$ & $8.78^{+0.03}_{-0.03}$ & 0.16 & $1.01\pm0.05$ \\
        202661$^{\dagger}$ & 53.14396 & $-27.79896$ & 26.06 & 3 & 5.6 & $9.24^{+0.02}_{-0.02}$ & $0.55^{+0.02}_{-0.02}$ & $6.97^{+0.03}_{-0.03}$ & $7.07^{+0.03}_{-0.02}$ & 0.31 & $1.31\pm0.02$ \\
        214417 & 53.19411 & $-27.75927$ & 29.78 & 8 & 7.6 & $7.87^{+0.21}_{-0.47}$ & $<-0.58$ & $7.20^{+0.21}_{-0.24}$ & $8.48^{+0.36}_{-0.43}$ & 0.00 & $1.22\pm0.68$ \\
        112304 & 53.15595 & $-27.79966$ & 28.26 & 8 & 4.8 & $8.48^{+0.06}_{-0.07}$ & $-0.35^{+0.19}_{-0.16}$ & $7.34^{+0.12}_{-0.12}$ & $7.84^{+0.07}_{-0.06}$ & 0.17 & $1.59\pm0.15$ \\
        138085$^{\dagger}$ & 53.19208 & $-27.76662$ & 28.07 & 1 & 7.6 & $8.37^{+0.06}_{-0.09}$ & $-0.09^{+0.06}_{-0.07}$ & $7.19^{+0.09}_{-0.13}$ & $7.26^{+0.08}_{-0.12}$ & 0.11 & $1.69\pm0.16$ \\
        310810$^{\dagger}$ & 53.20751 & $-27.79165$ & 27.87 & 2 & 5.4 & $8.88^{+0.02}_{-0.02}$ & $0.15^{+0.34}_{-0.25}$ & $7.88^{+0.05}_{-0.05}$ & $8.03^{+0.06}_{-0.05}$ & 0.18 & $1.84\pm0.19$ \\
        202141 & 53.14161 & $-27.80065$ & 25.08 & 7 & 4.4 & $9.66^{+0.10}_{-0.17}$ & $0.77^{+0.02}_{-0.02}$ & $6.95^{+0.04}_{-0.03}$ & $8.16^{+0.33}_{-0.40}$ & 0.32 & $1.60\pm0.02$ \\
        202852$^{\dagger}$ & 53.14295 & $-27.79829$ & 27.60 & 2 & 5.6 & $8.65^{+0.04}_{-0.08}$ & $-0.14^{+0.08}_{-0.07}$ & $7.00^{+0.05}_{-0.05}$ & $7.18^{+0.01}_{-0.11}$ & 0.28 & $1.41\pm0.06$ \\
        635752 & 53.15161 & $-27.80331$ & 27.26 & 7 & 4.8 & $8.94^{+0.16}_{-0.20}$ & $-0.07^{+0.08}_{-0.08}$ & $6.84^{+0.15}_{-0.13}$ & $8.62^{+0.26}_{-0.29}$ & 0.22 & $1.45\pm0.07$ \\
        136522$^{\dagger}$ & 53.20663 & $-27.76820$ & 27.97 & 2 & 7.6 & $8.58^{+0.15}_{-0.21}$ & $-0.10^{+0.09}_{-0.12}$ & $7.69^{+0.17}_{-0.18}$ & $8.26^{+0.30}_{-0.28}$ & 0.02 & $1.60\pm0.36$ \\
        191580 & 53.06781 & $-27.83540$ & 25.38 & 7 & 4.2 & $9.68^{+0.06}_{-0.08}$ & $0.51^{+0.02}_{-0.02}$ & $6.95^{+0.02}_{-0.02}$ & $8.37^{+0.18}_{-0.14}$ & 0.32 & $1.37\pm0.01$ \\
        632746 & 53.10722 & $-27.89060$ & 26.48 & 6 & 6.2 & $9.68^{+0.04}_{-0.05}$ & $0.77^{+0.02}_{-0.02}$ & $7.35^{+0.06}_{-0.06}$ & $8.93^{+0.01}_{-0.01}$ & 0.21 & $1.32\pm0.06$ \\
        \hline
        \multicolumn{12}{c}{\textbf{Photometric candidates above the main sequence}}\\
        \hline
        182899$^{\dagger}$ & 53.12536 & $-27.85617$ & 27.74 & 2 & 4.8 & $8.03^{+0.09}_{-0.13}$ & $-0.09^{+0.07}_{-0.07}$ & $6.90^{+0.05}_{-0.06}$ & $7.14^{+0.15}_{-0.13}$ & 0.20 & $1.67\pm0.14$ \\
        85427$^{\dagger}$ & 53.06866 & $-27.83498$ & 25.88 & 3 & 5.4 & $9.22^{+0.02}_{-0.03}$ & $0.99^{+0.01}_{-0.01}$ & $6.95^{+0.09}_{-0.09}$ & $7.07^{+0.05}_{-0.10}$ & 0.24 & $1.30\pm0.02$ \\
        \hline
    \end{tabular}

\end{table*}

\begin{figure*}
    \centering
    \includegraphics[width=0.8\linewidth]{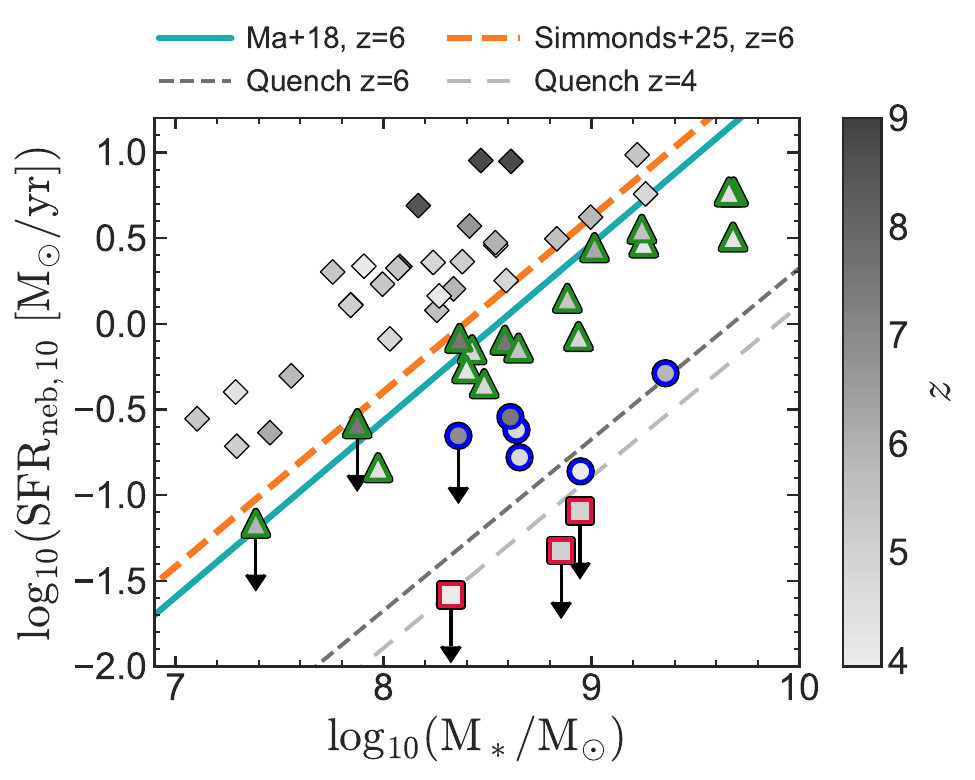}
    \caption{
\textbf{Three mini-quenched and five lulling galaxies lie far below the SFMS.} The SFR--\Mstar\ plane for the $z \simeq 4$--9 OASIS galaxies, with the mini-quenched and lulling populations highlighted.
Red squares mark the three mini-quenched galaxies, whose nebular SFR upper
limits place them below the redshift-dependent quenching threshold
(Eq.~\eqref{eq:quenching-threshold}; Figs.~\ref{fig:mq-181834},
\ref{fig:mq-168352} and \ref{fig:mq-206569}). Blue circles mark lulling
galaxies, which lie clearly below the main sequence but above the threshold
(Figs.~\ref{fig:lulling-114769} and \ref{fig:lulling_all4}). Green triangles mark nearly-MS galaxies, $<0.6$\,dex below the redshift-dependent \citet{Simmonds+2025}
SFMS (Fig.~\ref{fig:nearlyMS_spectra}), and gray diamonds the remainder of
the sample. Marker fills encode redshift. Downward arrows are $3\sigma$ upper
limits on the nebular SFR. Three of these systems have limits too shallow to
classify them securely (Appendix~\ref{appendix:faint_low_sfr}). The turquoise and orange lines
show the main-sequence relations of \citet{Ma+2018} and \citet{Simmonds+2025},
and the dark and light gray dashed lines the quenching threshold, all evaluated
at the fixed redshifts indicated for reference. Because $\Delta_{\rm MS}$ is evaluated at each galaxy's own redshift, a galaxy can lie above the lines drawn here while lying below its own main sequence. 
}
    \label{fig:SFR-mass}
\end{figure*}

One of the highest-priority photometric mini-quenched candidates turns out to be lulling rather than mini-quenched (Table~\ref{tab:oasis_ids_coords_m444w_pri_z_mass_sfr_age_redd_balmer_burst}): the Pri~1 galaxy ID~114769 at redshift $z=7.5$ was selected as a mini-quenched candidate based on visual inspection: the absence of emission-line excess and the presence of a putative Balmer break. However, the NIRSpec PRISM spectrum (Fig.~\ref{fig:lulling-114769}) reveals that while the galaxy is close to the mini-quenched regime and exhibits a pronounced Balmer break, we detect very weak H$\beta$ emission at $3.3\sigma$. We also detect weak \oiii\ emission ($3.5\sigma$). Such a weak \Hbeta\ line cannot be detected photometrically (Appendix~\ref{app:phot_spec}), yet it places 114769 at $\Delta_{\mathrm{MS}}=-0.96$, well below the SFMS but above the quenching threshold. More generally, the photometric lulling and mini-quenched candidates (Pri~1--3, marked with $^{\dagger}$ in Table~\ref{tab:oasis_ids_coords_m444w_pri_z_mass_sfr_age_redd_balmer_burst}) span more than $2$\,dex in $\Delta_{\mathrm{MS}}$, from $\Delta_{\mathrm{MS}}\leq-1.65$ (168352) to $+0.47$ (182899; Sect.~\ref{sec:phot_selection}). Only spectroscopy can quantify their star-formation level reliably.

Unlike the three mini-quenched galaxies discussed in Sect.~\ref{sec:Mini-quenched}, the reconstructed star-formation history of ID~114769 indicates a non-negligible contribution from younger stellar populations in the light-weighted SFH (Fig.~\ref{fig:SFH-114769}). This suggests that galaxies do not necessarily -- or at least not always -- cycle rapidly between intense starbursts and fully quenched phases. Instead, they can persist in a low-SFR ``lulling'' \citep[or ``smoldering'';][]{Trussler+2025} mode for extended periods of time (Sect.~\ref{sec:ages} and OASIS Paper~II).

We identify four additional galaxies with similar properties -- namely a pronounced Ly$\alpha$ drop, Balmer breaks, and weak nebular emission lines, including H$\beta$ and/or H$\alpha$ tracing recent star formation, and \oiii\ (see Figs.~\ref{fig:lulling-309077}--\ref{fig:lulling2-108487}). Their weak Balmer lines place them all above the quenching threshold, but clearly below the SFMS, see Fig.~\ref{fig:SFR-mass} (blue circles).

Like the mini-quenched galaxies, the lulling systems reside outside overdense environments and lack massive nearby companions, pointing again to internal rather than environmental causes.

In 309077, 120014, 202438 and 108487 (Figs.~\ref{fig:lulling-309077}, \ref{fig:lulling-120014},
\ref{fig:lulling-202438}, and \ref{fig:lulling2-108487}), the fit residuals around \Halpha\ show
structure that the single Gaussian component (yellow) does not capture. At PRISM resolution, a
broad component and \nii\ emission blended with \Halpha\ cannot be separated; we discuss a possible AGN contribution in OASIS Paper~II. Further, although all four galaxies are spatially extended, their morphologies cannot be adequately described by a single-component fit (see OASIS Paper~III for more details on morphology). This points to a complex internal structure, potentially including clumpy components, and leaves open the possibility that they may also host unresolved point-source components.

\section{Results: Accuracy and completeness of photometric selection}\label{sec:phot_selection}

The hand-curated tiers of the OASIS input catalog (Pri~1--3; Sect.~\ref{sec:target_selection}) rest entirely on the photometric signature of a flat
rest-frame optical continuum with weak or absent nebular emission. The
spectroscopy now allows us to test how reliable that signature is. Ten
visually selected mini-quenched and lulling candidates received a shutter
across the two pointings; all ten yielded secure spectroscopic redshifts, and
all ten fall at $z>4$. 

The photometric redshifts themselves proved accurate. For these ten targets
the median offset is $\Delta z/(1+z) = 0.023$, and nine of ten agree with the spectroscopic
redshift to better than $\Delta z/(1+z) = 0.05$; the largest deviation,
$0.085$, is that of ID~182899. Where the selection failed, it therefore failed
in diagnosing the star-formation state rather than the redshift.

As summarized in Table~\ref{tab:oasis_ids_coords_m444w_pri_z_mass_sfr_age_redd_balmer_burst}, where these candidates are marked with an $^\dagger$, that diagnosis yields mixed results. Three of the ten candidates were
confirmed as mini-quenched or lulling systems, as anticipated from visual
inspection of the photometry; in the remaining seven, the photometric
signature was not borne out. Although their photometric redshifts were
approximately correct, slight offsets, combined with emission lines -- namely
\oii, \Hbeta, \oiii\ and \Halpha -- conspired to mimic
a Balmer break with weak emission lines. The spectroscopy, however, shows that
at least part of the broad-band flux is attributable to emission lines rather
than to a genuine break. For most of these objects the failure is nonetheless
one of degree rather than of kind: five are nearly-MS systems that remain
marginally below the SFMS, one lies just above the SFMS ($\Delta_{\mathrm{MS}}=+0.23$), and one
further above it ($+0.47$). The last represents the only outright failure of the selection; in
the other six the photometry correctly identified a comparatively quiescent
system, but overestimated how strongly its star formation was suppressed.

Photometry alone, at least with our selection procedure and the data available
to us, was insufficient to establish the degree of suppression robustly. We
note that medium-band imaging can improve matters, since medium bands sample
the \oiii+\Hbeta\ complex and the Balmer break simultaneously and thus help to
separate line emission from a genuine break \citep[e.g.,][]{Endsley+2025}.
However, in our data set coverage was uneven: all ten candidates have F410M,
as the selection requires, but none of the seven unconfirmed ones is covered
by F250M, F300M, F430M, F460M or F480M, whereas the two best-characterized
confirmations are. \citet{Trussler+2025}, by contrast, imposed stronger
constraints on both $m_{\mathrm{F335M}}-m_{\mathrm{F356W}}$ and
$m_{\mathrm{F410M}}-m_{\mathrm{F444W}}$. Three of their candidates were
observed by OASIS, all three of which we had also selected by eye, and two are
confirmed here; among our seven remaining candidates only one is. Their
tighter criteria therefore appear to yield a purer sample, albeit on very
small numbers, and plausibly at some cost in completeness. Medium bands are not a complete remedy, however. ID~138085 (Fig.~\ref{fig:nearlyMS-138085}) has both
F335M and F410M photometry and is itself part of the \citet{Trussler+2025}
sample, yet it is among our unconfirmed candidates. Its grism redshift ($z=7.1$, against the spectroscopic $z=7.6$) placed \oiii\ on the edge of F410M, so that a flat F410M$-$F444W color mimicked weak line emission. Medium bands nonetheless shift the odds appreciably, which
motivated the filter set adopted for the first OASIS NIRCam parallel (Sect.~\ref{sec:nircam_parallel}), adding F250M, F300M and F480M over the JOF.

Conversely, and perhaps more importantly, we uncover a substantial number of
bona fide mini-quenched and lulling galaxies that had been assigned low
priority in the MSA input catalog. Of the eight mini-quenched and lulling
galaxies confirmed at $z>4$, only three were visually selected; the remaining
five were drawn from Pri~5, 7 and 8. In particular, two of the three
mini-quenched galaxies, IDs~181834 and 206569, came from Pri~8, the tier that
carries no selection criterion beyond inclusion in the parent catalog. That
such objects surface from tiers carrying no information on star-formation
state suggests that mini-quenched and lulling galaxies emerge naturally once
sufficiently deep spectroscopy is obtained, implying that they may be more
common at the faint end of the galaxy population than previously recognized.
It also highlights the need for further magnitude-selected spectroscopic
samples -- either alone or combined with redshift and stellar-mass cuts -- to
establish the number density of such objects. Given the various selection
effects, and the attendant purity and completeness uncertainties, we do not
attempt a number-density estimate from these candidates; an order-of-magnitude estimate from the
tiers without selection on star-formation state is given in Sect.~\ref{sec:duty_cycle}.

The same limitation appears in the physical parameters themselves. Our input
catalog carried Prospector stellar-population fits to the same JADES v0.9.5
photometry, restricted to reduced $\chi^2\leq1$ \citep{Simmonds+2024}, from which the
Pri~4 and Pri~5 tiers were drawn. Compared with our Balmer-line star-formation
rates these are accurate for the star-forming majority: the median offset $\log\mathrm{SFR_{neb}}-\log\mathrm{SFR_{phot}}$, with $\mathrm{SFR_{phot}}$ the Prospector rate averaged over 10\,Myr, is $-0.09$\,dex, with a normalized median absolute deviation of 0.21\,dex and $\rho=+0.80$. The agreement degrades systematically with suppression, however: the offset falls from $-0.05$\,dex above the main sequence to $-0.54$\,dex for galaxies more than 0.5\,dex below it, so that the photometric fits overestimate the star-formation rates of the suppressed galaxies by a factor of ${\sim}3.5$, and they place our mini-quenched galaxies 181834 and 168352 0.37--0.73\,dex above the $3\sigma$ upper limits from their Balmer lines.

There are two reasons. First, the lines that set the current state are weak: at the fluxes measured in the spectra, the Balmer and [O\,\textsc{iii}] lines of the lulling galaxies add only 7--11 per cent to the band that contains them, and the $3\sigma$ limits of the mini-quenched galaxies 2--8 per cent, comparable to realistic photometric uncertainties. The rest-frame equivalent widths of H$\alpha$+[N\,\textsc{ii}] of the lulling galaxies, 56--141\,\AA\ against a median of ${\approx}830$\,\AA\ above the SFMS, are comparable to the ${\sim}50$--100\,\AA\ to which medium-band colors are sensitive \citep{Trussler+2025}. Second, these excesses are comparable to the changes in color that a slightly older stellar population or more dust would produce in the same bands, so that a modest excess cannot be uniquely attributed to line emission. Photometry alone therefore cannot distinguish lulling from mini-quenched galaxies (Appendix~\ref{app:phot_spec}). The fits also miss the recent decline: for the lulling galaxies they predict a median change over the last 100\,Myr of only $-0.15$\,dex, against $-0.75$\,dex from the spectra (OASIS Paper~II). Photometric SED fitting therefore tends to
\emph{under}estimate suppression in precisely the objects of interest, which
plausibly contributes to so many genuinely suppressed systems being left in
the filler tiers.

Not even fixing the redshift to the spectroscopic value removes this bias. The JADES DR5 Prospector fits \citep{Duan+2026_DR5}, run at our spectroscopic redshifts and with up to eight medium bands, recover the mini-quenched state of 168352 and 181834, but still overestimate the star-formation rates of the galaxies more than 0.5\,dex below the SFMS by a median 0.53\,dex and place the mini-quenched galaxy 206569 on the SFMS and the lulling galaxy 309077 above it. They also recover the recent change in star formation only loosely ($\rho=0.67$ against the spectroscopic value; OASIS Paper~II; see also \citealt{Wan+2024}). The spectra, in contrast, resolve the lines directly, detecting H$\alpha$+[N\,\textsc{ii}] at $\mathrm{S/N}=16$--23 in four of the lulling galaxies and excluding Balmer emission at $3\sigma$ down to $(6$--$10)\times10^{-20}\,\mathrm{erg\,s^{-1}\,cm^{-2}}$ in the mini-quenched ones. We note that such non-detections alone, e.g., from grism or grating spectroscopy, can rule out photometric solutions with strong recent star formation \citep{DEugenio+2026_DarkHorse}.

Both the unconfirmed candidates and the unrecognized mini-quenched and lulling
systems underline the necessity of spectroscopy --- not only to establish the
star-formation state robustly at the epoch of observation, but also to recover
reliable and physically meaningful star-formation histories. We note, finally,
that the two failure modes pull in opposite directions and do not cancel: in
this sample the photometric selection both overestimated suppression in seven
of ten candidates and overlooked five of the eight suppressed galaxies that
the spectroscopy ultimately revealed.

\section{Results: Stellar populations across the SFR--$M_\star$ plane}\label{sec:pop_plane}

In this section, we relate the Balmer breaks, stellar ages and dust attenuation of our galaxies to their location in the SFR--$M_\star$ plane, and estimate the fraction and number density of the suppressed galaxies.

\subsection{Balmer breaks}\label{sec:bb_results}

\begin{figure}[t]
    \centering
    \includegraphics[width=\linewidth]{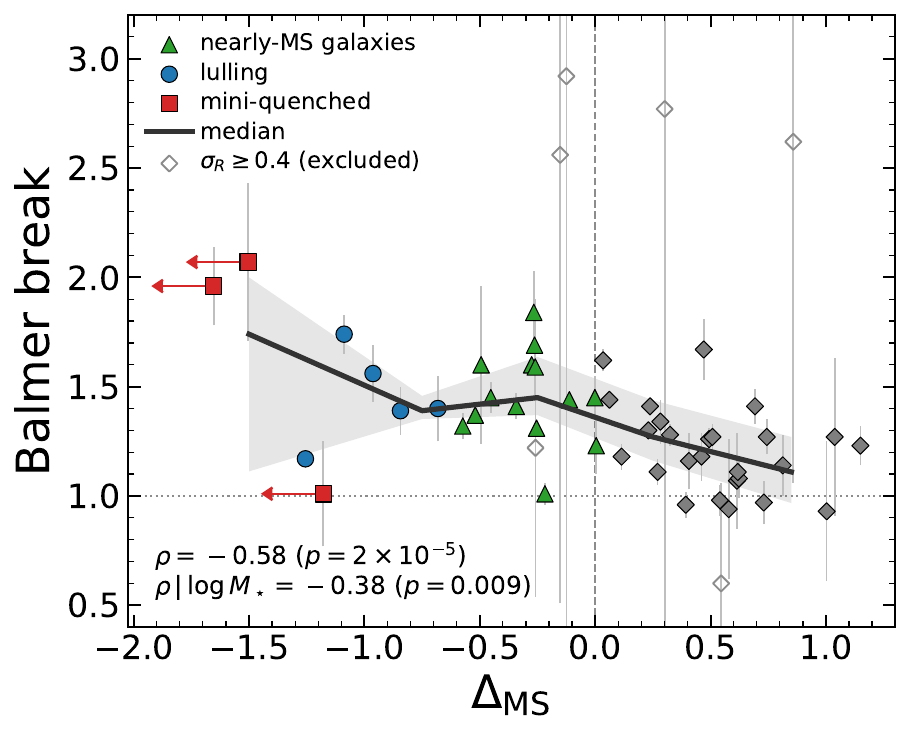}
    \caption{ \textbf{Galaxies below the SFMS have stronger Balmer breaks.} Balmer-break strength against distance from the star-forming main sequence, $\Delta_{\mathrm{MS}} = \log_{10}\mathrm{SFR} -
        \log_{10}\mathrm{SFR_{MS}}(M_\star,z)$, adopting the
        \citet{Simmonds+2025} relation evaluated at each galaxy's own
        redshift. Mini-quenched galaxies are shown as red squares, lulling
        galaxies as blue circles and nearly-MS galaxies as green
        triangles; gray diamonds are the remainder of the $z>4$ sample.
        Leftward arrows mark galaxies with only an upper limit on the
        Balmer-line SFR, retained here at their limit value. Open gray
        diamonds are objects whose break is measured to worse than
        $\sigma_R = 0.4$ and are excluded from the quoted statistics. The dark
        line is the running median of the well-measured sample, the shaded
        band spans the 16th--84th percentiles, and the dotted line marks
        $R=1$, i.e., no break.
    }
    \label{fig:BB_dMS}
\end{figure}

Figure~\ref{fig:BB_dMS} reveals a clear ordering of Balmer-break strength with
distance from the star-forming main sequence: galaxies above the SFMS
typically show weak breaks, systems on the sequence intermediate ones, and
galaxies below it pronounced breaks, most markedly the two more massive
mini-quenched objects. Analogous
trends have been reported spectroscopically, from stacked NIRSpec spectra at
$3<z<14$ in which break strength is tightly anti-correlated with sSFR
\citep{Langeroodi+2024}, and photometrically across $3.5<z<10$
\citep{Kuruvanthodi+2024}. Our measurements extend these results by resolving
the break strength for individual galaxies as a function of \emph{distance}
from the SFMS, in a sample that is less biased against faint systems with
suppressed star formation.

We quantify this ordering, and all other trends in this paper, with the Spearman rank
correlation coefficient $\rho$, which tests for a monotonic relation without
assuming a functional form and is less sensitive to outliers. To test whether a
correlation survives at fixed stellar mass, we compute the partial rank
correlation $\rho\,|\,M_\star$: all three quantities are replaced by their
ranks, the ranks of the two correlated quantities are each regressed linearly
on the rank of $M_\star$, and $\rho\,|\,M_\star$ is the Pearson correlation
coefficient of the two sets of residuals. It therefore measures the
association that remains once the part of each quantity that tracks stellar
mass has been removed. Of the 54 galaxies in the $z>4$ analysis sample, we restrict the
statistics to the 48 measured to better than $\sigma_R < 0.4$
(Sect.~\ref{sec:balmerbreak}). Break strength is anti-correlated with distance
from the main sequence, $\rho = -0.58$ ($p = 2\times10^{-5}$), and positively
correlated with stellar mass, $\rho = +0.47$ ($p = 7\times10^{-4}$). These two
are not independent, since the more massive galaxies in our sample lie
preferentially closer to or below the sequence. Partial rank correlations
separate them: the dependence on $\Delta_{\mathrm{MS}}$ survives controlling
for stellar mass ($\rho = -0.38$, $p = 0.009$), whereas the mass dependence
does not survive controlling for $\Delta_{\mathrm{MS}}$ ($\rho = +0.08$,
$p = 0.58$). Distance from the main sequence is therefore the primary
variable, and the apparent mass trend is largely induced by the correlation
between stellar mass and star-formation state. The break strength does not rise smoothly with
distance below the sequence, however: it increases from the galaxies above the SFMS to the nearly-MS galaxies (median $R=1.21$ and 1.44), but is the same for nearly-MS and lulling galaxies (1.44
and 1.40) and is higher only for the mini-quenched ones. Below the sequence alone the
correlation is therefore not significant ($\rho=-0.13$), whereas the light-weighted age still
orders the galaxies there ($\rho=-0.59$, $p=0.003$; Sect.~\ref{sec:ages}). Galaxies with only an upper
limit on the Balmer-line SFR are retained at their limit value.

The restriction to well-measured breaks is necessary because, particularly at
the low-mass end, the measurements are subject to substantial uncertainties
due to the intrinsic faintness of these systems: across our $z>4$ sample
the median formal uncertainty on the break strength rises from $0.02$ for
galaxies with $\log(M_\star/M_\odot)>9$ to $0.36$ below
$\log(M_\star/M_\odot)=8$, more than twice the median break excess
($R-1 = 0.22$) of that bin. Including the poorly measured objects weakens both
correlations above, to $\rho = -0.47$ ($p = 4\times10^{-4}$) and $+0.24$
($p = 0.08$) respectively (the latter no longer significant), as expected if the additional scatter is
observational rather than physical.

This ordering is the expected consequence of ongoing star formation
continuously replenishing young stars, whose light enhances the flux
immediately blueward of the Balmer break discontinuity and fills it in
\citep[e.g.,][]{Vikaeus+2024}; at the highest masses, however, the accumulated
intermediate-age population is large enough that a measurable break persists
even in actively star-forming systems. 

Shallower spectroscopic JWST surveys have predominantly captured only the tip of the iceberg, especially at the low-mass end \citep[e.g.,][]{Rinaldi+2022,Rinaldi+2025}: highly star-forming galaxies selected either because of their intrinsic brightness, which reflects their elevated star-formation state at the epoch of observation \citep{Sun+2023}, or through the prioritization of the highest-redshift sources \citep{curtis-lake+2023, arrabal-haro+2023}. Such galaxies occupy precisely the
region of the SFR--$M_\star$ plane where breaks are weakest and evolved
stellar populations are hardest to identify, while the suppressed population,
in which the breaks are strongest, has until now been largely absent from such
samples.

This behavior places important constraints on the duty cycle of star formation. The transition is not instantaneous: a Balmer break emerges only
once the O/early-B stars have faded and the late-B--F populations responsible for it come
to dominate the continuum, and it is erased again once renewed star formation
produces massive stars that outshine them. \citet{Trussler+2025} estimate that
$\sim40$~Myr are required to build a $\sim0.5$~mag break following a starburst,
corresponding to a break strength of $R\simeq1.6$ in our definition
(Sect.~\ref{sec:balmerbreak}), whereas the nebular emission tracing the most
recent star formation fades within $\sim10$~Myr
\citep{kennicutt_star_2012,FloresVelazquez+2021}. Break strength therefore
acts as a clock, setting a lower limit on the time a galaxy has spent in its
current state: the systematic ordering of break strength with distance from
the SFMS implies that these galaxies are not simply flickering between states
on the shortest timescales. For the mini-quenched systems, the two more
massive objects show pronounced breaks ($R=1.96\pm0.18$ and
$2.07\pm0.36$), and both have light-weighted ages exceeding 100~Myr and
mass-weighted ages exceeding 250~Myr, indicating correspondingly long
intervals since their last significant episode of star formation. We quantify these timescales in Sect.~\ref{sec:ages} and in OASIS Paper~II.

We also identify two galaxies on or above the SFMS with pronounced Balmer breaks (182899 and 212674, $R\geq1.6$, against a median of 1.21 above the SFMS), potentially indicating the very start of rejuvenation following a previous period of suppressed star formation. Similar
systems have been reported at higher redshift by \citet{Kuruvanthodi+2024},
who find that roughly half of their spectroscopically confirmed Balmer-break
objects at $7.1 < z < 9.6$ show both a break and nebular emission, which they
likewise attribute to recently restarted star formation. The combination of
high SFRs with strong breaks implies that the starburst is young enough to
dominate the Balmer emission lines (which mostly arise from the reprocessed
light of stars younger than $\lesssim 10$~Myr) but has not yet come to
dominate the stellar continuum, which occurs on timescales of a few tens of
Myr \citep{FloresVelazquez+2021}. \citet{Witten+2025a} estimate that a
young population amounting to $\sim30$ per cent of the total stellar mass is
sufficient to erase a Balmer break entirely, so the persistence of a
pronounced break in these systems places a corresponding upper limit on the
mass formed in the current episode. Conversely, two galaxies well below the SFMS show weak breaks ($R<1.2$): the lulling galaxy 202438, which may have been suppressed more recently, and the faint mini-quenched galaxy 206569, whose break is not significant ($R=1.01\pm0.24$) despite light-weighted ages above 100\,Myr.

\subsection{Stellar ages} \label{sec:ages}

\begin{figure*}[t]
    \centering
\begin{subfigure}[t]{0.48\textwidth}
        \centering
        \includegraphics[width=\linewidth]{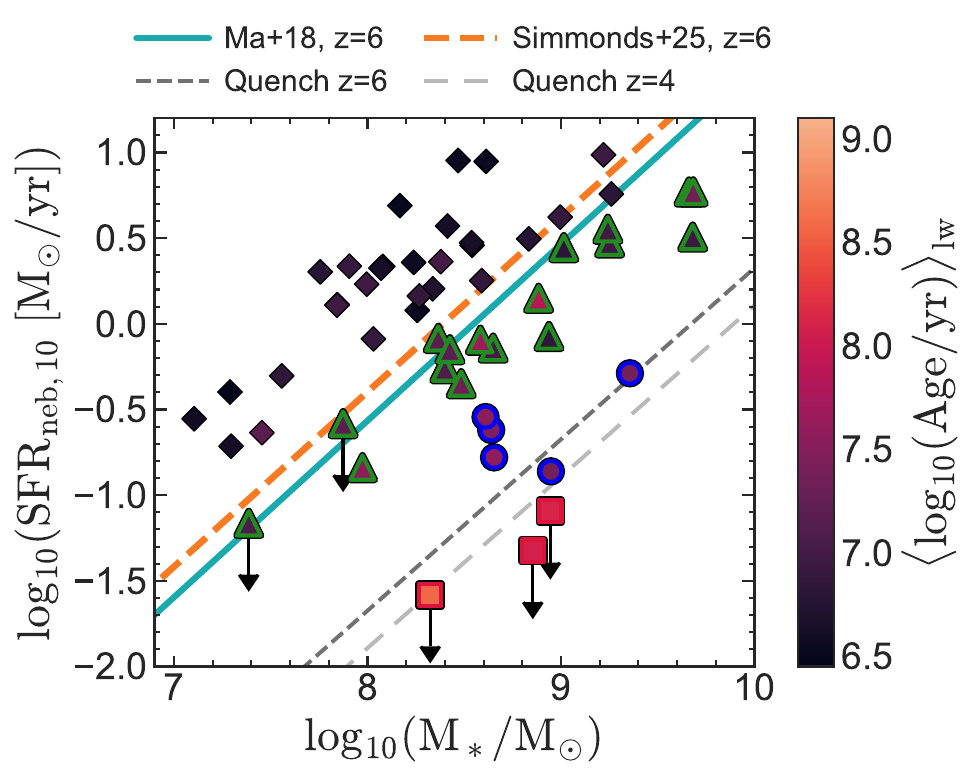}
        \caption{Light-weighted mean stellar age, which traces the
        population that dominates the continuum light.}
        \label{fig:mass_sfr_light_age}
    \end{subfigure}
    \hfill
    \begin{subfigure}[t]{0.48\textwidth}
        \centering
        \includegraphics[width=\linewidth]{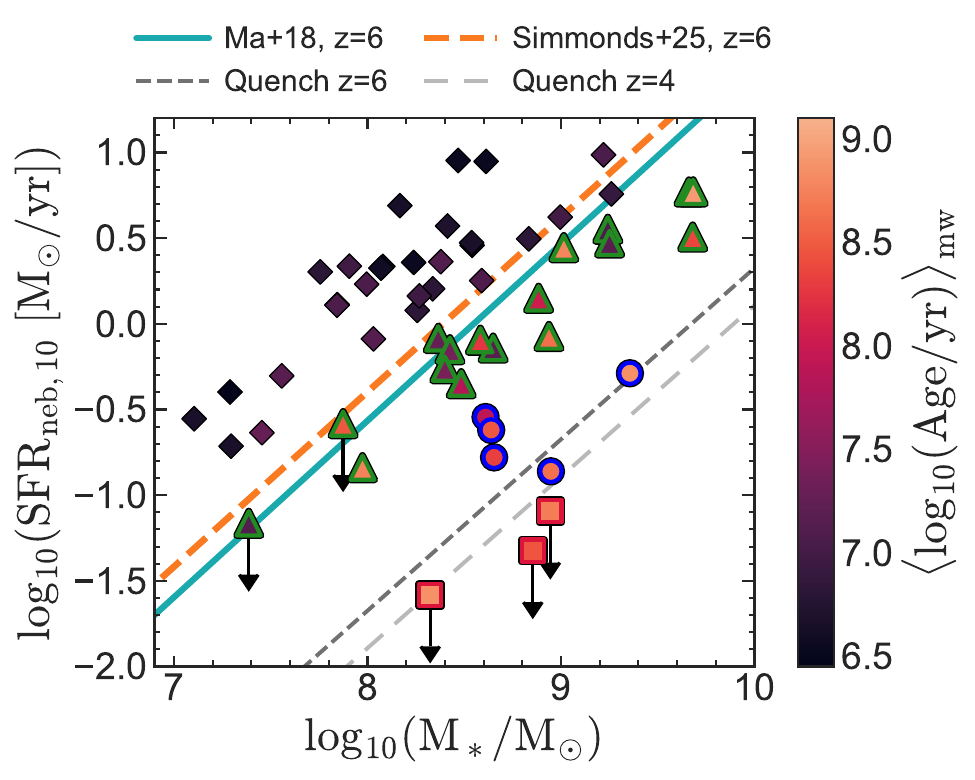}
        \caption{Mass-weighted mean stellar age, which traces the average
        build-up of stellar mass.}
        \label{fig:mass_sfr_mass_age}
    \end{subfigure}

    \caption{
        \textbf{Stellar ages rise with distance below the SFMS across the SFR--\Mstar\ plane.} Stellar ages across the SFR--\Mstar\ plane for the $z>4$ OASIS galaxies. Marker fills encode the \ppxf\ light-weighted (left)
        and mass-weighted (right) mean stellar age. The ages come from the
        stellar continuum alone, and the SFRs independently from the Balmer
        lines. Symbols follow
        Fig.~\ref{fig:SFR-mass}: red squares for mini-quenched, blue circles
        for lulling and green triangles for nearly-MS galaxies, with gray-edged
        diamonds for the remainder. Downward arrows are $3\sigma$ upper limits
        on the nebular SFR. The lines show the main sequences of
        \citet{Ma+2018} and \citet{Simmonds+2025} and the quenching threshold
        (Eq.~\eqref{eq:quenching-threshold}), evaluated at the redshifts
        indicated. The ages rise with distance below the main sequence, in both weightings. The mini-quenched galaxies are old even
        in light-weighted terms, with ages ${>}100$\,Myr.
    }
    \label{fig:mass_sfr_light_mass_age}
\end{figure*}

\begin{figure*}
    \centering
    \includegraphics[width=\textwidth]{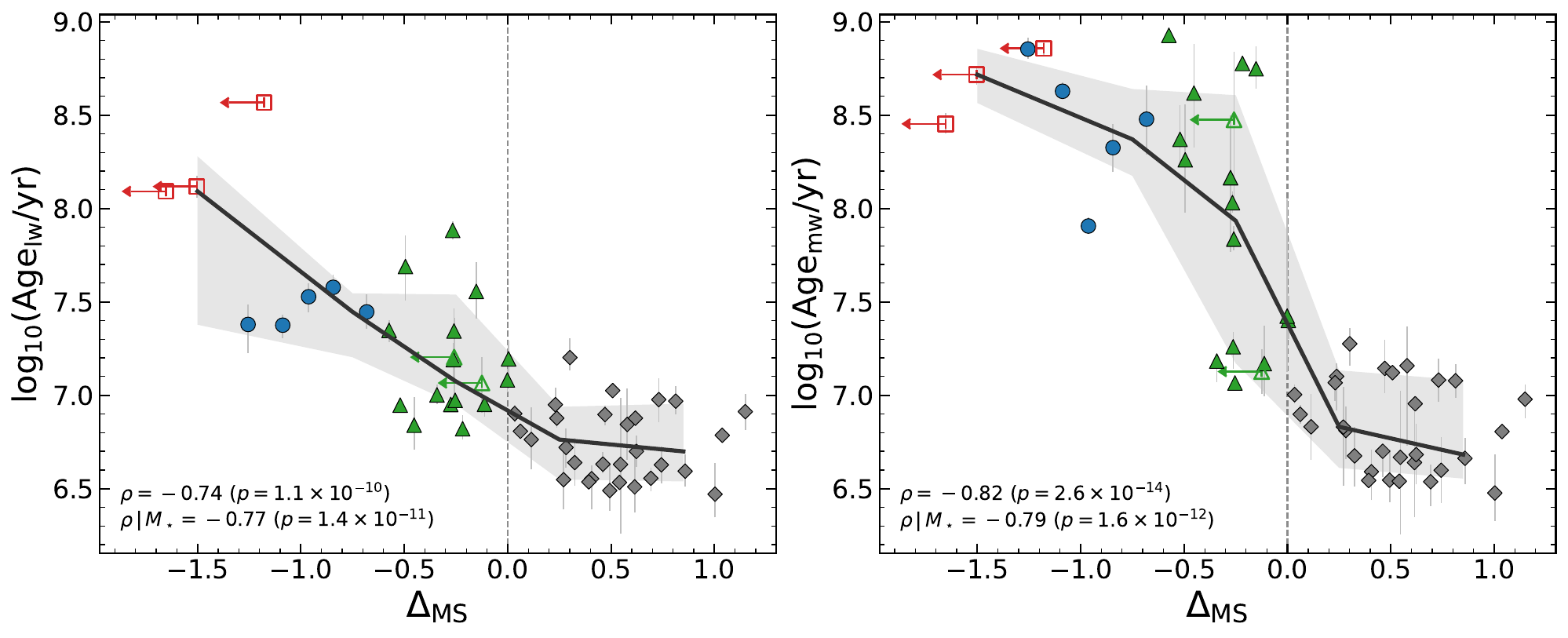}
    \caption{\textbf{Galaxies below the SFMS are older, in both light- and mass-weighted age.} Light-weighted (left) and mass-weighted (right) mean stellar ages from the \ppxf\ fits against distance from the star-forming main
    sequence, $\Delta_\mathrm{MS} = \log_{10}\mathrm{SFR} -
    \log_{10}\mathrm{SFR_{MS}}(M_\star, z)$, adopting the
    \citet{Simmonds+2025} relation. Symbols follow
    Fig.~\ref{fig:BB_dMS}: red squares for mini-quenched, blue circles for
    lulling and green triangles for nearly-MS galaxies, and gray diamonds
    for the remainder of the sample. Open symbols with leftward arrows mark
    galaxies with only an upper limit on the Balmer-line SFR. Error bars are the 16th--84th
    percentiles of the \ppxf\ bootstrap realizations. Both ages decline with increasing
    $\Delta_\mathrm{MS}$. In mass-weighted age the decline is close to a
    step at $\Delta_\mathrm{MS}\approx0$: above the SFMS, the light of
    young stars outshines the older population.}
    \label{fig:age_dMS}
\end{figure*}

Figure~\ref{fig:mass_sfr_light_mass_age} shows the SFR--$M_\star$ plane color-coded by the light- and mass-weighted mean stellar ages derived from the \ppxf fits to the stellar continuum. The two axes of
this comparison rest on different diagnostics: the ages are inferred from the
stellar continuum alone, whereas $\Delta_{\rm MS}$ is set by the
star-formation rate from the Balmer recombination lines. They share only the
stellar mass, which we control for below, and, for 15 galaxies, the
continuum-based dust correction; without these 15 the anti-correlations
persist ($\rho=-0.63$ and $-0.79$, $n=39$). 

Figure~\ref{fig:age_dMS} shows both ages directly against $\Delta_\mathrm{MS}$.
To guide the eye, we overplot the running median of
each age in bins of $\Delta_\mathrm{MS}$ with edges $-2$, $-1$, $-0.5$, $0$,
$0.5$ and $1.2$, the same as in Fig.~\ref{fig:BB_dMS}. The shaded band spans
the 16th--84th percentiles of the galaxies in each bin.

Both ages are strongly anti-correlated with $\Delta_\mathrm{MS}$, with
$\rho=-0.74$ ($p=1.1\times10^{-10}$) for the light-weighted and $\rho=-0.82$
($p=2.6\times10^{-14}$) for the mass-weighted age ($n=54$). The correlations
are essentially unchanged at fixed stellar mass ($\rho\,|\,M_\star=-0.77$
and $-0.79$) and at fixed stellar mass and redshift ($-0.78$ and $-0.81$).
They also persist without the mini-quenched and lulling galaxies
($\rho=-0.61$ and $-0.75$, $n=46$), so they are not driven by the suppressed
classes alone. From the bins above the SFMS to the lowest bin, the median
light-weighted age rises from $\sim$5 to $\sim$120\,Myr and the median
mass-weighted age from $\sim$5 to $\sim$500\,Myr. The two ages behave
differently near the SFMS. The light-weighted age rises steadily across the
whole range of $\Delta_\mathrm{MS}$. The mass-weighted median instead changes
by $\sim$1\,dex between the bins on either side of $\Delta_\mathrm{MS}=0$,
from $\sim$7 to $\sim$90\,Myr, and continues to rise more slowly below it.
This step is the signature of outshining discussed below.

The age trend also largely subsumes the Balmer-break trend of
Sect.~\ref{sec:bb_results}. At fixed light-weighted age the correlation of
the break with $\Delta_{\rm MS}$ falls from $-0.47$ to an insignificant
$-0.15$ ($p = 0.28$; for the 48 reliably measured breaks, from $-0.58$ to
$-0.31$, $p=0.03$), whereas the age correlation survives at fixed break
strength ($\rho = -0.67$). This is expected if the age draws on the full
continuum and the break on one feature of it. Galaxies below the SFMS are
therefore systematically older, and their suppressed star formation is not a
short-lived fluctuation but has left its imprint on the stellar population.

Outshining affects the two ages differently, and the difference matters
here. A young population can hide older stars that carry mass but little
light \citep[e.g.,][]{Papovich+2001, Narayanan+2024}, so mass-weighted ages
in strongly star-forming systems are lower limits. This is evident in our
sample: the galaxies above the SFMS have median light- and mass-weighted ages
of only 5 and 6\,Myr, under 1 per cent of the age of the Universe at their
redshifts. The effect is, however, far weaker in the suppressed galaxies.
Outshining depends on how strongly the young stars dominate the light, and
once star formation has declined their contribution fades, so that the
older population supplies a growing share of the continuum and is directly
constrained by the fit. The data bear this out: in the lulling galaxies
\ppxf\ recovers a mass-weighted age a full dex above the light-weighted one
(median 301 versus 28\,Myr), i.e., it detects the old population beneath
the residual young light rather than missing it, and the same galaxies
show Balmer breaks stronger than those of the galaxies above the SFMS. Both ages are therefore considerably more reliable in the suppressed galaxies than in those above the SFMS\footnote{The same holds for any other
stellar-population property derived from the continuum, such as the
star-formation histories themselves (cf.\ Sect.~\ref{sec:Mini-quenched}).}, and any residual outshining there would hide
further old mass, making the suppressed galaxies older still. The contrast
in mass-weighted age between the classes is thus, if anything, a lower
limit on the true one at the suppressed end, while above the SFMS it must be read with the outshining bias in mind. The light-weighted age,
finally, is by definition the age of the population that dominates the
light, and cannot be biased by stars the light does not reveal. The
light-weighted gradient is therefore robust against outshining, and it is
almost as strong as the mass-weighted one.

The mini-quenched galaxies are old even in light-weighted terms, with
median ages of 131\,Myr (124--369\,Myr), compared with 28\,Myr
(24--38\,Myr) for the lulling galaxies and 5\,Myr for the galaxies above the SFMS. The distinction is physically meaningful. O stars leave the
main sequence within ${\sim}10$\,Myr, so removing their light needs only a
brief pause, whereas B stars survive for 50--100\,Myr or longer. The
lulling galaxies still show Balmer emission, i.e., they still host some
ionizing O stars, and their light-weighted ages of a few tens of Myr are
consistent with B stars dominating the continuum after a lull in which
star formation was reduced but not halted; beneath this residual young
light lies the much older population traced by their mass-weighted ages.
The mini-quenched galaxies, by contrast, show no significant Balmer
emission, and a light-weighted age above 100\,Myr requires that the light
of the early- and mid-B stars, too, has faded, leaving late-B and A stars to dominate: star formation must have been suppressed
for ${\gtrsim}100$\,Myr.

Taken together, the age sequence across the SFR--$M_\star$ plane, from galaxies above the SFMS through nearly-MS and lulling to mini-quenched
galaxies, supports the interpretation that distance from the SFMS traces a
temporal sequence rather than purely stochastic fluctuations, and it
requires suppressed phases lasting tens of Myr in the lulling galaxies and
${\gtrsim}100$\,Myr in the mini-quenched ones.

\subsection{Dust attenuation} \label{sec:dust}

\begin{figure}[t]
  \centering
  \includegraphics[width=\linewidth]{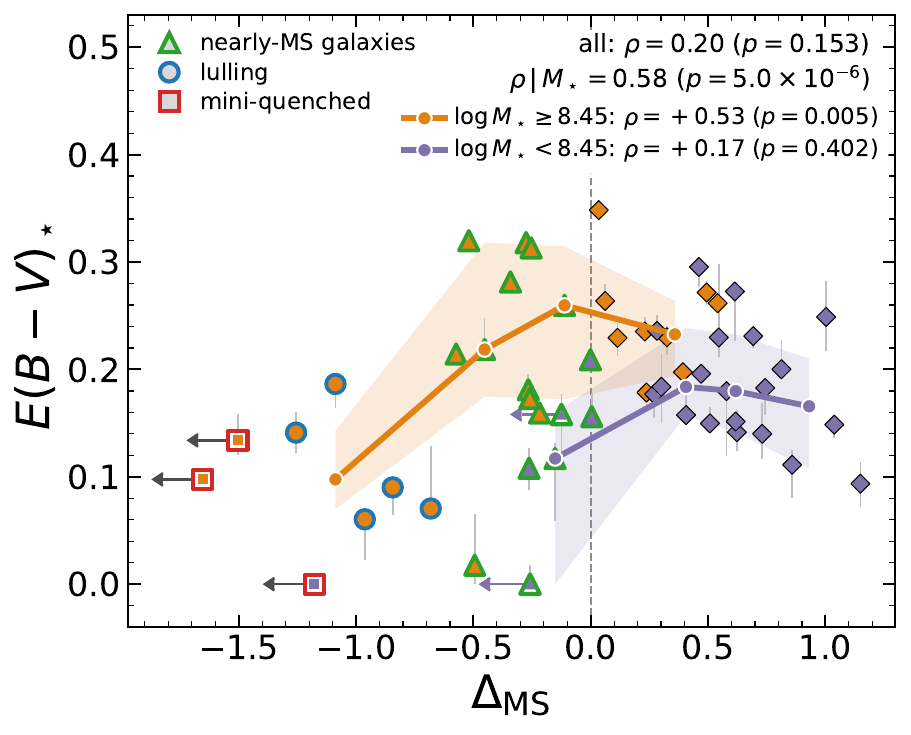}
  \caption{
    \textbf{The mini-quenched and lulling galaxies show little stellar reddening.} Stellar reddening against distance from the SFMS, split by stellar mass.
    $E(B-V)_\star$ is measured from the \ppxf\ continuum fits, with error bars giving the
    16th--84th percentiles of the bootstrap realizations. Fill color marks two stellar mass bins split at
    $\log_{10}(M_\star/M_\odot)=8.45$ (orange above, purple below); marker shape and
    edge color denote the class as in Fig.~\ref{fig:SFR-mass}. At fixed $\Delta_{\mathrm{MS}}$ the more massive
    galaxies are the more reddened, and within the high-mass sample the reddening rises
    with $\Delta_{\mathrm{MS}}$ ($\rho=+0.53$, $p=0.005$); the pooled correlation is weak
    because the suppressed galaxies lie almost exclusively in the high-mass bin.
    The lulling and mini-quenched galaxies show little reddening, consistent with,
    though not by itself evidence for, reduced gas reservoirs.
  }
  \label{fig:sfrmass-reddening}
\end{figure}

Figure~\ref{fig:sfrmass-reddening} shows the stellar reddening of the $z>4$ galaxies
against their distance from the SFMS, split in two mass bins. Three
features stand out. First, the suppressed galaxies are the least reddened: the
mini-quenched and lulling galaxies have median $E(B-V)_\star=0.10$ and $0.09$ (range
$0.00$--$0.19$), against $0.20$ for the rest of the sample (Mann--Whitney
$p=7\times10^{-4}$), whereas the nearly-MS galaxies are as reddened as the galaxies above the SFMS (median $0.18$, $p=0.67$). Second, at fixed
$\Delta_{\mathrm{MS}}$ the more massive galaxies are the more reddened, and within the
high-mass sample the reddening rises steadily with $\Delta_{\mathrm{MS}}$ ($\rho=+0.53$,
$p=0.005$). Third, pooled over all masses the trend with $\Delta_{\mathrm{MS}}$ is weak
($\rho=+0.20$), because mass and $\Delta_{\mathrm{MS}}$ are anti-correlated in our sample:
nine of the ten galaxies with $\Delta_{\mathrm{MS}}<-0.5$ lie above the median mass.

Quantitatively, holding either variable fixed lets the other emerge strongly
($\Delta_{\mathrm{MS}}$ at fixed mass, $\rho=+0.58$, $p=5\times10^{-6}$; mass at fixed
$\Delta_{\mathrm{MS}}$, $\rho=+0.61$, $p=10^{-6}$), and their combination, the SFR, is
the strongest single predictor ($\rho=+0.63$, $p=3\times10^{-7}$, $n=54$), while the
correlation with mass alone is modest ($\rho=+0.30$, $p=0.03$) and there is no
dependence on redshift. Both trends agree with previous studies
\citep[e.g.,][]{whitaker+2012, Pannella+2015, Shivaei+2020}. They are not an artifact
of the dust corrections applied to the SFRs. Of the 54 galaxies, 39 are dust corrected based on the Balmer decrement, and the remaining 15 are corrected with $E(B-V)_\star$ itself
(Sect.~\ref{sec:mass_sfr}). These corrections are small (median $0.15$\,dex): without
any, or with only the decrement-based ones, each of the three correlations changes by at most $0.06$. Nor is the low
reddening of the suppressed galaxies likely to have placed them below the sequence.
If the reddening of a galaxy is underestimated, both its SFR and its stellar mass are
underestimated, but only slightly and by similar amounts, so the galaxy moves largely
along the sequence rather than away from it. For the 39 galaxies with a measured
decrement the SFR does not depend on $E(B-V)_\star$ at all; for the 15 corrected with
$E(B-V)_\star$, SFR and mass respond to it at the same rate (1.13 and
1.14\,dex\,mag$^{-1}$), so that even an error of $0.05$\,mag shifts $\Delta_{\mathrm{MS}}$ by $<0.01$\,dex. Raising the SFRs of the
mini-quenched and lulling galaxies onto the SFMS would require nebular reddening
$0.8$--$1.9$\,mag higher than applied, beyond any decrement measured in our sample. Likewise, pushing the two most suppressed
mini-quenched galaxies (168352 and 181834) above the quenching threshold would require
$E(B-V)_\star\approx0.4$--$0.5$, more than twice the median of the galaxies above the SFMS; 206569 lies within $0.01$\,dex of the threshold, so its classification is
marginal irrespective of dust. The reddening itself is well determined, with a typical bootstrap
uncertainty of $0.01$, and it is uncorrelated with the Balmer-break strength
($\rho=-0.09$), so dust does not drive the trend of Sect.~\ref{sec:bb_results}.

The lower reddening of the mini-quenched and lulling galaxies below the SFMS is the
most interesting of these results, but also the hardest to interpret. These galaxies
are also the oldest, $0.67$\,dex older than the rest of the sample
(Sect.~\ref{sec:ages}), and across the sample low reddening goes with old ages
($\rho=-0.52$ between $E(B-V)_\star$ and light-weighted age). The question is whether
this is due, in part or entirely, to the well-known age--dust degeneracy, or whether
it has a physical meaning. Within the fits, \ppxf\ trades the two at a rate of
$0.11$\,mag per dex of light-weighted age, which alone could account for $\simeq0.07$\,mag
of the $0.10$\,mag deficit of the suppressed galaxies. Indeed, at fixed light-weighted
age they are no less reddened than the other galaxies ($p=0.59$). The same holds for
the trend with $\Delta_{\mathrm{MS}}$, which vanishes at fixed mass, redshift and age
(partial Pearson $r=+0.02$): the galaxies below the SFMS are both older and less
reddened, and the continuum fits alone cannot separate the two. Balmer decrements
provide the only independent test, and they are available for only part of the
suppressed population. Three lulling galaxies (108487, 309077 and 120014) have
decrements consistent with negligible nebular reddening. A fourth, 202438, is the
exception: H$\alpha$ is detected but H$\beta$ is not, implying $E(B-V)_{\rm neb}\gtrsim0.9$,
although [N\,\textsc{ii}] blended with H$\alpha$ or a broad component could also produce this (OASIS Paper~II). The mini-quenched galaxies
and 114769 have no usable decrement. More
generally, nebular and stellar reddening are only loosely coupled in our sample: for
the 39 galaxies with a decrement they correlate weakly ($\rho=+0.27$, $p=0.09$) and not
at all at fixed SFR ($\rho=+0.10$). The nebular reddening itself shows no trend with
$\Delta_{\mathrm{MS}}$ ($\rho=-0.03$), and at fixed nebular reddening older galaxies have
lower continuum reddening ($\rho=-0.54$). This is expected both from the fit
degeneracy and from an age-dependent attenuation in which young stars remain embedded
in their birth clouds while older populations are seen through the diffuse ISM alone
\citep{Charlot2000ApJ...539..718C}; our data cannot distinguish the two. We therefore regard the
low continuum reddening of the suppressed galaxies as consistent with, but not proof
of, a low dust content: it is confirmed by the Balmer lines in three lulling
galaxies, while for the others it cannot be separated from their older stellar
populations.

Whether the low reddening of the suppressed galaxies implies a low gas content, and what
this means for the mechanism of suppression, is discussed in OASIS Paper~II.

\subsection{SFMS scatter and number density}\label{sec:duty_cycle}

The results presented here bear directly on theoretical models of bursty
star formation, which predict markedly different fractions of galaxies at
different locations relative to the SFMS, and different stellar-population
properties for them. The observed spread is large: at $z>4$ our galaxies
span ${\sim}2.8$\,dex in $\Delta_\mathrm{MS}$, from at least 1.65\,dex below
the SFMS to 1.1\,dex above it, with a normalized median absolute deviation
of 0.63\,dex. Only 35 per cent of the galaxies lie within $\pm0.3$\,dex of
the SFMS, while 41 per cent lie more than 0.3\,dex above it and 24 per
cent more than 0.3\,dex below it. Galaxies at these redshifts therefore appear to spend a substantial fraction of their time in
elevated or suppressed states rather than evolving smoothly along the main
sequence.

The observed fractions must be read with the OASIS selection function in mind, which acts in opposite directions on the suppressed population. The hand-selected tiers (Pri 1--3) favor
mini-quenched and lulling candidates. Conversely, at fixed mass these
galaxies are fainter than star-forming ones, and are also more extended
(OASIS Paper~III), which reduces the flux captured by the
shutter. Their redshifts must moreover be secured from the continuum rather
than from emission lines.

Of the 54 galaxies at $z>4$, 15 per cent are lulling or
mini-quenched, and 6 per cent lie below the quenching threshold. All eight
suppressed galaxies lie at $\log(M_\star/\mathrm{M_\odot})=8$--9.5, where
they make up 21 per cent of the sample (8 per cent mini-quenched; 68 per
cent binomial intervals 14--30 and 4--15 per cent, respectively). These
fractions are not driven by the hand-selected tiers: in the priority tiers
that carry no information on star-formation state (Pri\,$\geq$\,4), the
suppressed fraction in this mass range is 19 per cent (5 of 27). Because
these tiers are flux-limited, and redshifts are more easily secured for
galaxies with emission lines, the observed fractions are more likely to
underestimate the suppressed population than to overestimate it. The
mini-quenched galaxy 206569, at $\log(M_\star/\mathrm{M_\odot})=8.3$ but
$m_\mathrm{F444W}=29.6$, entered the sample only through the filler tier
(Pri\,8).

In the tiers without selection on star-formation state (Pri\,$\geq$\,4), the five suppressed galaxies make up 29 per cent of the galaxies at $z\leq5.5$, the range of the $z=4$ and 5 mass functions used below, and $10^{8}$--$10^{9.5}$\,M$_\odot$ (5 of 17; 68 per cent interval 17--44 per cent). Combined with the stellar mass function of \citet{Weibel+2024}, this corresponds to a number density of ${\sim}1.6\times10^{-3}$\,Mpc$^{-3}$\,dex$^{-1}$ (68 per cent binomial interval 0.9--2.4$\times10^{-3}$; 0.6--2.6$\times10^{-3}$ including the uncertainty of the mass-function normalization and the choice of redshift and mass range) at $z\approx4$--5. Because only the fraction is taken from our ${\sim}24$\,arcmin$^2$ footprint, while the normalization comes from a much larger area, cosmic variance largely cancels, unless the suppressed fraction itself depends on environment, which the isolation of our galaxies does not suggest. The selection effects described above act in one direction: at fixed mass, suppressed galaxies are fainter and lack strong emission lines, so the flux limit of these tiers and the difficulty of securing their redshifts both work against them, and two of the five entered only through the filler tier. The estimate is therefore more likely a lower limit than an overestimate, and, given the small number of galaxies, it should be read as an order of magnitude.

In the same redshift and mass range, \citet{Khullar+2026} find 5--15 per cent of galaxies in a ``napping'' phase, with weak H$\alpha$ and a Balmer break, in medium-band photometry of the Abell~2744 cluster, and a number density of 4--5$\times10^{-4}$\,Mpc$^{-3}$ at $z=4$--5 over $10^{7.5}$--$10^{10}$\,M$_\odot$, i.e., ${\sim}2\times10^{-4}$\,Mpc$^{-3}$\,dex$^{-1}$. Applying their criteria to our spectra gives 8 per cent, consistent with their 5--15 per cent, but a number density about three times higher, $7\times10^{-4}$\,Mpc$^{-3}$\,dex$^{-1}$ (68 per cent interval 3--15$\times10^{-4}$), possibly because of the different mass ranges and mass functions adopted; our suppressed population as a whole, which also includes galaxies with residual H$\alpha$ or weaker breaks, is about 2.5 times more abundant. Densities counted directly in photometric samples without completeness corrections are lower ($10^{-5}$--$10^{-4}$\,Mpc$^{-3}$\,dex$^{-1}$ at $M_\star\sim10^{8}$--$10^{9}$\,M$_\odot$ and $z\sim6$--7; \citealt{Trussler+2025}), although their suppressed fractions, 5--35 per cent, bracket ours. 

Integrated over $10^{8}$--$10^{9.5}$\,M$_\odot$, the suppressed galaxies reach ${\sim}2\times10^{-3}$\,Mpc$^{-3}$, one to two orders of magnitude more than the massive quiescent galaxies at the same redshifts \citep[cumulative ${\sim}10^{-5}$--$10^{-4}$\,Mpc$^{-3}$ at $4<z<5$;][]{carnall+2023b, Baker+2025_quiescent, Baker+2025_quiescent_b, Merlin+2025, Zhang+2025}. Even the mini-quenched galaxies alone, which meet the same sSFR threshold as the quiescent sample of \citet{Baker+2025_quiescent_b}, although with SFRs averaged over ${\sim}10$\,Myr rather than 100\,Myr, as in most quiescent samples, reach ${\sim}1\times10^{-3}$\,Mpc$^{-3}$, nearly two orders of magnitude above their cumulative $(1$--$2)\times10^{-5}$\,Mpc$^{-3}$ at $M_\star>10^{9.5}\,\mathrm{M_\odot}$ and $4<z<5$. The contrast fits a transient phase that is common among low-mass galaxies, rather than the permanent quenching of a rare massive population. These number densities remain order-of-magnitude estimates. A full determination, corrected for completeness and purity, will be the subject of a forthcoming study of photometric samples, in which the selection is calibrated with the spectroscopic classifications presented here (Sect.~\ref{sec:phot_selection}).

\section{Discussion: Comparison with simulations}\label{sec:sims}
The fractions of Sect.~\ref{sec:duty_cycle} discriminate between cosmological simulations of bursty star formation at $z>4$, which differ both in how often galaxies are suppressed and in the masses at which this occurs. The simulations underpredict the occupancy of the suppressed regime: in
IllustrisTNG and an empirical halo model, the fraction of galaxies lying
0.75\,dex or more below the main sequence is 1--2 per cent at $z=5$, ${\lesssim}1$ per cent at $z=6$, and
2--4 per cent at $z=4$ for $M_\star=10^7$--$10^9$\,M$_\odot$, and it
decreases with stellar mass, to ${<}2$ per cent at
$10^{8.5}$--$10^{9.5}$\,M$_\odot$ \citep[][]{Dome2023}.
In OASIS, before any correction for selection, and with the caveats of Sect.~\ref{sec:duty_cycle}
in mind, 13 per cent of the $z>4$
sample and 18 per cent of the galaxies at $10^{8}$--$10^{9.5}$\,M$_\odot$
meet this criterion. Restricted to the tiers that carry no information on star-formation state
(Pri\,$\geq$\,4), the fraction at these masses is still 15 per cent (4 of 27; 68 per cent
binomial interval 8--25 per cent), an order of magnitude above the simulated values. A like-for-like comparison will require the simulated fractions and masses to be forward-modeled into observational space, including the sample biases and observational limitations, and analyzed with the same definitions and methods.

The mass of the suppressed galaxies poses a further tension. In MEGATRON, mini-quenched galaxies
occur almost exclusively at $M_\star\lesssim10^{7.5}$\,M$_\odot$ \citep{Katz+2025}. In IllustrisTNG and the empirical halo model, the suppressed fraction likewise decreases with stellar mass (see above). By
contrast, all eight mini-quenched and lulling galaxies in OASIS have
$\log(M_\star/\mathrm{M_\odot})=8.3$--9.4. Selection effects may cause us
to miss suppressed galaxies at lower masses, but they cannot account for
the presence of those we find at higher masses. Our results therefore
support the conclusion that simulations struggle to form mini-quenched
galaxies at these masses \citep{Dome2023, Katz+2025}, although the MEGATRON volume contains few such
massive galaxies at $z\geq8.5$ to begin with \citep{Katz+2025}. Our results also support the suggestion
that stronger feedback, or mechanisms beyond supernova feedback, operate in
this regime \citep{Gelli+2025}.

\section{Conclusions} \label{sec:Summary}

JWST has revealed that star formation in early galaxies is bursty, but the galaxies caught in the quiet phases of the burst cycle had been found mostly by chance. OASIS (GO~5997) was designed to find them deliberately, and we have presented the first systematically selected and spectroscopically confirmed sample of mini-quenched and lulling galaxies, at $z\sim4$--7.5. By targeting galaxies across the full burst cycle, crucially including those below the star-forming main sequence (SFMS), OASIS reduces the bias of early JWST spectroscopic programs toward UV-bright starbursts and delivers a more representative spectroscopic census of galaxies at these epochs. Of the 332 galaxies observed with NIRSpec/PRISM, 65 have secure redshifts $z\gtrsim4$, and 54 of these, at $z\approx4$--8.8 and $\log_{10}(M_\star/M_\odot)\approx7.1$--9.7, form our analysis sample. Three are mini-quenched and five lulling, i.e., 15 per cent of the sample, or 21 per cent at $\log_{10}(M_\star/M_\odot)=8$--9.5, before any correction for selection; a further 17 lie just below the SFMS and 29 above it.

Our main conclusions are as follows:

\begin{itemize}

\item \emph{Three new low-mass mini-quenched galaxies are identified.} They have $\log_{10}(M_\star/M_\odot) \sim 8.3$--$8.9$ at $z=4.0$--5.1 and substantially expand the handful of spectroscopically confirmed mini-quenched galaxies reported to date; none shows detectable Balmer emission lines, and the two more massive ones show pronounced Balmer breaks. These systems exhibit little to no ongoing star formation at the epoch of observation, demonstrating that temporary quenching episodes occur even in low-mass galaxies within the first $\sim1.5$~Gyr of cosmic history.

\item \emph{Lulling galaxies, with reduced but non-zero star formation, are more common still.} Five lulling systems display weak Balmer emission combined with significant Balmer breaks, consistent with declining star-formation activity. Despite light-weighted ages of only ${\sim}30$\,Myr, \ppxf\ recovers mass-weighted ages of ${\sim}300$\,Myr (medians), revealing an old population beneath the residual star formation.

\item \emph{Finding them requires spectroscopy.} Photometric selection can misjudge the degree of suppression: seven of the ten visually selected candidates turned out to be nearly-MS or above the SFMS, whereas five of the eight suppressed (lulling or mini-quenched) galaxies came from tiers without selection on star-formation state. Prospector fits to the photometry underestimate the suppression by 0.54\,dex for galaxies more than 0.5\,dex below the SFMS, and by 0.53\,dex even at the spectroscopic redshift. Photometry alone also cannot distinguish lulling from mini-quenched galaxies (Sect.~\ref{sec:phot_selection} and Appendix~\ref{app:phot_spec}).

\item \emph{They are common.} In the tiers without selection on star-formation state, the suppressed galaxies make up 29 per cent of the galaxies at $z\leq5.5$ and $\log_{10}(M_\star/M_\odot)=8$--9.5, corresponding to a number density of ${\sim}2\times10^{-3}$\,Mpc$^{-3}$, one to two orders of magnitude above that of massive quiescent galaxies at the same redshifts.

\item \emph{They have been suppressed long enough to build up Balmer breaks.} A strong anti-correlation is found between Balmer-break strength and offset from the main sequence ($\rho=-0.58$ for the reliably measured breaks). Galaxies below the SFMS generally exhibit stronger Balmer breaks, including the four strongest reliably measured ones, whereas those above it are dominated by young massive stars that outshine older populations. This implies that galaxies must remain sufficiently long in suppressed phases to build up Balmer breaks, tracing intermediate-age stellar populations, and sufficiently long in burst phases to dilute them.

\item \emph{They are old.} The stellar ages rise steeply with distance below the SFMS ($\rho=-0.74$ and $-0.82$ for the light- and mass-weighted ages), also at fixed stellar mass and redshift. The mini-quenched galaxies are old even in light-weighted terms (${>}100$\,Myr), which requires star formation to have been suppressed for ${\gtrsim}100$\,Myr.

\item \emph{They are dust-poor.} Mini-quenched and lulling galaxies show little to no stellar reddening ($E(B-V)_\star\lesssim0.2$), in contrast to the galaxies above the SFMS, although part of this deficit may reflect the age--dust degeneracy of the continuum fits. If it reflects a genuinely low dust content, it is consistent with depleted gas reservoirs (OASIS Paper~II).

\item \emph{Simulations underpredict them.} Suppressed galaxies are an order of magnitude more common than in simulations: 18 per cent of our galaxies at $\log_{10}(M_\star/M_\odot)=8$--9.5 lie 0.75\,dex or more below the SFMS, the criterion used for the simulations, against ${<}2$ per cent in IllustrisTNG and an empirical halo model. All eight suppressed galaxies have $\log_{10}(M_\star/M_\odot)=8.3$--9.4, higher than the masses at which most simulations produce suppressed galaxies (Sect.~\ref{sec:sims}).

\item \emph{Bursty star formation is ubiquitous.} The coexistence of galaxies above the SFMS with nearly-MS, lulling, and mini-quenched galaxies and their broad scatter across the SFR--$M_\star$ plane (${\sim}2.8$\,dex in $\Delta_\mathrm{MS}$, with only 35 per cent of the galaxies within $\pm0.3$\,dex of the SFMS) within a single spectroscopic sample designed to cover the full burst cycle demonstrates that stochastic star formation is ubiquitous at $z>4$.

\end{itemize}

Together, these results establish that low-mass galaxies in the early Universe undergo frequent fluctuations in star-formation activity, including, probably temporary, suppression below our detection limits. The suppressed galaxies are common, have remained in this phase for tens to more than 100\,Myr, as their stellar populations show, and lie at masses where most simulations produce few such systems. The quiet phases of galaxy evolution therefore appear not as rare exceptions, but as an integral component of early stellar mass assembly, and galaxies observed in fading or recently faded states expose the fossil record of burst-driven growth. 

Because predictions diverge most strongly in the suppressed regime, galaxies below the SFMS provide particularly strong observational leverage for distinguishing between competing feedback prescriptions in cosmological simulations. Constraining how long galaxies remain below the SFMS, how rapidly they return to it, how efficiently feedback evacuates their gas, and how quickly gas is re-accreted will ultimately determine the power spectrum of star-formation variability at cosmic dawn \citep{caplar+tacchella2019,Wan+2024}. In this sense, the suppressed systems presented here do not merely complement the luminous starbursts discovered by JWST; they provide a missing piece required to transform burstiness from a qualitative paradigm into a quantitatively testable theory of early galaxy evolution. OASIS Paper~II takes the next step, measuring how fast star formation declines in these galaxies and why. Our Cycle~5 program LULLs (GO~10149) will then test the mechanisms behind the suppression with deep medium-resolution spectroscopy, resolving the weak and blended emission lines and the gas kinematics beyond the reach of the PRISM.

\begin{acknowledgements}
TJL gratefully acknowledges support from the Swiss National Science Foundation through a SNSF Mobility Fellowship and from the NASA/JWST Program OASIS (PID \#5997). FDE acknowledges support by the Science and Technology Facilities Council (STFC), by the ERC through Advanced Grant 695671 ``QUENCH'', and by the UKRI Frontier Research grant RISEandFALL.
WMB gratefully acknowledges support from DARK via the DARK fellowship. This work was supported by a research grant (VIL54489) from VILLUM FONDEN.
SA acknowledges grants PID2021-127718NB-I00 and PID2025-169195NB-I00 funded by the Spanish Ministry of Science and Innovation/State Agency of Research (MICIN/AEI/10.13039/501100011033).
AJB and JC acknowledge funding from the ``FirstGalaxies'' Advanced Grant from the European Research Council (ERC) under the European Union's Horizon 2020 research and innovation programme (Grant agreement No.\ 789056).
SC acknowledges support by the European Union's HE ERC Starting Grant No.\ 101040227 -- WINGS.
QD acknowledges support from a PhD studentship awarded by Trinity College, University of Cambridge.
JMH acknowledges support from the Evolving Universe Fellowship, which is made possible by a generous donation from Dr.\ Keiko Miwa Ross; JMH also acknowledges support from JWST Program \#8544.
YI is supported by JSPS KAKENHI Grant No.\ 24KJ0202.
BER and BDJ acknowledge support from the JWST/NIRCam Science Team contract to the University of Arizona, NAS5-02105, and JWST Program 3215.
PR acknowledges support from the University of Texas at Austin Cosmic Frontier Center.
ST acknowledges support by the Royal Society Research Grant G125142.
JAAT acknowledges support from the Simons Foundation and JWST programs 2959 and 3215. Support for programs 2959 and 3215 was provided by NASA through a grant from the Space Telescope Science Institute, which is operated by the Association of Universities for Research in Astronomy, Inc., under NASA contract NAS 5-03127.
H\"U acknowledges support by the Max Planck Society through the Lise Meitner Excellence Program. H\"U acknowledges funding by the European Union (ERC APEX, 101164796). Views and opinions expressed are however those of the authors only and do not necessarily reflect those of the European Union or the European Research Council Executive Agency. Neither the European Union nor the granting authority can be held responsible for them.
JW gratefully acknowledges support from the Cosmic Dawn Center through the DAWN Fellowship. The Cosmic Dawn Center (DAWN) is funded by the Danish National Research Foundation under grant No.\ 140.
The authors acknowledge use of the lux supercomputer at UC Santa Cruz, funded by NSF MRI grant AST 1828315.
This work is based on observations made with the NASA/ESA/CSA \textit{James Webb Space Telescope}. The data were obtained from the Mikulski Archive for Space Telescopes at the Space Telescope Science Institute, which is operated by the Association of Universities for Research in Astronomy, Inc., under NASA contract NAS 5-03127 for \jwst. These observations are associated with program GO~5997. Target selection used JADES NIRCam imaging.
\end{acknowledgements}

\bibliographystyle{aa}
\bibliography{Bib_TL}

\begin{appendix}

\section{Faint low-SFR galaxies and a lulling galaxy at $z=2.9$}
\label{appendix:faint_low_sfr}

\begin{figure}[h!]
  \centering
  \includegraphics[width=\linewidth]{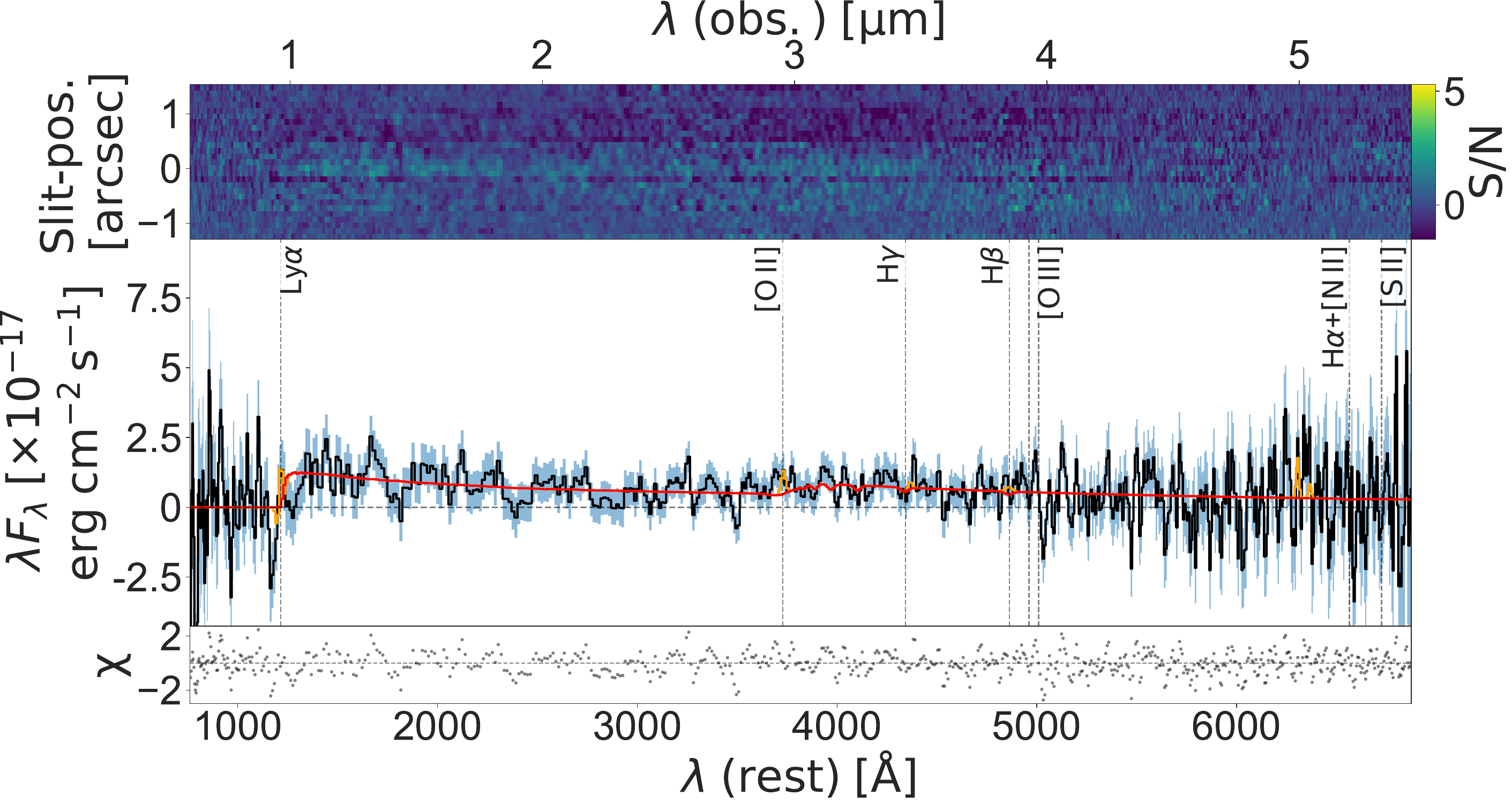}
  \caption{NIRSpec/PRISM spectrum of ID~216135 ($z=6.9$). No emission line
  is significantly detected; see the text for more details.}
  \label{fig:mq-216135}
\end{figure}

\begin{figure}[t]
  \centering
  \includegraphics[width=\linewidth]{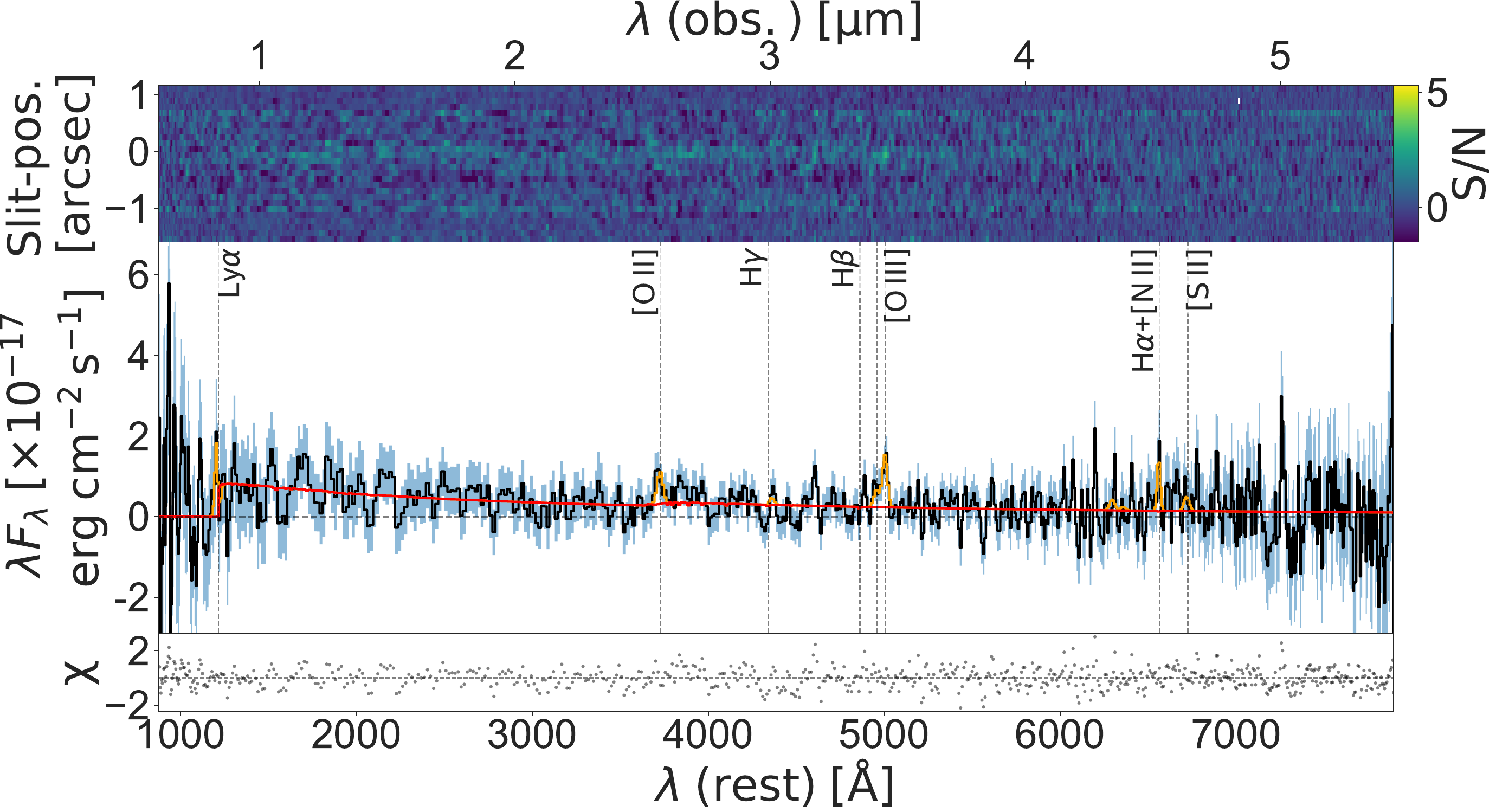}
  \caption{NIRSpec/PRISM spectrum of ID~112587 ($z=5.9$).
  [O\,\textsc{iii}]\,$\lambda5007$ is detected at $5\sigma$, whereas neither
  H$\alpha$ nor H$\beta$ is significant. The resulting SFR upper limit
  remains consistent with the main sequence, precluding a mini-quenched
  classification.}
  \label{fig:mq-112587}
\end{figure}

\begin{figure}[t]
  \centering
  \includegraphics[width=\linewidth]{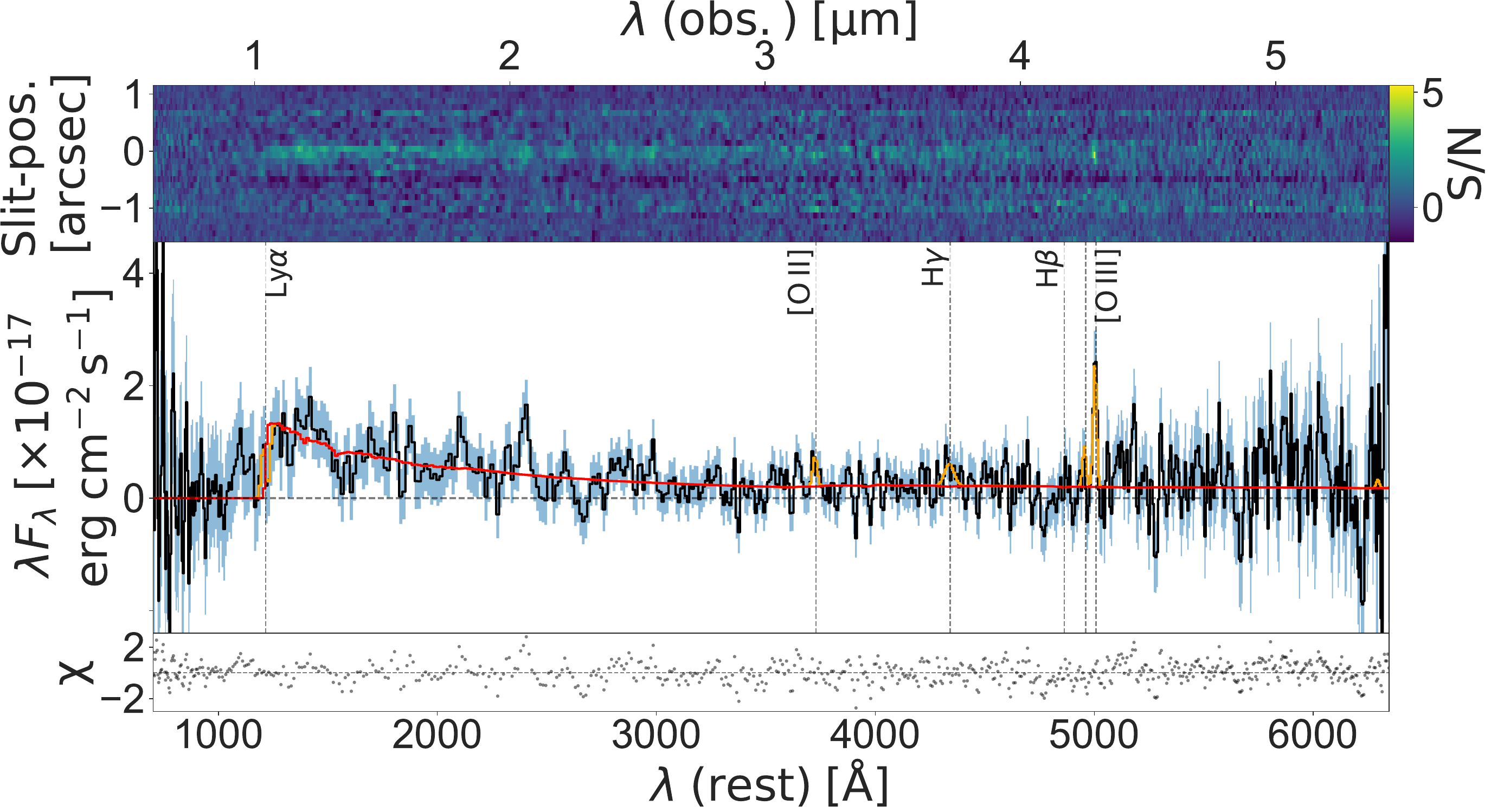}
  \caption{NIRSpec/PRISM spectrum of ID~214417 ($z=7.6$).
  [O\,\textsc{iii}]\,$\lambda5007$ is detected at $6\sigma$, but H$\beta$ is
  not, and H$\alpha$ lies beyond the PRISM wavelength range at this
  redshift. As for 112587, the resulting SFR upper limit is too shallow to
  classify the galaxy as quenched.}
  \label{fig:mq-214417}
\end{figure}

This appendix presents three $z>4$ galaxies without significant Balmer-line
emission whose star-formation state cannot be established securely
(Figs.~\ref{fig:mq-216135}--\ref{fig:mq-214417}), and one lulling galaxy at
lower redshift. All three $z>4$ systems are faint ($m_{\rm F444W} =
28.9$--$30.6$), low-mass and at $z = 5.9$--$7.6$, where the Balmer lines
are hard to constrain: two of them show [O\,\textsc{iii}]\,$\lambda5007$
but no Balmer line, and the third shows no emission line at all.

The absence of Balmer lines is not by itself sufficient to classify a
galaxy as mini-quenched. That class requires the SFR upper limit to fall
below the quenching threshold of Eq.~\eqref{eq:quenching-threshold}
(Sect.~\ref{sec:intro}), and for none of these three galaxies is the limit
deep enough. For 112587 and 214417 the limits, $\Delta_{\rm MS} < -0.12$
and $< -0.26$, remain consistent with the main sequence; both formally fall
in the nearly-MS class, which they are assigned in the analysis of
Sect.~\ref{sec:pop_plane}. For 216135 the limit places the galaxy at least
0.7\,dex below the main sequence, in the lulling regime, but up to
0.58\,dex above the quenching threshold, so it may be either lulling or
quenched. We therefore do not assign these galaxies to the mini-quenched
class. Their key properties are summarized in Table~\ref{tab:oasis_faint_low_sfr}.

Uncertainties are as described in Sect.~\ref{sec:Mini-quenched}. The one
exception is the stellar mass of 216135, whose continuum is too faint for a
reliable \ppxf\ fit: for this galaxy we adopt the photometric \prospector\
mass and its 16th--84th percentile range \citep{Simmonds+2025}, noting that
it was fitted at the photometric redshift ($z_{\rm phot}=7.4$) rather than
the spectroscopic one ($z=6.9$).

\begin{table*}
    \centering
    \caption{OASIS faint low-SFR galaxies at $z>4$ and the lower-redshift
    lulling galaxy 306637: coordinates (RA, Dec; deg), F444W magnitudes
    ($m_{\mathrm{F444W}}$; AB), target priorities (Pri), redshifts ($z$),
    stellar masses ($\log_{10}(M_\star/M_\odot)$), and nebular star-formation
    rates ($\log_{10}(\mathrm{SFR}/[M_\odot\,\mathrm{yr}^{-1}])$), with
    $3\sigma$ upper limits marked by $<$. Uncertainties are statistical
    16th--84th percentiles (see text). $^{a}$\,\prospector\ stellar mass.}
    \label{tab:oasis_faint_low_sfr}
    \renewcommand{\arraystretch}{1.4}
    \begin{tabular}{lllcccll}
        \hline\hline
        ID & RA & Dec & $m_{\mathrm{F444W}}$ & Pri & $z$ & $M_\star$ & SFR \\
        \hline
        \multicolumn{8}{c}{\textbf{Faint low-SFR galaxies}}\\
        \hline
        112587 & 53.19885 & $-27.79920$ & 30.58 & 8 & 5.9 & $7.39^{+0.11}_{-0.15}$ & $<-1.16$ \\
        214417 & 53.19411 & $-27.75927$ & 29.78 & 8 & 7.6 & $7.87^{+0.21}_{-0.47}$ & $<-0.58$ \\
        216135$^{a}$ & 53.16073 & $-27.75334$ & 28.94 & 2 & 6.9 & $8.36^{+0.14}_{-0.17}$ & $<-0.65$ \\
        \hline
        \multicolumn{8}{c}{\textbf{Lower-redshift lulling galaxy}}\\
        \hline
        306637 & 53.20409 & $-27.81345$ & 28.60 & 8 & 2.9 & $8.27^{+0.05}_{-0.06}$ & $-1.44^{+0.12}_{-0.17}$ \\
        \hline
    \end{tabular}
\end{table*}

\begin{figure}
    \centering
    \includegraphics[width=\linewidth]{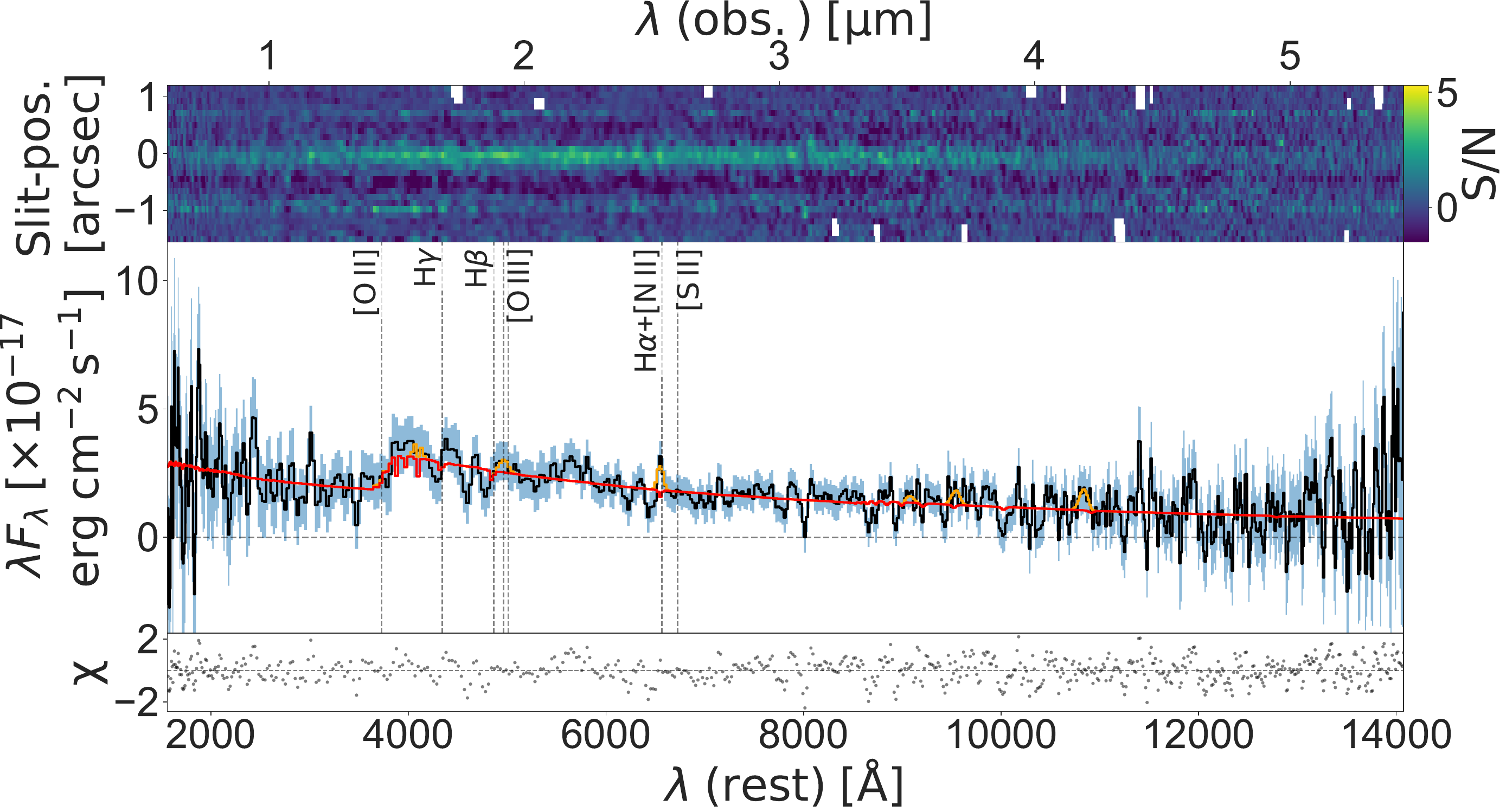}
    \caption{NIRSpec/PRISM spectrum of ID~306637, a faint lulling galaxy at
    $z=2.9$, below the redshift range of this paper
    ($m_{\rm F444W}=28.6$, $\log_{10}(M_\star/M_\odot)=8.3$). \Halpha+\nii\
    is only weakly detected ($3\sigma$), giving
    $\log_{10}(\mathrm{SFR}/[M_\odot\,\mathrm{yr}^{-1}])=-1.4$, 0.7\,dex below
    the main sequence and 0.35\,dex above the quenching threshold at its
    redshift. The spectrum shows a tentative Balmer break
    ($R = 1.66 \pm 0.44$). The galaxy was a Pri~8 filler target, selected on
    no criterion related to its star-formation state.}
    \label{fig:lulling-306637}
\end{figure}

Figure~\ref{fig:lulling-306637} shows a further lulling galaxy, 306637 at
$z=2.9$. Although it lies below the redshift range of this paper, it is
worth reporting, since it shows that low-mass galaxies with suppressed star
formation are not confined to $z>4$ but are found at lower redshift too,
once spectra are deep enough to detect their weak nebular emission, as in
OASIS. Its SFR uncertainty in Table~\ref{tab:oasis_faint_low_sfr} is
statistical, propagated from the H$\alpha$+[N\,\textsc{ii}] flux error.
The nebular dust correction adds a systematic $+0.18/{-0.06}$\,dex for the
range of $f_{\rm neb}$ considered in Sect.~\ref{sec:mass_sfr}, and any
[N\,\textsc{ii}] contribution to the blend would lower the SFR further.

\section{The limits of photometry for suppressed galaxies}\label{app:phot_spec}

\begin{figure}
    \centering
    \includegraphics[width=\linewidth]{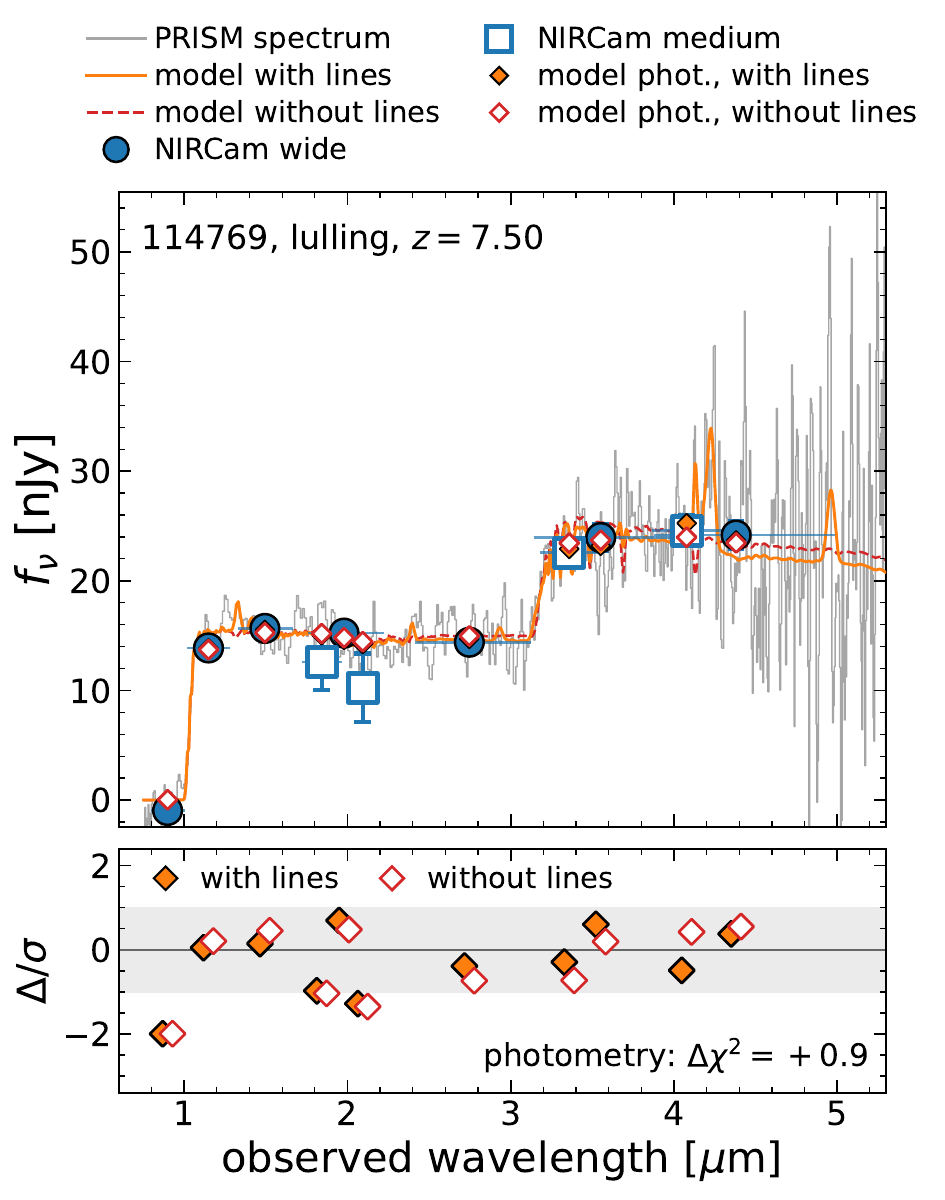}
    \caption{Photometry and spectroscopy of the lulling galaxy 114769. Top: NIRSpec/PRISM spectrum (gray)
    and JADES DR5 NIRCam photometry (circles: wide, squares: medium bands; horizontal bars: band FWHM;
    errors include a 5 per cent floor),
    with the \ppxf\ model including its emission lines (orange) and the same model without them (red
    dashed), each fitted to the photometry with a free tilt $A\,(\lambda/3\,\mu\mathrm{m})^b$; diamonds
    give their band-integrated fluxes. The spectrum is shown with the normalization and tilt fitted for the model with lines. Bottom: photometric residuals, $\Delta/\sigma$ = (photometry $-$ model)/$\sigma$, of both models, which the photometry
    cannot distinguish ($\Delta\chi^2=0.9$), whereas the spectrum prefers the model with lines by
    $\Delta\chi^2=29$.}
    \label{fig:phot_spec_114769}
\end{figure}

\begin{figure}
    \centering
    \includegraphics[width=\linewidth]{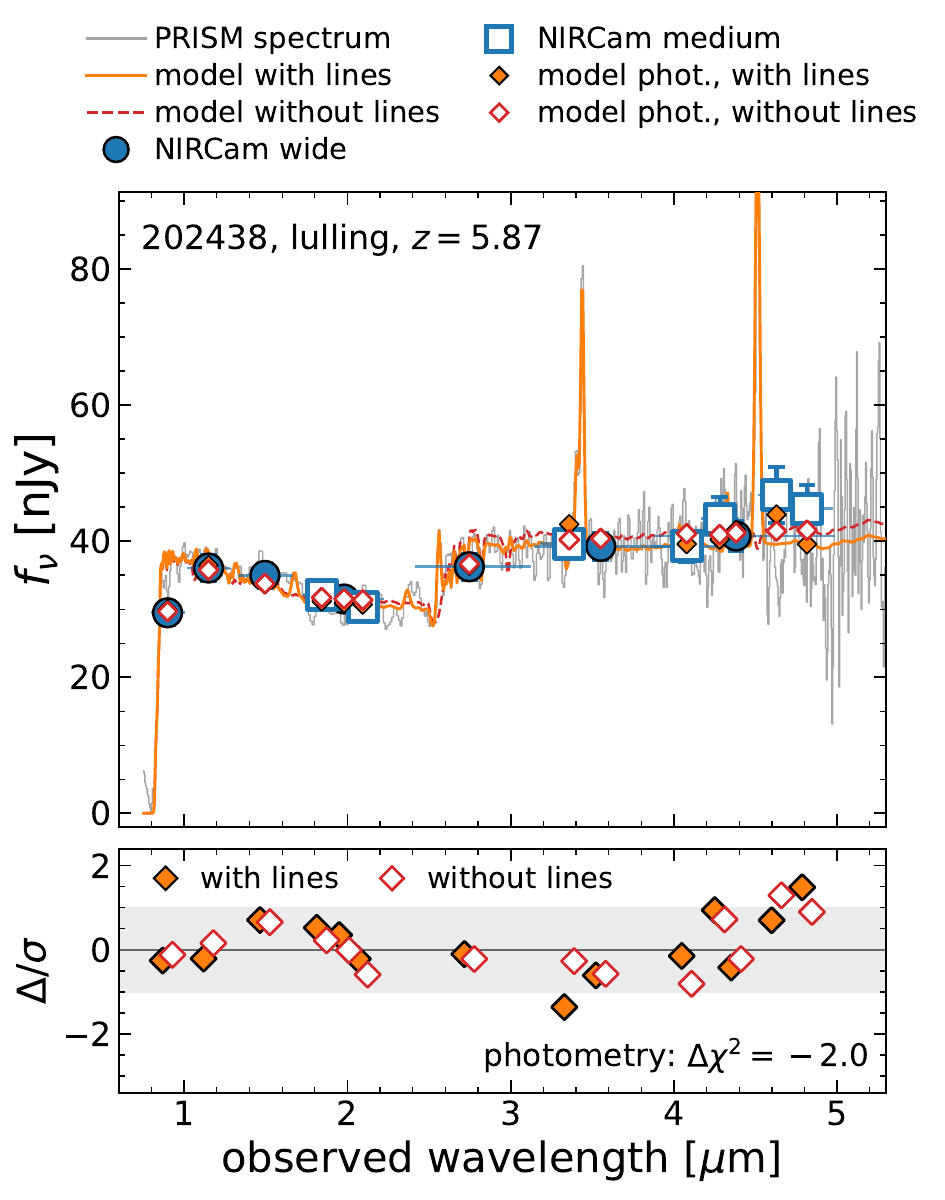}
    \caption{As Fig.~\ref{fig:phot_spec_114769}, for the lulling galaxy 202438 (fourteen NIRCam bands, seven of
    them medium bands). The photometry slightly prefers the model without lines
    ($\Delta\chi^2=-2.0$), whereas the spectrum prefers the model with lines by $\Delta\chi^2=573$.}
    \label{fig:phot_spec_202438}
\end{figure}

Figures~\ref{fig:phot_spec_114769} and \ref{fig:phot_spec_202438} show why photometry cannot establish the current star-formation state of the lulling galaxies 114769 and 202438 (Sect.~\ref{sec:phot_selection}). We compare the \ppxf\ fit to the spectrum, including its emission lines, with the same model without them. We fit both to the JADES DR5 NIRCam photometry \citep{Robertson+2026_DR5} with a free normalization, which absorbs the different apertures, and a free power-law tilt, which absorbs the wavelength-dependent slit losses, equivalent to a first-order calibration polynomial \citep{Carnall2019MNRAS.490..417C, Johnson2021ApJS..254...22J}.

For 114769, the two models reproduce the eleven bands equally well ($\Delta\chi^2=\chi^2_{\rm without}-\chi^2_{\rm with}=0.9$), although four of them are medium bands and F410M contains H$\beta$ and [O\,\textsc{iii}] at full throughput, whereas the spectrum prefers the model with lines by $\Delta\chi^2=29$ over the spectral pixels around the lines (rest-frame 4800--5060 and 6470--6660\,\AA). 202438 is covered by seven medium bands, among them F335M, which contains [O\,\textsc{iii}] at full throughput, yet the photometry even slightly prefers the model without lines ($\Delta\chi^2=-2.0$), whereas the spectrum, dominated by the strong H$\alpha$+[N\,\textsc{ii}] line, prefers the model with lines by $\Delta\chi^2=573$. The photometry thus cannot tell these lulling galaxies from mini-quenched ones with the same continuum, which the spectra separate clearly. Conversely, adding the lines of the lulling galaxies, at the same equivalent widths, to the models of the mini-quenched galaxies 168352 and 181834 changes their photometric $\chi^2$ by only $-0.3$ to $+6.3$, but their spectroscopic $\chi^2$ by up to ${\approx}300$.

This test is conservative: the photometry is given the continuum and the redshift from the spectrum, and only a smooth tilt is allowed to mimic older stellar populations or more dust, whereas a photometric SED fit must infer the redshift, star-formation history, age and dust from the photometry alone and therefore has considerably more freedom to trade a modest line excess against the shape of the continuum. This agrees with \citet{Khullar+2026}, whose medium-band fits recover the spectroscopic star-formation histories of only half of their spectroscopically confirmed `napping' galaxies. Photometric estimates of lines and recent star-formation histories may become robust only with considerably denser medium-band coverage than the two to eight medium bands available for our galaxies.

\end{appendix}
\end{document}